\documentclass{aa}  

\usepackage{rotating, graphicx}
\usepackage{amsmath}	% Advanced maths commands
\usepackage{amssymb}	% Extra maths symbols
\usepackage{txfonts}
\usepackage{natbib}
\usepackage[normalem]{ulem} % To strike out text
\usepackage[dvipsnames]{xcolor}
\definecolor{grey}{gray}{0.7}
\definecolor{darkgreen}{rgb}{0.0,0.7,0.0}
\usepackage[nointegrals]{wasysym}
\usepackage[colorlinks=true,linkcolor=blue,citecolor=blue]{hyperref}
\usepackage{longtable,booktabs}
\usepackage{etoolbox}
\makeatletter
\patchcmd\@combinedblfloats{\box\@outputbox}{\unvbox\@outputbox}{}{%
   \errmessage{\noexpand\@combinedblfloats could not be patched}%
}%
 \makeatother
 
\makeatletter
\newlength\oriarrayrulewidth  
\newcount\orilowpenalty
\newcommand\nobreakmidrule{%
 \noalign{\global\oriarrayrulewidth\arrayrulewidth\relax
          \global\orilowpenalty\@lowpenalty\relax  
          \global\@lowpenalty=\numexpr-10000\relax%
          \global\arrayrulewidth\lightrulewidth\relax}
 \hline
 \noalign{\global\@lowpenalty=\orilowpenalty\relax%
          \global\arrayrulewidth\oriarrayrulewidth\relax}}
\makeatother

\newcommand{\eg}[0]{$\textnormal{e.g. }$}
\newcommand{\ie}[0]{$\textnormal{i.e. }$}
\newcommand{\Msun}[0]{\,\textnormal{M}_{\textnormal{\astrosun}}}
\newcommand{\asun}[0]{_{\textnormal{\astrosun}}}

\newcommand{\tn}[1]{\textnormal{#1}}
\newcommand{\sub}[1]{_{\textnormal{#1}}}
\newcommand{\bs}[1]{\boldsymbol{#1}}
\newcommand{\error}[1]{\tn{\scriptsize{$\pm #1$}}}
\newcommand{\Te}[0]{$T\sub{e}$}
\newcommand{\TOII}[0]{$T\sub{e}(\textnormal{\textsc{Oii}})$}
\newcommand{\TOIII}[0]{$T\sub{e}(\textnormal{\textsc{Oiii}})$}
\newcommand{\TSIII}[0]{$T\sub{e}(\textnormal{\textsc{Siii}})$}
\newcommand{\TNII}[0]{$T\sub{e}(\textnormal{\textsc{Nii}})$}
\newcommand{\ZTe}[0]{$Z_{\textnormal{Te}}$}

\newcommand{\OII}[0]{$[\textnormal{\textsc{Oii}}]$}
\newcommand{\OIII}[0]{$[\textnormal{\textsc{Oiii}}]$}
\newcommand{\SII}[0]{$[\textnormal{\textsc{Sii}}]$}
\newcommand{\SIII}[0]{$[\textnormal{\textsc{Siii}}]$}
\newcommand{\NII}[0]{$[\textnormal{\textsc{Nii}}]$}
\newcommand{\ROII}[0]{R(\textsc{Oii})}
\newcommand{\ROIII}[0]{R(\textsc{Oiii})}

\newcommand{\RNII}[0]{R(\textsc{Nii})}
\newcommand{\OpH}[0]{$\tn{O}^{+}\hspace{-0.03in}/\tn{H}^{+}$}
\newcommand{\OppH}[0]{$\tn{O}^{++}\hspace{-0.03in}/\tn{H}^{+}$}
\newcommand{\Op}[0]{$\tn{O}^{+}$} %\hspace{-0.03in}
\newcommand{\Opp}[0]{$\tn{O}^{++}$}
\newcommand{\HII}[0]{H\textsc{ii}}

\newcommand{\NumMaNGAs}[0]{12}

\newcommand{\NumMaNGABlobs}[0]{23}

\newcommand{\NumLowZGals}[0]{118}
\newcommand{\NumMSGals}[0]{55}
\newcommand{\NumSDSSNSAGals}[0]{41}
\newcommand{\NumLowZHIISystems}[0]{264}
\newcommand{\NumDirectHIISystems}[0]{130}
\newcommand{\NumIndirectHIISystems}[0]{134}
\newcommand{\MaxRedshift}[0]{0.25}

\begin{document}

   \title{Present-day mass-metallicity relation for galaxies\\ using a new electron temperature method}

    \author{R. M.~Yates\inst{1,2}\thanks{E-mail: robyates@mpa-garching.mpg.de}, P.~Schady\inst{1,3}, T.-W.~Chen\inst{1}, T.~Schweyer\inst{1,4}, P.~Wiseman\inst{1}}

   \institute{Max-Planck-Institut f{\"u}r Extraterrestrische             Physik, Giessenbachstra\ss{}e 1, 85748,                    Garching, Germany
         \and
             Max-Planck-Institut f{\"u}r Astrophysik, Karl-Schwarzschild-Stra\ss{}e 1, 85741, Garching, Germany
         \and
             Department of Physics, University of Bath, Bath, BA2 7AY, United Kingdom
         \and
             Department of Astronomy, The Oskar Klein Center, Stockholm University, AlbaNova, 10691 Stockholm, Sweden
             }

   \date{Received September 15, 1996; accepted March 16, 1997}
 
  \abstract
   {} %leave it empty if necessary  
   {We investigate electron temperature (\Te{}) and gas-phase oxygen abundance (\ZTe{}) measurements for galaxies in the local Universe ($z<\MaxRedshift{}$). Our sample comprises spectra from a total of \NumLowZHIISystems{} emission-line systems, ranging from individual \HII{} regions to whole galaxies, including \NumMaNGABlobs{} composite \HII{} regions from `star-forming main sequence' galaxies in the MaNGA survey.}
   {We utilise \NumDirectHIISystems{} of these systems with directly measurable \TOII{} to calibrate a new metallicity-dependent \TOIII{} -- \TOII{} relation that provides a better representation of our varied dataset than existing relations from the literature. We also provide an alternative \TOIII{} -- \TNII{} calibration. This new \Te{} method is then used to obtain accurate \ZTe{} estimates and form the mass -- metallicity relation (MZR) for a sample of \NumLowZGals{} local galaxies.}
   {We find that all the \TOIII{} -- \TOII{} relations considered here systematically under-estimate \ZTe{} for low-ionisation systems by up to 0.6 dex. We determine that this is due to such systems having an intrinsically higher \Op{} abundance than \Opp{} abundance, rendering \ZTe{} estimates based only on \OIII{} lines inaccurate. We therefore provide an empirical correction based on strong emission lines to account for this bias when using our new \TOIII{} -- \TOII{} and \TOIII{} -- \TNII{} relations. This allows for accurate metallicities ($1\sigma = 0.08$ dex) to be derived for any low-redshift system with an \OIII{}$\lambda$4363 detection, regardless of its physical size or ionisation state. The MZR formed from our dataset is in very good agreement with those formed from direct measurements of metal recombination lines and blue supergiant absorption lines, in contrast to most other \Te{}-based and strong-line-based MZRs. Our new \Te{} method therefore provides an accurate and precise way of obtaining \ZTe{} for a large and diverse range of star-forming systems in the local Universe.}
   {}

               
  \titlerunning{MZR using a new \Te{} method}
  \authorrunning{Yates et al.}

   \maketitle
%
%________________________________________________________________

\section{Introduction}
   Galaxies are complicated systems. In particular, the bright \HII{} regions in their interstellar medium (ISM) are subject to a number of complex and interrelated astrophysical processes and span a range of sizes, morphologies, luminosities, temperatures, and spatial distributions (\eg \citealt{Hodge&Kennicutt83,Kennicutt88,Osterbrock89}). This complexity is no less evident when studying the metal content of \HII{} regions. The standard diagnostics used to measure gas-phase metallicities rely on strong, collisionally-excited nebular emission lines which are sensitive not only to metallicity but also a host of other phenomena, such as nebular excitation, shocks, ionsing radiation field strength, gas pressure, electron density, temperature structure, dust content, diffuse ionised gas (DIG) contamination, and the N/O abundance ratio (\eg \citealt{Brinchmann+08,Stasinska10,Kewley+13b,Shirazi+14,Steidel+14,Kruehler+17,Sanders+17,Strom+17,Pilyugin+18}). Even when only considering the local Universe, some strong-line diagnostics are known to be prone to both metallicity-dependent (\eg \citealt{Kewley&Dopita02,Erb+06a,Yates+12,Andrews&Martini13}) and scale-dependent \citep{Kruehler+17} biases. This all means that the true chemical composition of the ISM in nearby galaxies is still not well understood, and warrants continued investigation.

The most direct way to measure the metallicity, or more precisely, the oxygen abundance $Z = 12+\tn{log(O/H)}$, in \HII{} regions is via metal recombination lines (RLs). This method allows for an estimate of the abundance of a given ionic species just from measurements of the relevant RLs (\eg \OII{}$\lambda$4651 and H$\beta$, for \Opp{}) and their effective recombination coefficients, without significant concern for gas temperatures or the properties of the ionising sources (\eg \citealt{Osterbrock89,Peimbert+93,Esteban+02}). However, optical metal RLs are extremely weak and require high-resolution spectroscopy of very nearby systems to be detected (\eg \citealt{Esteban+09,Esteban+14}). Alternatively, the absorption lines measured in the photospheres of individual blue supergiant stars have been used to obtain similarly direct estimates of gas-phase metallicities (\eg \citealt{Urbaneja+05,Kudritzki+08}). This method relies on the fact that blue supergiants are relatively bright ($M\sub{V} \sim -9.5$ mag, \citealt{Bresolin03}) and relatively young ($\sim 10$ Myr old), making their chemical composition both measureable and a fair representation of the recently star-forming gas. This method is, however, also limited to the very nearby Universe currently ($z \lesssim 0.0025$, \ie{} $\lesssim 10$ Mpc). Although, in the era of E-ELT, the use of brighter red supergiant stars \citep{Lardo+15,Davies+17} will push the method out to $z \sim 0.025$, and super star clusters \citep{Gazak+14} out to $z \sim 0.2$.

A more practical, yet still relatively direct, alternative is the electron-temperature (\Te{}) method, which relies on measurements of collisionally-excited auroral lines such as \OIII{}$\lambda$4363 (\eg \citealt{Peimbert67,Osterbrock89,Bresolin+09b}). This method works because the \Te{} of the line-emitting gas is strongly anti-correlated with its metallicity, due to the important role metal ions play in radiative cooling. However, \Te{}-based metallicities themselves suffer from certain limitations. For example, the auroral lines required are still relatively weak, making measurements difficult to accomplish currently in both more distant ($z \gtrsim 0.3$) and more metal-rich ($Z \gtrsim 0.5\,Z\asun$) systems (\eg \citealt{Bresolin08}). Additionally, at high metallicities ($Z \gtrsim Z\asun$), saturation of the \OIII{}$\lambda$4363 line as well as temperature gradients and fluctuations within the \HII{} regions are expected to bias \Te{} measurements high (and therefore abundance estimates low) \citep{Stasinska78a,Stasinska05,Kewley&Ellison08}. Moreover, \Te{} studies rely on simplified multi-zone models for the temperature structure in ionised nebulae, and often lack information on the temperature of the \Op{} zone, due to the required auroral emission line doublet at \OII{}$\lambda\lambda$7320,7330 being either too weak for detection or beyond the wavelength coverage of the spectrograph used. This means that \OpH{}, which can be a significant fraction of the total oxygen abundance (see \S \ref{sec:Accuracy of TT relations}), has to be inferred either from a close proxy temperature such as \TNII{} or from empirical relations linking \TOII{} to \TOIII{}. This renders such \Te{}-based metallicity estimates more `semi-direct' than direct.

Therefore, in this work we compile a dataset of \NumDirectHIISystems{} low-redshift individual and composite \HII{} regions for which both \TOIII{} and \TOII{} are directly measured. These systems are used to investigate the \TOIII{} -- \TOII{} relation, and derive a new empirical calibration which accounts for the apparent over-estimates in \TOII{} at low \Opp{}/\Op{} that we find. This new relation is then used to obtain accurate measurements of \ZTe{} for a further \NumIndirectHIISystems{} systems, providing a new insight into the mass -- metallicity relation (MZR) of local galaxies.

This paper is organised as follows: In \S{}\ref{sec:MaNGA sample}, we outline the new MaNGA sample utilised in this work. In \S{}\ref{sec:Sample selection effects}, we assess the possible biases present in our dataset, given its heterogeneous selection. In \S{}\ref{sec:Derived properties}, we describe how stellar masses, electron temperatures, and oxygen abundances are obtained, as well as how dust corrections are uniformly applied across our dataset. In \S{}\ref{sec:TT_relation}, we present our analysis of the \TOIII{} -- \TOII{} relation, including a new metallicity-dependent calibration and an empirical correction for low-\Opp{}/\Op{} systems. In \S{}\ref{sec:Local MZR}, we present the MZR, and compare it with those formed from other direct and indirect methods for obtaining metallicity. Finally, in \S{}\ref{sec:Conclusions} we provide our conclusions. In Appendix \ref{sec:Appendix A}, we investigate \OpH{} estimates based on \TNII{} via the \NII{}$\lambda 5755$ auroral line, which is a possible alternative to the \OII{} auroral line quadruplet. In Appendix \ref{sec:Appendix B}, we provide EW(H$\alpha$) maps for our MaNGA systems, as well as tables detailing their flux measurements and derived properties. In Appendix \ref{sec:Appendix C}, we provide tables containing the properties we derive from the additional literature samples considered in this work. And in Appendix \ref{sec:FittingMethods}, we describe the statistical methods used to fit the key relations presented.

In this work, we make the following distinction between the two ways \Te{}-based metallicities are obtained: `Direct \ZTe{} systems' are those for which \OppH{} and \OpH{} can be directly determined via measurements of both the \OIII{} and \OII{} auroral lines (see \S{}\ref{sec:Te_and_ZTe}). `Semi-direct \ZTe{} systems' are those for which only \OppH{} can be directly determined, and therefore require an assumed relation between \TOIII{} and \TOII{} in order to determine \OpH{} (see \S{}\ref{sec:TT_relation}).

Throughout, we assume a \citet{Chabrier03} stellar IMF, and a dimensionless Hubble parameter of $h=0.68$ as determined by the \textit{Planck} collaboration from combined CMB and lensing data \citep{Planck14}.

%__________________________________________________________________

\section{MaNGA sample}\label{sec:MaNGA sample}
Here, we outline the new MaNGA sample used in this work.

The basis of our dataset is formed of \NumMaNGAs{} galaxies selected from the Mapping Nearby Galaxies at APO (MaNGA) survey ($\bar{z} = 0.02$). MaNGA utilises integral-field units (IFUs) mounted on the Sloan Foundation 2.5m Telescope at the Apache Point Observatory to obtain fibre-based spatially-resolved spectroscopy of nearby galaxies. The MaNGA sample is taken from the NASA Sloan Atlas catalogue of the SDSS Main Galaxy Legacy Area with a spatial sampling of $\sim{}$1-2 kpc and typical resolution of $\sim{}2000$ \citep{Bundy+15}.

Our subset of targets was taken from the `MaNGA Product Launches-5' (MPL-5), which contains data for a total of 2778 galaxies observed with MaNGA that are fully reduced and vetted as of May 24, 2016 \citep{Law+16}. These data cubes are sky background subtracted using a a super-sampled sky model made from all of the sky fibers, resulting in a typical accuracy of 10\% in the skyline regions of the red camera, out to $\sim 8500$\AA, and an accuracy of $10-20$\% at longer wavelengths. This uncertainty was added in quadrature to extracted spectra in regions where skylines were present.

All galaxies are from the main MaNGA survey, as at the time of our data reduction, no systems from the `dwarf galaxies ancillary project' (Cano D\'{i}az et al., in prep.) had been observed yet. Reduced, calibrated and sky-subtracted data cubes were downloaded from the SDSS Science Archive Server. The median spatial and spectral resolution of these datacubes is 2.54 arcsec (2.60 arcsec for our MaNGA sample) at FWHM and 72 km/s, respectively \citep{Law+16}.

Galaxies were initially selected to have $\tn{log}(M_{*}/h_{68}^{-2}\Msun) < 10.0$. This upper mass limit is imposed due to the likely inaccuracy of \Te{}-based metallicities at super-solar metallicities. In metal-rich \HII{} regions, strong temperature gradients can cause the measured line-ratio temperature of a given species to deviate significantly from the mean ionic temperature to which the oxygen abundance is actually related \citep{Stasinska78a,Stasinska05,Bresolin08}. This can lead to an under-estimation of the gas-phase metallicity when using the \Te{} method by a factor of typically 2 or 3 \citep{Bresolin07}.

We then carried out two separate analyses on these \NumMaNGAs{} MaNGA galaxies. The first analysis uses global spectra of each galaxy, only considering spaxels with H$\alpha$ equivalent widths of EW(H$\alpha$) $>30$\AA{}, to minimise contamination from diffuse ionised gas (DIG) emission. The second analysis utilises the IFU capabilities of MaNGA to study distinct regions within these galaxies. We study large H$\alpha$-emitting regions which we term `\HII{} blobs' because their effective spatial resolution is in the range 0.8 to 1.4~kpc (2.3 to 2.9"), which is several times larger than that of typical \HII{} regions. We select spaxels with H$\alpha$ equivalent width (EW) of $>50$\AA{} and H$\alpha$ signal-to-noise ratio (S/N) of $>50$, identifying distinct high-EW(H$\alpha$) blobs from the resulting maps by eye (see Appendix \ref{sec:Appendix B}), and then extracting spectra from elliptical regions over each blob.

To be able to accurately measure the nebular emission lines without loss of flux from stellar absorption, we removed the stellar continuum from the total emission spectra by fitting each pixel in the MaNGA data cube with the spectral synthesis code {\sc starlight}. Subtracting the best-fit stellar component model from the total emission spectrum at each pixel then left us with the gas-only emission spectrum. Emission line fluxes were then measured using Gaussian fits, where in the case of line doublets, the line widths were constrained so that both lines in the doublet had the same velocity width.\footnote{For all line doublets, checks were also made to verify that the best-fit fluxes agreed with the expected ratios from atomic physics, which in all cases they did.} All fits were checked by eye to verify that the procedure was not fitting a Gaussian to noise. In particular, in the case of weak lines, the best-fit line peak (and thus implied redshift) and line widths were compared to the fits to stronger lines. In those cases where the line was not detected or the best-fit Gaussian was fitting noise, the fit was further constrained by fixing the line width and peak position to the best-fit parameters from fits to stronger lines of the same element. The resulting line fluxes are provided in Table \ref{tab:MaNGA_HIIblob_data} for our \HII{} blob spectra and Table \ref{tab:MaNGA_global_data} for our global spectra.

In both analyses, we limit our study to only those systems with S/N(\OIII{}$\lambda$4363) and S/N(\OII{}$\lambda\lambda$7320,7330) $\geq 3$. This criterion was applied to ensure that well-constrained \Te{}-based oxygen abundances could be obtained. For our sample of 23 \HII{} blobs, the mean S/N of the \OIII{}$\lambda$4363 line is 6.7, and for the \OII{}$\lambda\lambda$7320,7330 doublet is 23.1. All our MaNGA galaxies exhibit strong H$\alpha$ lines, with $F\sub{obs}(\tn{H}\alpha) \geq 1.0 \times{}10^{-15}\,\tn{erg}\,\tn{s}^{-1}\,\tn{cm}^{-2}$.

Our MaNGA sample provides an interesting and relatively new perspective on the low-redshift mass-metallicity relation because all its galaxies lie on or very close to the main sequence of star formation (see \S{}\ref{sec:Sample selection effects}), whereas most other studies of individual galaxies with $\tn{\ZTe{}}\gtrsim8.0$ are focused on higher-SFR systems (\eg{}\citealt{Izotov+06,Hirschauer+15,Hirschauer+18}).

\section{Selection effects}\label{sec:Sample selection effects}
Our combined dataset comprises systems of various different physical sizes, from individual \HII{} regions (\eg{}the \citealt{Bresolin+09b} sample), to composite ISM regions (\eg{}our MaNGA sample), to integrated galaxy spectra (\eg{}the \citealt{Ly+16a} sample). It is also assembled from various different studies, each with different selection criteria. Therefore, it is important that selection effects are assessed.

One concern is that our \Te{} measurements could be dependent on the physical size of the system. It has been established that the blending of emission from multiple \HII{} regions of different temperatures within one spectrum can bias estimates of \TOII{} high when inferring it from \TOIII{} measurements \citep{Kobulnicky+99,Pilyugin+10,Sanders+17}. However, we show in \S \ref{sec:Composite HII regions} that our new \TOIII{}-\TOII{} relation performs equally well for both individual and composite \HII{}-region spectra, in part because we have included a range of system sizes in our calibration sample.

Similarly, the possible inclusion of systems with contaminated \OII{}$\lambda\lambda$7320,7330 auroral lines could affect our results. We have therefore run our analysis on a number of sub-samples, including (a) only systems with high spatial resolution (as described above), (b) only systems with the theoretically expected ratio of the \OII{} auroral line doublet, $r' = \tn{\OII{}}\lambda7320/\tn{\OII{}}\lambda7330$, (c) only systems with similar \TOII{} and \TNII{} measurements, (d) using the method from \citet{Izotov+06} for obtaining \OpH{} that does not require \OII{} auroral lines, and (e) discarding \TOII{} measurements altogether and using \TNII{} measurements instead. These sub-samples are discussed in \S \ref{sec:OII line contamination} and Appendix \ref{sec:Appendix A}. We find our results hold for all of these sub-samples, indicating that our new calibration of the \TOIII{} --\TOII{} relation (see \S \ref{sec:New TT relation}) is not significantly affected by contamination of key auroral lines.

\begin{figure}
\centering
\includegraphics[angle=0,width=\linewidth]{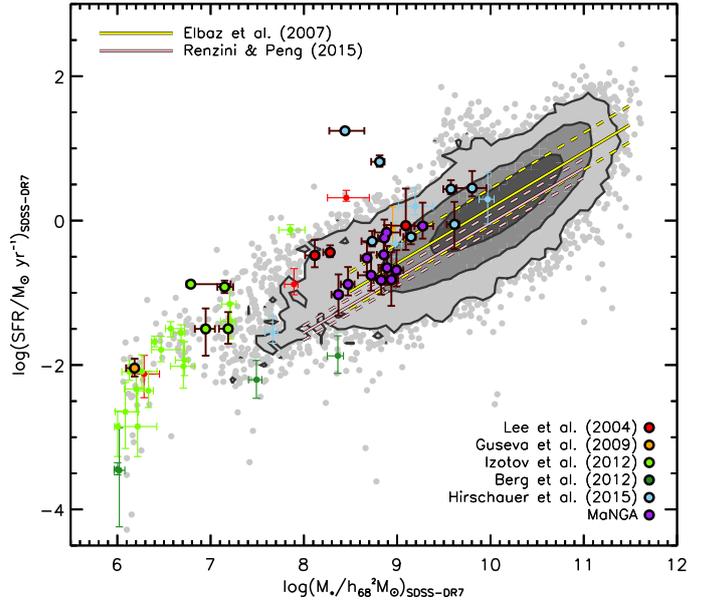}  \\
\caption{$M_{*}$-SFR relation for \NumMSGals{} galaxies in our low-redshift dataset with counterparts in the SDSS-DR7 catalogue. Local `main sequence' relations from \citet{Elbaz+07} and \citet{Renzini&Peng15} are also plotted for comparison, alongside a star-forming galaxy sample from the SDSS-DR7 (grey contours and points, \citealt{Yates+12}). Points with black outlines represent systems with direct \ZTe{} estimates, whereas those without represent those with semi-direct \ZTe{} estimates.}
\label{fig:Mstar-SFR_relations}
\end{figure}

A potential concern for our analysis of the mass-metallicity relation is that datasets such as ours, which require auroral-line detections, can be biased towards starburst galaxies. To assess this, Fig. \ref{fig:Mstar-SFR_relations} shows the $M_{*}$-SFR relation for our dataset (coloured points) alongside a typical star-forming sample of 109,678 SDSS-DR7 galaxies below $z=0.3$ from the study of \citet{Yates+12} (grey contours). To facilitate a fair comparison, only the \NumMSGals{} galaxies from our dataset with a counterpart in the SDSS-DR7 spectroscopic catalogue are considered, and their stellar masses and SFRs are taken from the SDSS-DR7 for this plot. While we do have some systems which are highly star-forming for their mass, there are also a significant number of systems that lie within the $1\sigma$ dispersion of the star-forming main sequence of \citet{Elbaz+07} below $\tn{log}(M_{*}/\Msun) = 10.0$ ($\sim{}45$\% of our SDSS-matched systems). These galaxies are predominantly from our new MaNGA sample, highlighting its important role in this analysis. We therefore conclude that our dataset is relatively representative of the low-redshift star-forming population.

Finally, we also checked that we are not preferentially selecting the lowest-metallicity regions within our MaNGA galaxies by requiring S/N(\OIII{}$\lambda$4363) $ > 3.0$. To do this, we measured the metallicity via the $R_{23}$ strong-line diagnostic for every \HII{} blob within MaNGA systems 8259-9101 and 8459-9102, in order to determine their rough, relative oxygen abundances. Both these systems contain 7 blobs each, allowing a more insightful comparison than for those systems with fewer distinct regions. We use the theoretically-derived $R_{23}$ diagnostic provided by \citet{Kewley&Dopita02} and an iterative procedure to obtain both the ionisation parameter, $U$, and $Z\sub{R23}$ (see \citealt{Kewley&Dopita02}, section 6). We find that those \HII{} blobs which make it into our \Te{}-based sample do not form a special sub-set of the lowest-metallicity regions. In fact, we are able to detect \OIII{}$\lambda$4363 in some of the most metal-rich regions in these galaxies.

\section{Derived properties} \label{sec:Derived properties}
\subsection{Stellar masses} \label{sec:Stellar masses}
The majority of the systems in our dataset have stellar masses taken either from the Sloan Digital Sky Survey MPA-JHU data release 7 (SDSS-DR7, \citealt{Abazajian+09})\footnote{wwwmpa.mpa-garching.mpg.de/SDSS/DR7/} or the NASA-Sloan Atlas v0\_1\_2 (NSA, \citealt{Blanton+11}).\footnote{nsatlas.org/}

Given that stellar masses are presented in units of $h_{68}^{-2}\Msun$ in this work, we have multiplied all SDSS-DR7 masses by the small factor $0.7^{2}/0.68^{2}=1.06$ and all NSA masses by the larger factor $1^{2}/0.68^{2}=2.16$ to maintain consistency throughout.

When stellar masses are available from both catalogues, we adopted those provided by the NSA as they do not suffer from the systematic flux under-estimation for extended systems that SDSS-DR7 stellar masses do, due to improved background subtraction \citep{Blanton+11}. This is particularly important for dwarf galaxies such as those in our study, as these tend to be nearby and therefore more extended on the sky.

The systems with the largest discrepancies in mass estimate between these two catalogues tend to have low specific SFRs ($\tn{sSFR}\equiv{}\tn{SFR}/M_{*}$). Galaxies with typical or high SFRs for their mass (such as those in our dataset) tend to have discrepancies of less than 1 order of magnitude. We have therefore corrected all SDSS-DR7 stellar masses used in this work, by fitting the $M_{*\tn{SDSS-DR7}}$-$M_{*\tn{NSA}}$ relation for the \NumSDSSNSAGals{} galaxies in our low-redshift dataset that are present in both these catalogues with a second-order polynomial given by:
\begin{equation}\label{eqn:SDSS_mass_correction}
\tn{log}(M_{*}/h_{68}^{-2}\Msun)\sub{NSA} = 2.81707 + 0.542641x + 0.0181975x^{2}\;\;,
\end{equation}
in the range $6.0<x<10.0$, where $x = \tn{log}(M_{*}/h_{68}^{-2}\Msun)\sub{SDSS-DR7}$. This yields a range of correction factors between 1.01 for high-mass systems to 1.12 for low-mass systems.

Masses for the remaining 46 per cent of systems (from NGC300, SLSN hosts, and the \citealt{Berg+12} and \citealt{Ly+16b} samples) are taken directly from the literature.

\subsection{Dust corrections} \label{sec:dust_corrections}
In order to make our dataset as homogeneous as possible, we apply exactly the same reddening corrections to all line fluxes for all systems. For those literature samples where uncorrected line fluxes are not provided, we first re-redden the fluxes by reversing the particular correction used in that work, and then correct all observed fluxes using the following two-step process. Firstly, we correct all lines for Milky Way dust extinction using the \citet{Cardelli+89} extinction law, an extinction factor of $R\sub{V} = 3.1^{+2.7}_{-1.0}$, and Milky Way reddening along the line of sight from the \citet{Schlafly&Finkbeiner11} Galactic reddening map. A fit to the wavelength-dependent uncertainty on the extinction law provided by \citet{Cardelli+89}, as well as on their measured value of $R\sub{V}$, is also folded-in to the error propagation for our dust corrections.

Secondly, we correct for internal attenuation within the host system using the \citet{Calzetti+00} attenuation law for star-forming galaxies, $k'(\lambda,R'\sub{V})$, and
\begin{equation}\label{eqn:dust_correction}
F\sub{cor}(\lambda) = F\sub{obs}(\lambda)\,10^{0.176\,E(B-V)\sub{gas}\,k'(\lambda,R'\sub{V})}\;\;,
\end{equation}
where $R'\sub{V} = 4.05\error{0.80}$, and $E(B-V)\sub{gas}$ is the colour excess of the ionised gas (see \citealt{Calzetti97b}). An intrinsic Balmer decrement of 2.86 is assumed, which is suitable for case B recombination in local star-forming galaxies with $N\sub{e} \sim{} 100\, \tn{cm}^{3}$ and $T\sub{e} \sim{} 10000$ K \citep{Osterbrock89}.

We note that this method always returns a non-zero correction to the observed fluxes, even when the internal extinction is determined to be zero, due to the ever-present Milky Way extinction to the redshifted line. Applying uniform dust reddening corrections in this way decreases slightly the scatter in the final electron temperature and oxygen abundance distributions.

\subsection{Electron temperatures and oxygen abundances}\label{sec:Te_and_ZTe}
In order to calculate electron temperatures (\Te{}) and \Te{}-based metallicities (\ZTe{}) for our dataset, we adopt the formalism developed by \citet{Nicholls+13,Nicholls+14a,Nicholls+14b}. Those works utilised the MAPPINGS IV photoionisation code \citep{Dopita+13} to fit relations between collisionally excited line (CEL) flux ratios from observed spectra to key physical properties of the gas. These relations allow a dependence on the electron density and electron energy distribution, and incorporate updated atomic data, including revised collision strengths for many ionic species. The atomic datasets used in MAPPINGS IV are listed in table 1 of \citet{Nicholls+13}. The collision strengths for the key oxygen ions used in this work are taken from \citet{Tayal07} for \Op{}, and from \citet{Aggarwal93,Lennon&Burke94,Aggarwal&Keenan99}; and \citet{Palay+12} for \Opp{} (with those for the $^{1}D_{2}$ and $^{1}S_{0}$ levels from which the $\lambda\lambda$4959,5007 and $\lambda$4363 lines originate taken from \citealt{Palay+12}).

The absolute accuracy of collision strength and transition probability estimates is very difficult to constrain, due to the complex calculations and assumptions involved in their determination. Therefore, we have included an additional error on all the electron densities, electron temperatures, and ionic abundances we calculate in this work, to account for the uncertainty in atomic data. This could be particularly important for our \TOIII{} and \OppH{} estimates, as the \Opp{} collision strengths provided by \citealt{Palay+12} are known to return electron temperatures for \HII{} regions that are up to $\sim{}600$ K lower than other Breit-Pauli R-matrix based methods from the literature (see \citealt{Storey+14,Izotov+15,Tayal&Zatsarinny17}).

We account for such atomic data discrepancies by adopting the typical uncertainties found for \HII{} regions with $N\sub{e} < 1000\ \tn{cm}^{-3}$ by \citet{JuanDeDios+17} when comparing 52 different atomic datasets from the literature, including that of \citet{Palay+12}. This leads to the following additional errors, which are propagated through all the calculations we make hereafter: $\sigma(N\sub{e}) = 0.16$ dex, $\sigma[\tn{\TNII{}}] = 0.02$ dex, $\sigma[\tn{\TOII{}}] = 0.02$ dex, $\sigma[\tn{\TOIII{}}] = 0.025$ dex, $\sigma[\tn{\OpH{}}] = 0.1$ dex, $\sigma[\tn{\OppH{}}] = 0.075$ dex. These additional uncertainties increase the typical error on our \ZTe{} estimates by 0.035 dex.

In this work, we assume thermal equilibrium in the gas (\ie a Maxwell-Boltzmann electron energy distribution), adopting the following expression for the electron temperature in Kelvin;
\begin{equation}\label{eqn:Te}
T\sub{e} = a \left[-\tn{log}\left(\frac{R\sub{obs}}{1+d\left(N\sub{e}/T^{1/2}\sub{e}\right)}\right)-b\;\right]^{-c}\;\;,
\end{equation}
where $R\sub{obs}$ is the reddening-corrected flux ratio of the particular ionic species, and $N\sub{e}$ is the electron density in cm$^{-3}$. The values of the coefficients $a$, $b$, $c$, and $d$ are also dependent on the ionic species and electron density, and have been calculated by \citeauthor{Nicholls+13} (2013, their tables 4 and 5) by fitting to the output from the MAPPINGS IV code. The flux ratios we consider in this work for $R\sub{obs}$ are:
\begin{align}\label{eqn:line_ratios}
R({\textsc{Oii}}) & =\tn{\OII}\lambda\lambda7320,7330\;/\;\tn{\OII}\lambda\lambda3726,3729\;\;, \nonumber \\
R({\textsc{Oiii}}) & =\tn{\OIII}\lambda4363\;/\;\tn{\OIII}\lambda\lambda4959,5007\;\;,\nonumber \\
R({\textsc{Nii}}) & =\tn{\NII}\lambda5755\;/\;\tn{\NII}\lambda6584\;\;,
\end{align}
where the nitrogen ratio, $\RNII{}$, is only used for a subset of systems for which the particularly weak \NII{}$\lambda5755$ auroral line was detected (see Appendix \ref{sec:Appendix A}). We then solve Eq. \ref{eqn:Te} iteratively to obtain \Te{}. Convergence is typically achieved within three iterations for \TOIII{}, and five iterations for \TOII{}.

We checked that the electron temperatures obtained from this procedure are very similar to those obtained by numerically solving the complete statistical equilibrium equations using the latest version of the \texttt{pyneb} package \citep{Luridiana+13}, which is a revised version of the \texttt{nebular/temden} routines provided by IRAF \citep{Shaw&Dufour95}. When assuming the same atomic data, we find the median difference in direct \TOIII{} is only $\sim{}40$ K for our dataset, and the median difference in direct \TOII{} is only 132 K. This indicates that analytic expressions such as Eq. \ref{eqn:Te} above (or eq. 5.4 from \citealt{OF06}) are valid in the temperature and density regime studied here.

We did not make any additional corrections to flux ratios to account for Eddington bias \citep{Eddington13} in this work, as these are believed to be negligible and depend sensitively on the instrument and type of observation made. However, we note that when applying the corrections to $\ROIII$ provided by \citet{Ly+16a} from their MMT and Keck data, the mean \ZTe{} for our sample increases by only 0.026 dex, due to the reduction in the assumed \OIII{}$\lambda4363$ line strength for systems with S/N(\OIII{}$\lambda4363$) $\lesssim{} 7$.

An estimate of $N\sub{e}$ is obtained from the ratio of [S\textsc{ii}] lines, as provided by \citet{O'Dell+13} based on the work of \citet{OF06};
\begin{equation}\label{eqn:Ne}
\tn{log}(N\sub{e}/\tn{cm}^{-3}) = 4.705 - \left(1.9875\;\frac{\tn{[S\textsc{ii}]}\lambda6716}{\tn{[S\textsc{ii}]}\lambda6731}\right)\;\;.
\end{equation}
Although \TOIII{} exhibits a negligible dependence on the electron density for the typical range of $N\sub{e}$ observed in local H\textsc{ii} regions, the MAPPINGS IV code does infer an important effect for \TOII{}, such that electron densities above $100\;\tn{cm}^{-3}$ would lead to an over-estimate in \TOII{} by up to 2000 K if not accounted for.

Following \citet{Nicholls+14a}, and making the common assumption of a constant collision strength and a uniform electron temperature in each ionic zone, we then obtain singly- and  doubly-ionised oxygen abundances as follows:
\begin{align}\label{eqn:OpH}
\tn{\OpH} \;= & \;\frac{\tn{\OII}\lambda\lambda3726,3729}{\tn{H}_{\beta}}\;g_{1}\;\alpha\sub{H$\beta$}\;\sqrt{\tn{\TOII}} \nonumber\\
 & \cdot \tn{exp}\left[E_{12}/k\tn{\TOII}\right]\;\times\;\frac{\beta}{E_{12}\Upsilon_{12}}%\;\;.
\end{align}
\begin{align}\label{eqn:OppH}
\tn{\OppH} \;= & \;\frac{\tn{\OIII}\lambda\lambda4959,5007}{\tn{H}_{\beta}}\;g_{1}\;\alpha\sub{H$\beta$}\;\sqrt{\tn{\TOIII}} \nonumber\\
 & \cdot \tn{exp}\left[E_{12}/k\tn{\TOIII}\right]\;\times\;\frac{\beta}{E_{12}\Upsilon_{12}}\;\;,
\end{align}
where $g_{1}$ is the statistical weight of the ground state ($g_{1} = 4$ for \Op{} and 9 for \Opp{}), $\alpha\sub{H$\beta$}$ is the \TOIII{}-dependent effective emissivity for H$_{\beta}$ assuming Case B recombination, $E_{12} = hc/\lambda\sub{avg}$ is the energy difference between the collisionally-excited state and the ground state (where $\lambda\sub{\tn{OII{}}} = 3727$ \AA{} and $\lambda\sub{\tn{OIII{}}} = 4997$ \AA{} are the flux-weighted average wavelengths for the \OII{} and \OIII{} lines considered here), $k$ is Boltzmann's constant, $\Upsilon_{12}$ is the temperature-dependent net effective collision strength for the transition in question, $\beta=(h^{4}/8\pi^{3}m\sub{e}^{3}k)^{-1/2}=115885.4\;\tn{K}^{-1/2}\,\tn{cm}^{-3}\,\tn{s}$ is the constant factor from the collision rate coefficient, $h$ is Planck's constant, and $m\sub{e}$ is the mass of an electron. We refer the reader to \citeauthor{Nicholls+14a} (2014a, section 4.1) for more details, including the equations used to determine $\alpha\sub{H$\beta$}$ and $\Upsilon_{12}$.

An estimate of the total oxygen abundance, \ZTe{}, can then be obtained by summing the abundances of these two ionic species;
\begin{equation}\label{eqn:direct_ZTe}
\tn{\ZTe} \equiv 12 + \tn{log}(\tn{\OpH} + \tn{\OppH})\;\;,
\end{equation}
assuming that higher and lower ionised states of oxygen have a negligible contribution (\eg \citealt{Stasinska+12}). We note here that this simple addition of the two measured oxygen abundances assumes that either the \Op{} and \Opp{} zones are co-spatial, or the amount of H$^{+}$ in each zone is the same.

For many systems, the weak \OII{}$\lambda\lambda7320,7330$ line doublet is either not detected or is redward of the wavelength range of the spectrograph used. In such cases, a direct measurement of the \OII{} temperature is not possible using Eq. \ref{eqn:Te}, and one must instead rely upon either an alternative ionic temperature (see Appendix \ref{sec:Appendix A}) or an approximation inferred from the measured \OIII{} temperature (\eg \citealt{Campbell+86,Garnett92,Pagel+92,Izotov+06,Pilyugin07,Pilyugin+09,Lopez-Sanchez+12,Andrews&Martini13,Nicholls+14a}). Such \TOIII{} -- \TOII{} relations are discussed in the following section.

\section{The \texorpdfstring{$\bs{\tn{\TOIII{}}-\tn{\TOII{}}}$}{TOIII-TOII} relation}\label{sec:TT_relation}
\begin{figure}
 \includegraphics[angle=0,width=1.0\linewidth]{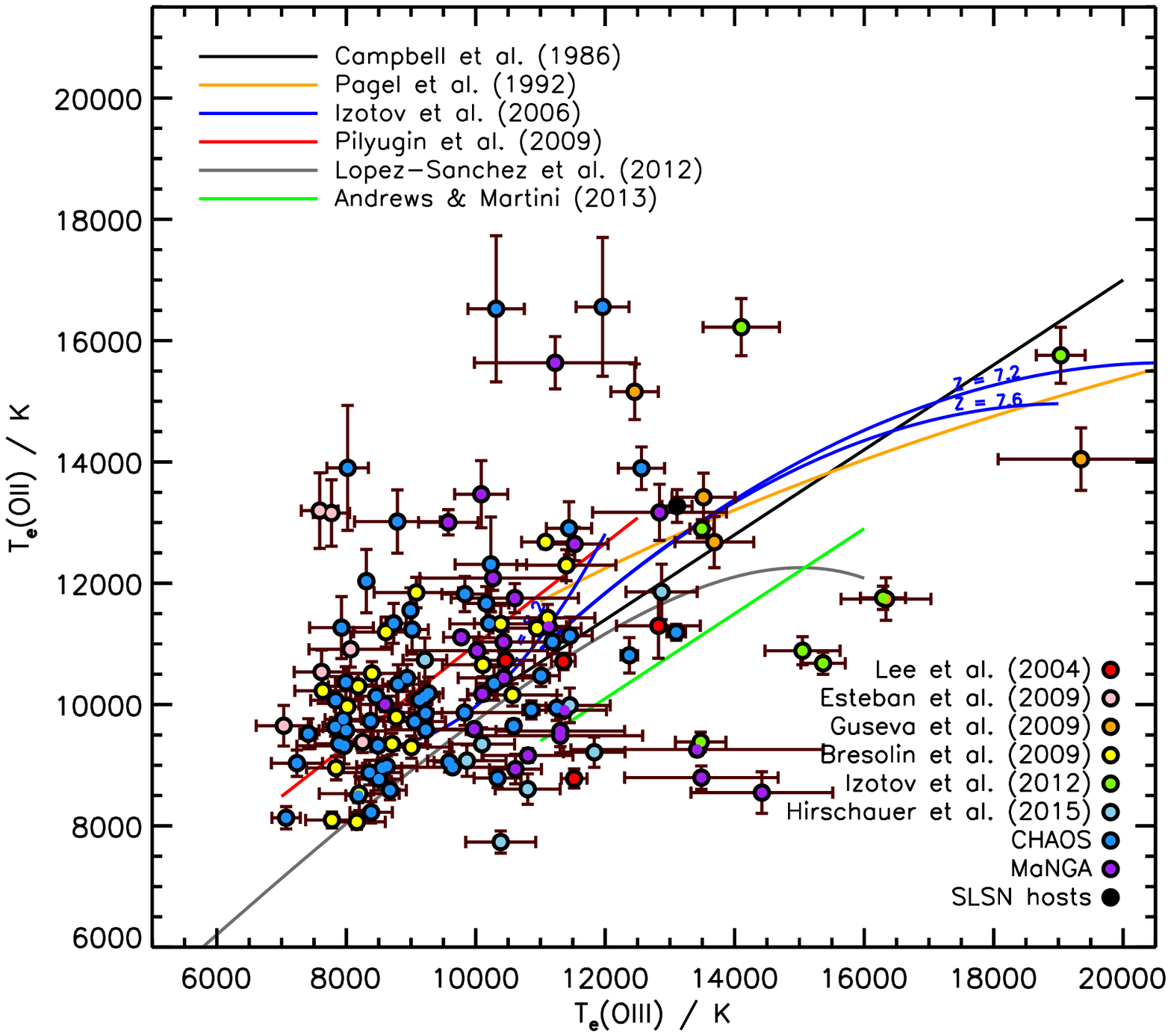}
 \caption{\TOIII{}-\TOII{} plane containing our \NumDirectHIISystems{} systems with direct \ZTe{} estimates [\ie{} those for which both \TOIII{} and \TOII{} can be calculated from auroral lines]. Six empirically- or theoretically-derived \TOIII{} -- \TOII{} relations from \citeauthor{Campbell+86} (1986, black), \citeauthor{Pagel+92} (1992, orange), \citeauthor{Izotov+06} (2006, blue, for metallicities of 7.2, 7.6, and 8.2), \citeauthor{Pilyugin+09} (2009, red), \citeauthor{Lopez-Sanchez+12} (2012, grey), and \citeauthor{Andrews&Martini13} (2013, green, as defined in \citealt{Ly+16a}), are also plotted for comparison. Our dataset reveals a much broader distribution of \TOII{}/\TOIII{} ratios than is expected from the literature relations.}
 \label{fig:TOIII-TOII_old_relations}
\end{figure}

Fig. \ref{fig:TOIII-TOII_old_relations} shows \TOIII{} versus \TOII{} for our \NumDirectHIISystems{} direct-\ZTe{} systems. Each system is coloured to indicate the base sample to which it belongs. Six empirically- or theoretically-derived \TOIII{} -- \TOII{} relations from the literature are also plotted for comparison.

The mean S/N of the auroral lines used here are S/N$(\tn{\OIII}\lambda{}4363) = 18.2$ and S/N$(\tn{\OII}\lambda{}\lambda{}7320,7330) = 22.5$. The S/N of the nitrogen auroral line for the 53 systems with reported detections is S/N$(\tn{\NII}\lambda{}5755) = 8.2$ (see Appendix \ref{sec:Appendix A}).

We first note that the distribution of our dataset across the \TOIII{} -- \TOII{} parameter space is quite broad, including a significant fraction of systems exhibiting lower \TOII{} than \TOIII{} (see also \citealt{Kennicutt+03}). Consequently, there appears to be no clear one-to-one correlation between \TOIII{} and \TOII{} in Fig. \ref{fig:TOIII-TOII_old_relations}, indicating that none of the common \TOIII{} -- \TOII{} relations plotted is particularly representative of the true range of \TOII{}/\TOIII{} ratios in this dataset. A similar conclusion can be drawn from the samples considered by \citeauthor{Izotov+06} (2006, fig. 4a) and \citeauthor{Andrews&Martini13} (2013, fig. 3), and is also discussed by \citet{Yan+18a} in reference to their \textit{Cloudy 17.00} \citep{Ferland+17} modelling.

Motivated by this issue, in the following sections we develop a new empirical \TOIII{} - \TOII{} relation, which allows for a broader range of \TOII{}/\TOIII{} ratios.

\begin{figure}
\centering
 \includegraphics[angle=0,width=0.93\linewidth]{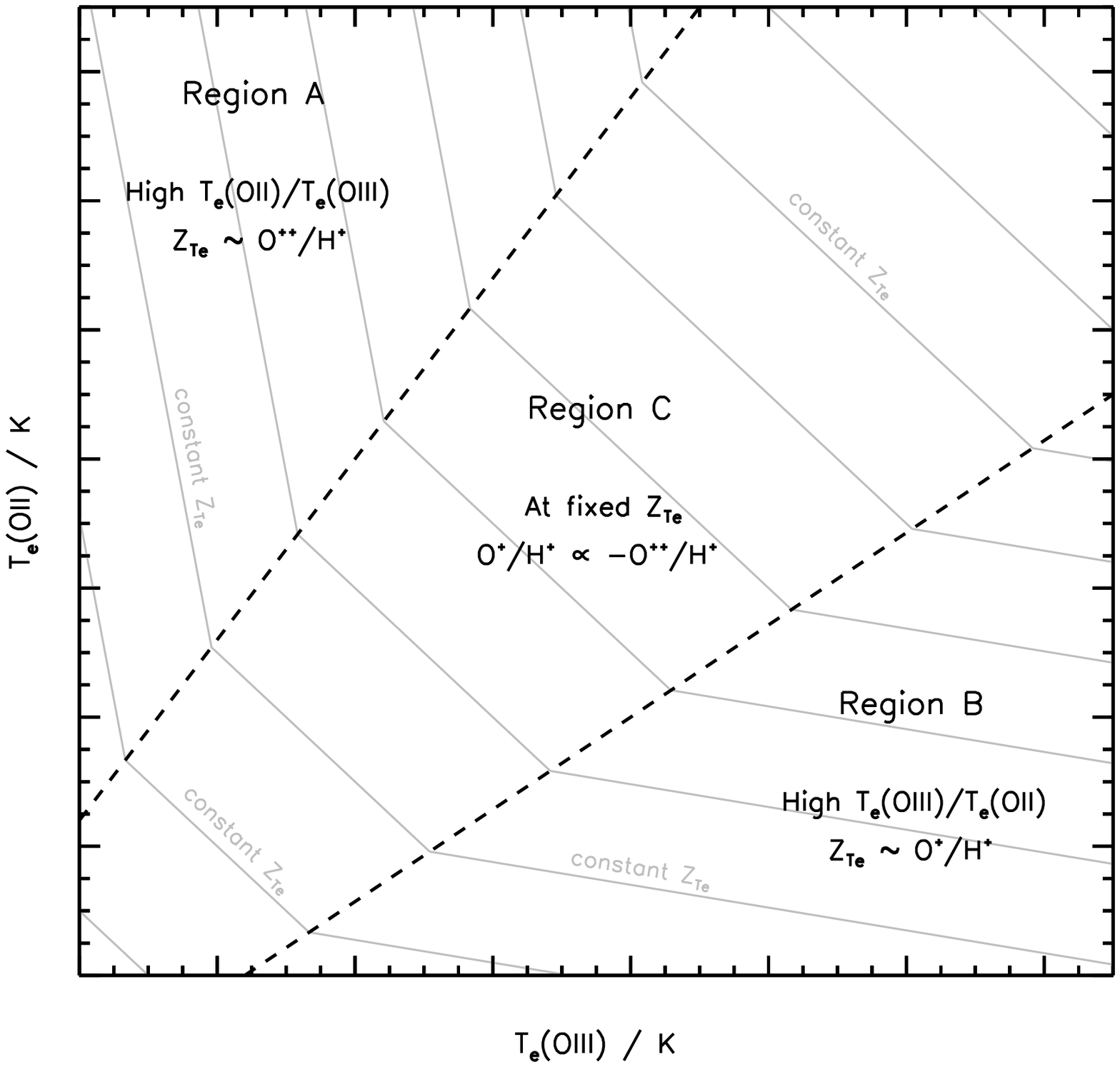}
 \caption{Schematic diagram qualitatively illustrating the expected relation between \TOIII{} and \TOII{} at fixed \ZTe{} in different regions of the \TOIII{} -- \TOII{} plane. Grey lines illustrate arbitrary tracks of constant \ZTe{}. Regions A and B denote the approximate regimes where \ZTe{} is likely dominated by only one ionisation state of oxygen (\OppH{} or \OpH{}, respectively). Region C denotes the approximate regime where both ionisation states are likely significant, and a uniform anti-correlation between \TOIII{} and \TOII{} at fixed \ZTe{} is expected. We stress that the simplified picture illustrated here may change if, for example, the nebulae are not radiation bounded, or if the density distribution in the \Opp{} and \Op{} zones are very different.}
 \label{fig:TOIII-TOII_schematic}
\end{figure}

\subsection{A new empirical \texorpdfstring{\TOIII{} -- \TOII{}}{TOIII-TOII} relation}\label{sec:New TT relation}
The form of our new \TOIII{} -- \TOII{} relation is motivated by two important considerations. First, that electron temperature and oxygen abundance are anti-correlated. Consequently, we would expect systems with \TOIII{} $<$ \TOII{} to have \OppH{} $>$ \OpH{}, and for their total \ZTe{} to be relatively insensitive to \TOII{}. This is illustrated  by Region A in the schematic in Fig. \ref{fig:TOIII-TOII_schematic}. Likewise, we would expect systems with \TOIII{} $>$ \TOII{} to have \OppH{} $<$ \OpH{}, and therefore their total \ZTe{} to be relatively insensitive to \TOIII{} (see Region B in Fig. \ref{fig:TOIII-TOII_schematic}). Second, that there is an empirical anti-correlation between \TOIII{} and \TOII{} at fixed \ZTe{} for our dataset, (see coloured points in Fig. \ref{fig:TOIII-TOII_our_relation} below). This is also expected theoretically from the equations laid-out in \S \ref{sec:Te_and_ZTe}, as $\tn{\OpH{}} \propto -\tn{\OppH{}}$ at fixed \ZTe{}. This is in contrast to what is typically assumed for other metallicity-dependent \TOIII{} -- \TOII{} relations in the literature (\eg{}\citealt{Izotov+06,Nicholls+14a}). The anti-correlation we find is illustrated by Region C in Fig. \ref{fig:TOIII-TOII_schematic} by grey lines of constant \ZTe{}.

\begin{figure}
\centering
\begin{tabular}{c}
 \includegraphics[angle=0,width=1.0\linewidth]{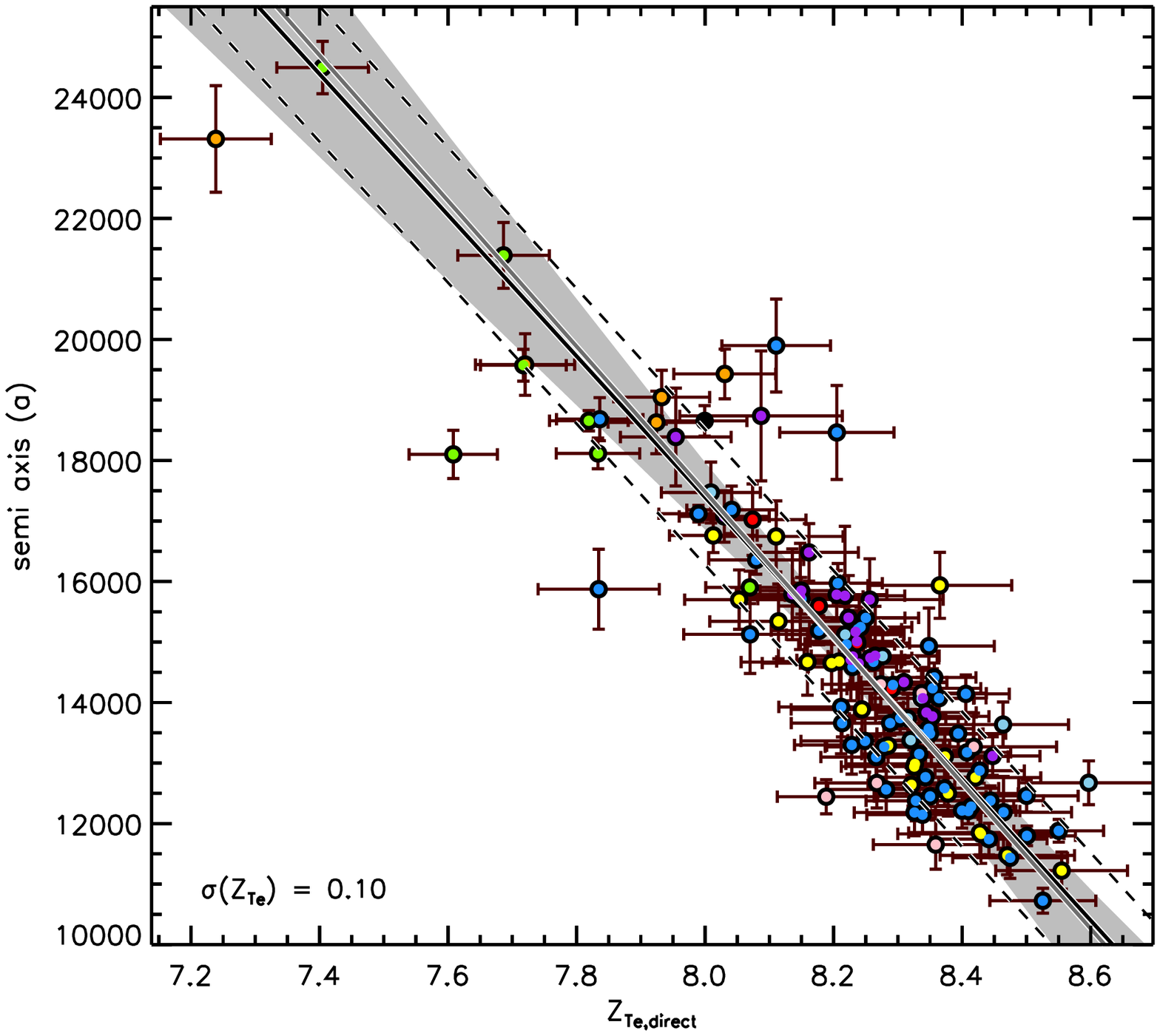} \\
 \includegraphics[angle=0,width=1.0\linewidth]{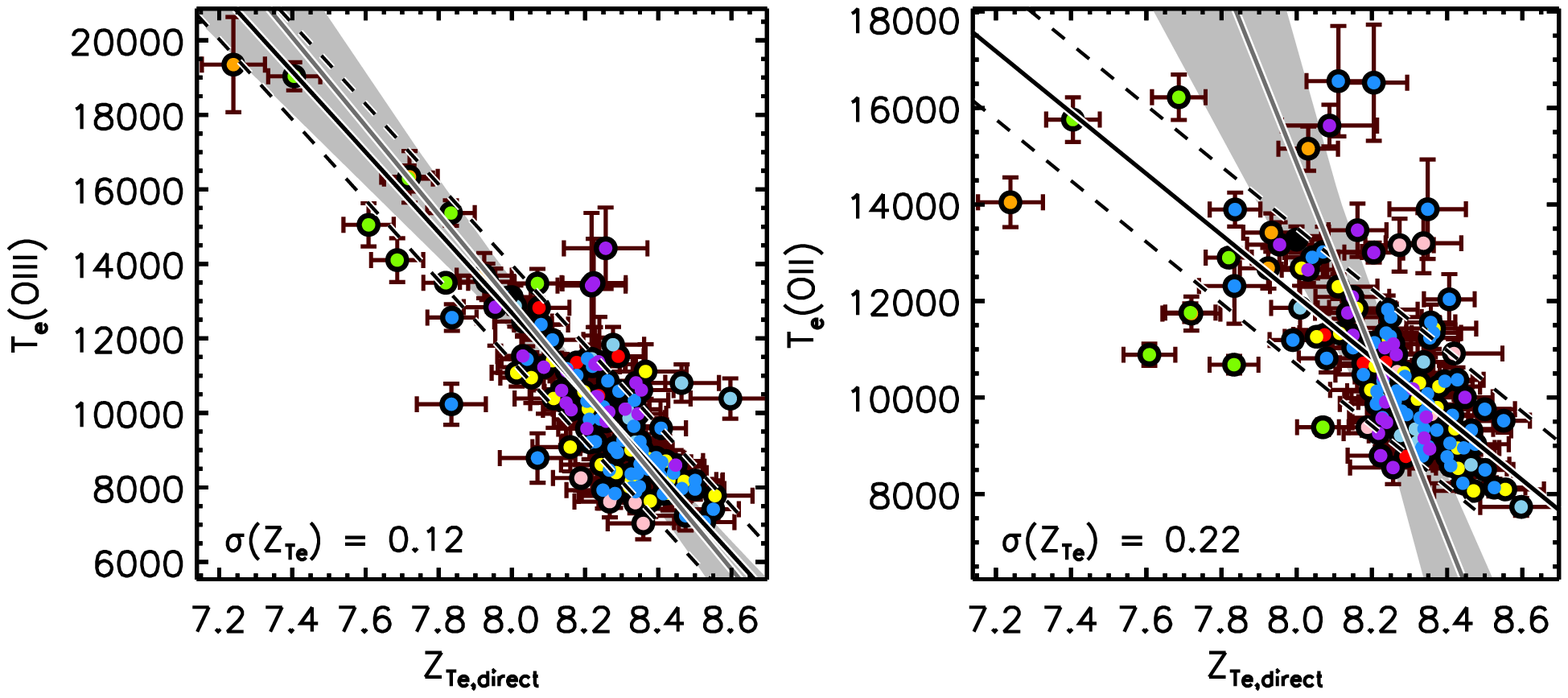} \\
\end{tabular}
 \caption{\textit{Top panel:} Relation between \ZTe{} and the hyperbolic semi-axis $a$ for our direct dataset. Linear fits to the data (Eq. \ref{eqn:semiaxis}) are shown for our Bayesian analysis (dark grey solid line) and least squares fitting (black solid line). Dashed lines indicate the 1$\sigma$ scatter around the distribution of \ZTe{} values for the least-squares fit. The grey area around the Bayesian best-fit indicates the range of possible fits considering the full covariance matrix. \textit{Bottom panels:} \ZTe{}-\TOIII{} and \ZTe{}-\TOII{} relations for the same dataset. The large spread in the \ZTe{}-\TOII{} relation leads to quite discrepant fits being returned by our two statistical fitting methods.}
 \label{fig:ZTe-rT}
\end{figure}

Therefore, we propose a functional form for our new \ZTe{}-dependent \TOIII{} -- \TOII{} relation which broadly embodies these two considerations, and allows for an unrestricted range of possible \TOII{}/\TOIII{} ratios. To do this, we adopt the equation of a rectangular hyperbola, centered on 0 Kelvin, given by
\begin{equation}\label{eqn:hyp_definition}
\tn{\TOII} = \frac{a(\tn{\ZTe{}})^{2}}{2}\, \frac{1}{\tn{\TOIII}}\;\;,
\end{equation}
where $a$ is the hyperbolic semi axis. We then fit $a$ to the directly-measured \ZTe{} values from our dataset, finding the following linear dependence (see Appendix \ref{sec:FittingMethods} for details on our fitting methods):
\begin{align}\label{eqn:semiaxis}
a = & -11657.85\error{434.08}\;\tn{\ZTe{}} + 110655.66\error{3567.40} & \tn{\footnotesize{(Least squares)}}\;,\nonumber\\
\nonumber\\
a = & -12030.22\error{499.37}\;\tn{\ZTe{}} + 113720.75\error{562.50} & \tn{\footnotesize{(Bayesian)}}\;.
\end{align}

We find that \ZTe{} is more tightly correlated with $a$ than with \TOIII{} or \TOII{} alone. This is illustrated in Fig. \ref{fig:ZTe-rT}, which shows the \ZTe{} -- $a$ relation for our direct dataset in the top panel, and the \ZTe{} -- \TOIII{} and \ZTe -- \TOII{} relations in the bottom panels. The standard deviation from least-squares fitting is slightly lower for the \ZTe{} -- $a$ relation [with $\sigma(\tn{\ZTe{}}) = 0.10$, 0.12, and 0.22, respectively], and that this relation is marginally favoured over the \TOIII{} or \TOII{} relations by our Bayesian analysis, with comparative Bayes factors of 1.2 and 2.3, respectively.

Eq. \ref{eqn:hyp_definition} is plotted for discrete values of \ZTe{} in Fig. \ref{fig:TOIII-TOII_our_relation}, alongside our direct \ZTe{} systems which are coloured by their direct $Z\sub{Te}$. The correspondence between our new relation and the measured \ZTe{} for our dataset is good across the whole \TOIII{} -- \TOII{} plane. Eq. \ref{eqn:hyp_definition} can then be solved for both \ZTe{} and \TOII{} using fixed-point iteration, similar to the approaches used by \citet{Izotov+06}, \citet{Pilyugin07}, and \citet{Nicholls+14a}, except that we are explicitly accounting for the interdependence of \ZTe{} and electron temperature here.

Eqs. \ref{eqn:hyp_definition} and \ref{eqn:semiaxis} can be combined to provide the following equivalent expression for \ZTe{},
\begin{align}\label{eqn:ZTe}
\tn{\ZTe{}} = &\ 9.49\error{0.05} - \frac{\left[2\,\tn{\TOIII{}}\cdot{}\tn{\TOII{}}\right]^{1/2}}{11657.85\error{434.08}} & \tn{\footnotesize{(Least squares)}}\;,\nonumber\\
\nonumber\\
\tn{\ZTe{}} = &\ 9.45\error{0.04} - \frac{\left[2\,\tn{\TOIII{}}\cdot{}\tn{\TOII{}}\right]^{1/2}}{12030.22\error{499.37}} & \tn{\footnotesize{(Bayesian)}}\;.
\end{align}

An alternative fit to the \ZTe{} -- $a$ relation, when using \TNII{} rather than \TOII{} to determine \Op{}, is discussed in Appendix \ref{sec:Appendix A}.

\begin{figure}
 \includegraphics[angle=0,width=1.0\linewidth]{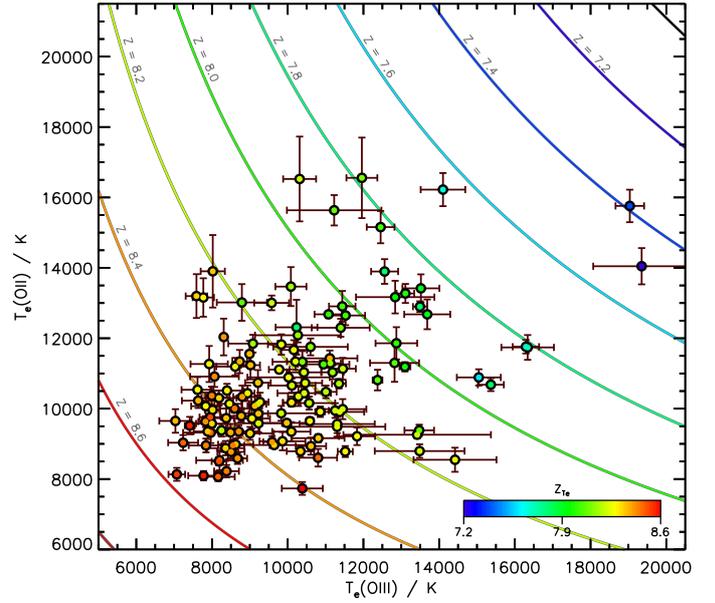}
 \caption{\TOIII{}-\TOII{} plane containing our \NumDirectHIISystems{} systems with direct \ZTe{} estimates, as shown in Fig. \ref{fig:TOIII-TOII_old_relations}. Here, each system is coloured by its direct \ZTe{}. Our new empirically-derived \TOIII{} -- \TOII{} relation  (Eq. \ref{eqn:hyp_definition}) is plotted for discrete values of \ZTe{} for comparison. There is a good correspondence seen between the metallicites of the data and the relation across this plane.}
 \label{fig:TOIII-TOII_our_relation}
\end{figure}

\subsection{A semi-direct abundance deficit at low \texorpdfstring{\Opp{}/\Op{}}{O++/O+}}\label{sec:Accuracy of TT relations}
\begin{figure}
 \includegraphics[angle=0,width=1.0\linewidth]{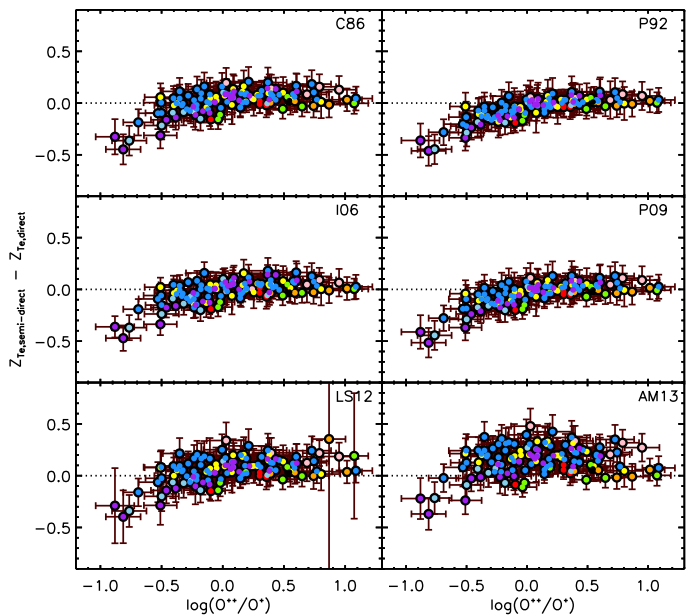}
 \caption{Semi-direct \ZTe{} deficit as a function of \Opp{}/\Op{} when semi-direct \ZTe{} is calculated using each of the six literature \TOIII{}-\TOII{} relations considered in this work. Labels in each panel denote which \TOIII{} -- \TOII{} relation is considered. A similar (although often weaker) anti-correlation is also seen for these literature relations.}
 \label{fig:Ofrac-Zdiff_all_relations}
\end{figure}

First, we compare the direct and semi-direct \ZTe{} estimates for our dataset, using each of the six literature \TOIII{} -- \TOII{} relations considered here. Significantly, we find that all the literature relations under-estimate \ZTe{} by up to 0.6 dex for low-ionisation systems. This is illustrated in Fig. \ref{fig:Ofrac-Zdiff_all_relations}, where the difference between the semi-direct and direct \ZTe{} estimates is plotted against \Opp{}/\Op{}. We hereafter refer to this \ZTe{} under-estimation as the `semi-direct \ZTe{} deficit'.

This deficit at low \Opp{}/\Op{} is also present for our new \TOIII{} --\TOII{} relation, as shown in panel (a) of Fig. \ref{fig:TOII-Zdiff}. The apparent ubiquity of the semi-direct \ZTe{} deficit suggests that it is an intrinsic issue for all semi-direct methods that rely on only \TOIII{} measurements. We have also checked this analysis by using the equations provided by \citet{Izotov+06} to calculate \TOIII{}, \OpH{}, and \OppH{} (their eqs. 1 to 5) rather than Eqs. \ref{eqn:Te}, \ref{eqn:OpH}, and \ref{eqn:OppH} above, and find that our results are unchanged.

Physically speaking, systems with low \Opp{}/\Op{} ostensibly have a larger \OpH{} abundance than \OppH{} abundance, meaning that singly-ionised oxygen dominates their total oxygen budget. \citet{Andrews&Martini13} have also determined that a significant fraction of higher-mass galaxies appear to have their overall \ZTe{} dominated by \OpH{}. They find that log(\Opp{}/\Op{}) typically drops below 0.0 at masses above $\tn{log(}M_{*}/\Msun) \sim{}8.2$ for stacks of SDSS-DR7 galaxies (their fig. 5c). Similarly, \citet{Curti+17} found that higher-metallicity galaxies have log(\Opp{}/\Op{}) below 0.0. In what follows, we discuss various possible explanations for the semi-direct \ZTe{} deficit at low \Opp{}/\Op{} that we find.

\subsubsection{Weak auroral lines}\label{sec:Weak auroral lines}
Systematic effects on \TOIII{} due to inaccurate \OIII{}$\lambda4363$ measurements are unlikely to play a significant role as we only consider spectra with S/N(\OIII{}$\lambda4363$) $\geq 3.0$. Similarly, a systematic over-estimation of \OII{}$\lambda7320,7330$ fluxes due to poor line fitting is unlikely, as we have also enforced a S/N lower limit of 3.0 for these lines and find an average S/N(\OII{}$\lambda7320,7330$) of 22.5. There is also no trend present between  S/N(\OII{}$\lambda\lambda7320,7330$) and the measured semi-direct \ZTe{} deficit.

\subsubsection{\texorpdfstring{\OII{}}{[OII]} line contamination}\label{sec:OII line contamination}
The \OII{}$\lambda\lambda7320,7330$ auroral line quadruplet lies in a region of the optical spectrum which is populated by a large number of skylines. Additionally, collisional de-excitation, reddening effects, the telluric emission of OH bands, absorption of water bands, and absorption features in the underlying stellar continuum can affect their flux measurements (see \eg{}\citealt{Kennicutt+03,Pilyugin+09,Croxall+15}).

When considering our own dataset, we note that collisional de-excitation is already accounted for in the \citet{Nicholls+13} photoionisation models, via an explicit dependence of \Te{} on electron density. This means that the \TOII{} and \TNII{} temperatures we derive for each system are actually quite similar for most of our dataset, as shown in Fig. \ref{fig:ROII-RNII_and_TOII-TNII}. We have also been particularly careful to accurately and consistently correct all emission line fluxes for reddening (see \S \ref{sec:dust_corrections}), and our MaNGA spectra have been corrected for stellar absorption (see \S \ref{sec:MaNGA sample}). Also, the majority of our dataset lies at redshifts above $z \sim{}0.006$, meaning the \OII{} auroral lines are redshifted away from strong potential contamination from the OH Meinel band emission in Earth's lower atmosphere.

\begin{figure}
\centering
 \includegraphics[angle=0,width=1.0\linewidth]{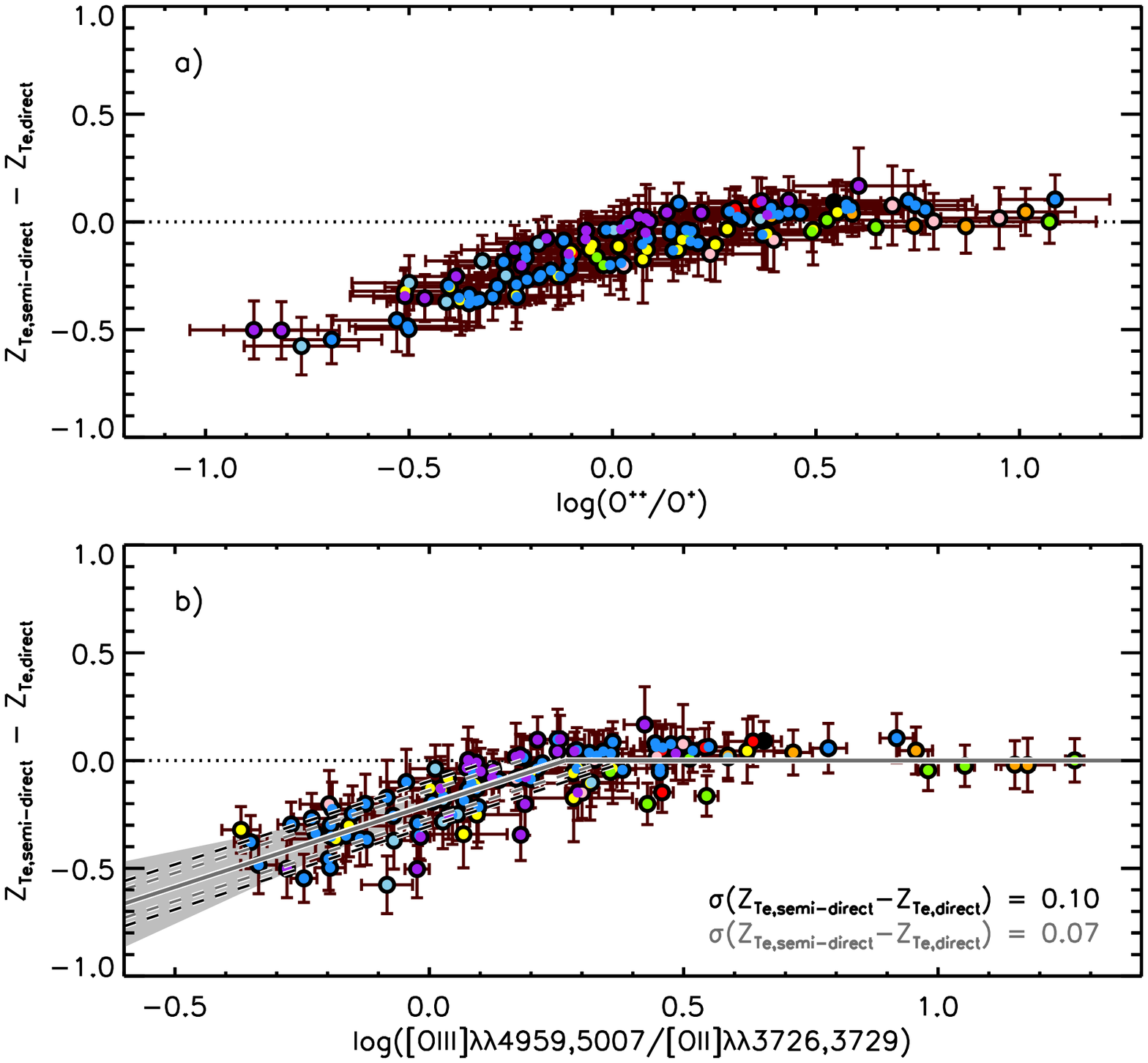} \\
 \caption{\textit{Panel (a):} Semi-direct \ZTe{} deficit (as defined in \S \ref{sec:Accuracy of TT relations}) as a function of \Opp{}/\Op{}. There is a tight anti-correlation between these two properties at low values of \Opp{}/\Op{}. \textit{Panel (b):} Semi-direct \ZTe{} deficit as a function of the nebular line ratio \OIII{}/\OII{}. As in Fig. \ref{fig:ZTe-rT}, the Bayesian fit to this relation is given by the solid dark grey line, and the least-squares fit is given by the solid black line. These fits can be used to help correct semi-direct \ZTe{} estimates for the \Opp{}/\Op{} bias seen in panel (a).}
 \label{fig:TOII-Zdiff}
\end{figure}

We also expect contamination from recombination emission to only have a minor effect. This effect in the \OII{} auroral lines should be weakest at low \Opp{}/\Op{}, indicating that the semi-direct \ZTe{} deficit we find at low-\Opp{}/\Op{} is not caused by such contamination. \citet{Liu+00} detect recombination emission in the planetary nebula NGC6153. However, the densities and oxygen abundances of NGC6153 are much higher than for any of the \HII{} regions in our dataset, with $N\sub{e} > 2000\ \tn{cm}^{-3}$ and \OppH{} $> 4.4\times10^{-4}$. For our full dataset, the mean values are $112\ \tn{cm}^{-3}$ and $7.0\times10^{-5}$, respectively. Also, for their intermediate \OppH{} estimate of $5.61\times10^{-4}$, \citet{Liu+00} state that the contamination from recombination excitation is roughly the same for the \OII{}$\lambda\lambda3726,3729$ nebular lines and the \OII{}$\lambda\lambda7320,7330$ auroral lines, so that their ratio is relatively unaffected. Similarly, \citet{Garcia-Rojas+18} studied nine planetary nebula with $n\sub{e} > 3500\ \tn{cm}^{-3}$ and found \TOII{} was over-estimated by only a few hundred Kelvin due to recombination excitation. We therefore expect our lower-density \HII{} regions to be even less affected.

To test for any other residual contamination of the \OII{}$\lambda\lambda7320,7330$ lines, we have re-run our analysis using the equations provided by \citeauthor{Izotov+06} (2006, their eqs. 3 and 4), which allow independent estimates of \OpH{} to be made using either the auroral \OII{}$\lambda\lambda7320,7330$ lines or the nebular \OII{}$\lambda3727$ lines [along with their semi-direct \TOIII{} -- \TOII{} relation in both cases]. We find that the \OpH{} estimates obtained from these two methods differ on average by only $\sim{}9$ per cent (\ie{}$\sim{}0.04$ dex), and removing the very few systems with a difference greater than 0.06 dex has no impact on our results. This indicates that the \OII{}$\lambda\lambda7320,7330$ lines in our dataset are reliable for obtaining direct estimates of \TOII{} and \OpH{}.

Alternatively, specific narrow emission (or absorption) features could cause contamination of one of the \OII{}$\lambda\lambda7320,7330$ doublets relative to the other, decreasing the sensitivity of our flux estimates. To test if such contamination is significant, we have measured the $r' = \tn{\OII}\lambda7320 / \tn{\OII}\lambda7330$ ratio for the 102 systems in our dataset where these doublets are resolved. We find a mean $r'$ of 1.19, which is in good agreement with the expected theoretical value of $r' \sim{} 1.24$ for systems of similar electron density and temperature \citep{Seaton&Osterbrock57,DeRobertis+85}. However, the scatter in $r'$ we find is quite large, with $\sigma(r') = 0.22$. We therefore create a sub-sample of 82 systems with $r'$ values within the typical range accurately measured for nearby \HII{} regions and planetary nebulae: $1.0 < r' \leq 1.6$ \citep{Kaler+76,Keenan+99}. We find both our calibration of the \TOIII{} -- \TOII{} relation and the semi-direct \ZTe{} deficit at low \Opp{}/\Op{} are unchanged when using this sub-sample, indicating that neither of these results are driven by inaccurate \OII{}$\lambda\lambda7320,7330$ measurements.

Despite these checks all suggesting line contamination does not significantly affect our metallicity calculations, in Appendix \ref{sec:Appendix A} we also provide a separate \Te{} calibration using nitrogen lines, which allows an estimate of \Op{} to be made without relying on the \OII{} auroral lines at all. This alternative \NII{}-based calibration has qualitatively the same features as our \OII{}-based calibration.

\subsubsection{Dust extinction}\label{sec:Dust extinction}
Higher metallicity systems are expected to contain more dust, making any differences between the assumed and actual attenuation curve more significant when measuring their emission line fluxes. Additionally, the emission lines required to calculate \TOII{} are widely separated in wavelength, meaning that \TOII{} would be more sensitive than \TOIII{} to such dust effects. This could lead to an under-estimation of \ZTe{} if, for example, galaxy attenuation curves were to systematically flatten with increasing \OpH{}. Such a scenario would imply that it is not the empirical \TOIII{} -- \TOII{} relations that are inaccurate, but rather the directly measured \OII{} temperatures.

However, we find no systematic correlation between the internal extinction, $A\sub{\textsc{V}}$, and the observed semi-direct \ZTe{} deficit for our systems with direct \TOII{} measurements. Also, the tight and systematic anti-correlation between the semi-direct \ZTe{} deficit and \Opp{}/\Op{} we find suggests that general variations in attenuation curves among systems are not significant here.

\subsubsection{Composite \HII{} regions}\label{sec:Composite HII regions}
\citet{Kobulnicky+99} found that \TOIII{} can be over-estimated by up to $\sim{}1600$ K for composite spectra, and consequently that \ZTe{} can be under-estimated by up to 0.2 dex. Similarly, \citet{Pilyugin+10} demonstrated that composite spectra can return higher \TOIII{} [and/or lower \TOII{}] than the mean electron temperature of their constituent \HII{} regions. Though we note that this effect was found to decrease significantly when the composite spectrum contained more than 2 \HII{} regions. Additionally, \citet{Sanders+17} have shown that emission from diffuse ionised gas (DIG) can combine with the above bias to produce a total over-estimation of \TOIII{} and under-estimation of \TOII{} of up to $\sim{} 2000$ K for global galaxy spectra.

When considering this issue for our dataset, we first note that the semi-direct \ZTe{} deficit we find is present for both composite and individual \HII{} region spectra. For example, the high-resolution spectra analysed by \citet{Bresolin+09b} and the CHAOS team should not be prone to the effects attributed to composite spectra, however we find that they also suffer \ZTe{} discrepancies of up to 0.35 and 0.5 dex, respectively [see panel (a) of Fig. \ref{fig:TOII-Zdiff}, light blue and yellow points]. Indeed, \citet{Pilyugin+10} also found the same result for some of the high-resolution spectra from the \citet{Bresolin+09b} sample (their fig. 5). Moreover, the composite spectra from our MaNGA samples have been selected to have at least EW(H$\alpha$) $> 30$ \AA{}, which is well above the level expected for DIG regions (\eg{}\citealt{Belfiore+16}), and all the spaxels considered in our global analysis lie within the star-forming or composite regions of the \citeauthor{Baldwin+81} (1981, BPT) diagram. Yet some of these systems are also affected by \ZTe{} discrepancies.

We have further checked the issue of composite-spectra bias by splitting our \NumDirectHIISystems{} direct-\ZTe{} systems into three distinct sub-samples: an `individual \HII{} region sub-sample' containing spectra of 83 very nearby extragalactic \HII{} regions (\ie{}\citealt{Esteban+09,Bresolin+09b}; CHAOS), a `composite \HII{} region sub-sample' containing 27 composite HII regions (\ie{}\citealt{Guseva+09}; MaNGA), and an `integrated galaxy sub-sample' containing integrated spectra from 20 galaxies (\ie{}\citealt{Lee+04,Izotov+12,Hirschauer+15}; SLSN hosts). We find that the relations shown in Fig. \ref{fig:TOII-Zdiff} are very similar for all three of these sub-samples, as are their fits to the \ZTe{} -- $a$ relation. This indicates that our new \Te{} method is robust to differences in spatial resolution, and that the semi-direct \ZTe{} deficit we find is a real feature of low-\Opp{}/\Op{} systems.

\begin{figure}
 \includegraphics[angle=0,width=1.0\linewidth]{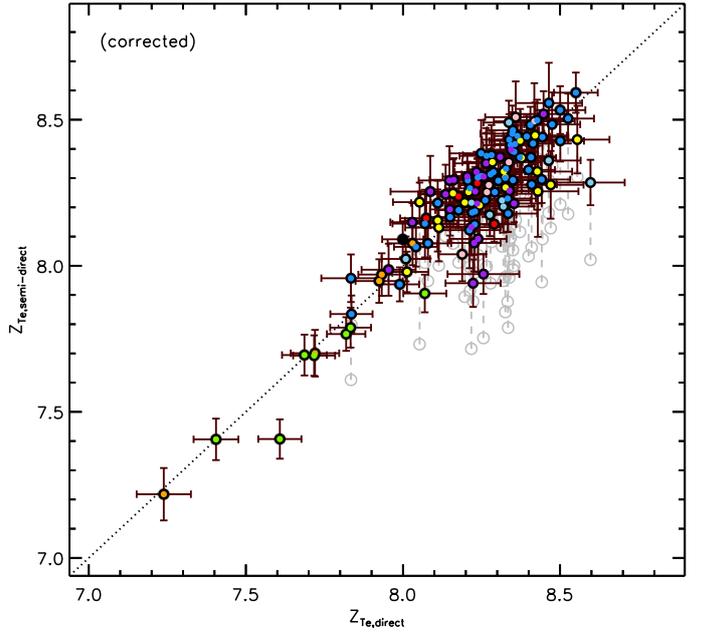}
 \caption{Comparison between the direct \ZTe{} and the semi-direct \ZTe{} from our new \TOIII{} -- \TOII{} relation corrected for the \Opp/\Op{} bias using our least-squares-derived $f\sub{cor}$ (Eq. \ref{eqn:correction_factor}). The uncorrected semi-driect estimates for these systems are shown as grey empty circles connected to their corrected values by a grey dashed line.}
 \label{fig:Zcomp_new_relation_corrected}
\end{figure}

\subsubsection{Ionisation factor}\label{sec:Ionisation factor}
The final effect we consider here is that of the ionisation state of the gas. This can be defined by the ionisation parameter, $U$, which is typically approximated by the emission-line ratio \OIII{}$\lambda\lambda4959,5007$/\OII{}$\lambda3727$ or \SIII{}$\lambda9067$/\SII{}$\lambda\lambda6717,6731$  \citep{Kewley&Dopita02}.

Panel (b) of Fig. \ref{fig:TOII-Zdiff} shows that systems with low \OIII{}/\OII{} indeed tend to have large \ZTe{} discrepancies when using our new \TOIII{}-\TOII{} relation. This relation is not as tight as that for \Opp{}/\Op{} in panel (a), but there is a clear trend of increasing semi-direct \ZTe{} deficit with decreasing \OIII{}/\OII{} for all systems below $\tn{log(\OIII{}/\OII{})} \sim{}0.25$. A similar finding was discussed by \citet{Kobulnicky+99}, who concluded that semi-direct methods using only \OIII{}$\lambda$4363 measurements will inevitably under-estimate \ZTe{} for low-ionisation nebulae, such as those with extended ionised filaments or shells.

We note that inconsistencies in the relationship between \OIII{}/\OII{} and $U$ could affect the interpretation of Fig. \ref{fig:TOII-Zdiff} and that a detailed investigation into the link between ionisation state, its proxies using various strong emission lines, and semi-direct \ZTe{} estimations is beyond the scope of this current work. Therefore, we only conclude here that the estimated \ZTe{} is particularly sensitive to \TOII{} for low-\Opp{}/\Op{} systems because \Op{} dominates their oxygen budget. This physical effect presents a problem for all semi-direct methods that rely only on \OIII{}$\lambda$4363 measurements, as they do not have a direct handle on the dominant \Op{} ion.

\begin{figure}
 \includegraphics[angle=0,width=1.0\linewidth]{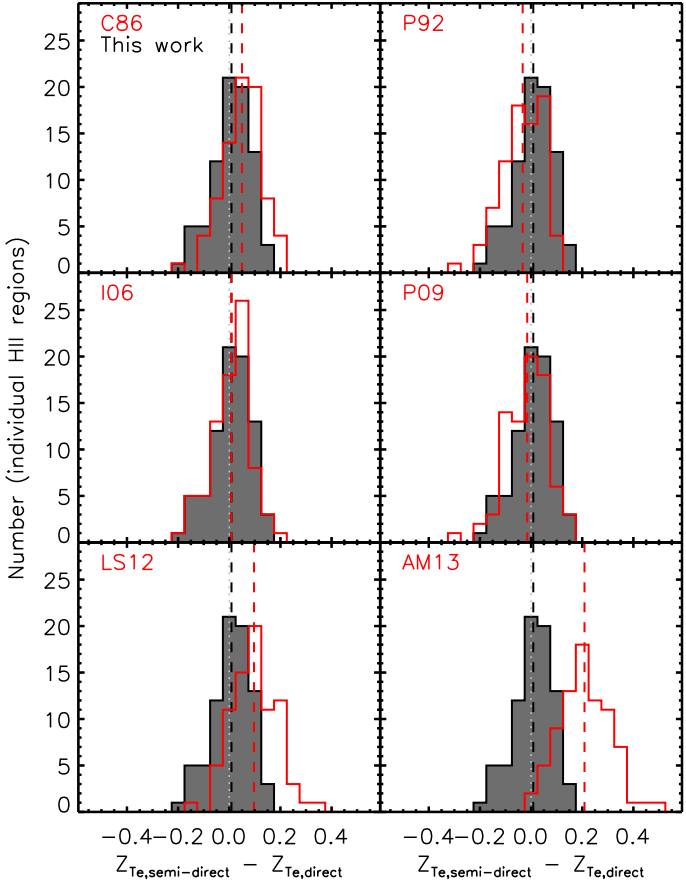}
 \caption{Distribution of \ZTe{} discrepancies for individual \HII{} regions (as defined in \S \ref{sec:Composite HII regions}) when using each of the six literature \TOIII{} -- \TOII{} relations considered (red histograms). The distribution for our new semi-direct method (dark grey historgram) when corrected for the observed \Opp{}/\Op{} bias (Eq. \ref{eqn:correction_factor}) is also shown in each panel. The mean of each distribution is shown as a vertical dashed line.}
 \label{fig:Zdiff_dists_highres}
\end{figure}

\begin{figure}
 \includegraphics[angle=0,width=1.0\linewidth]{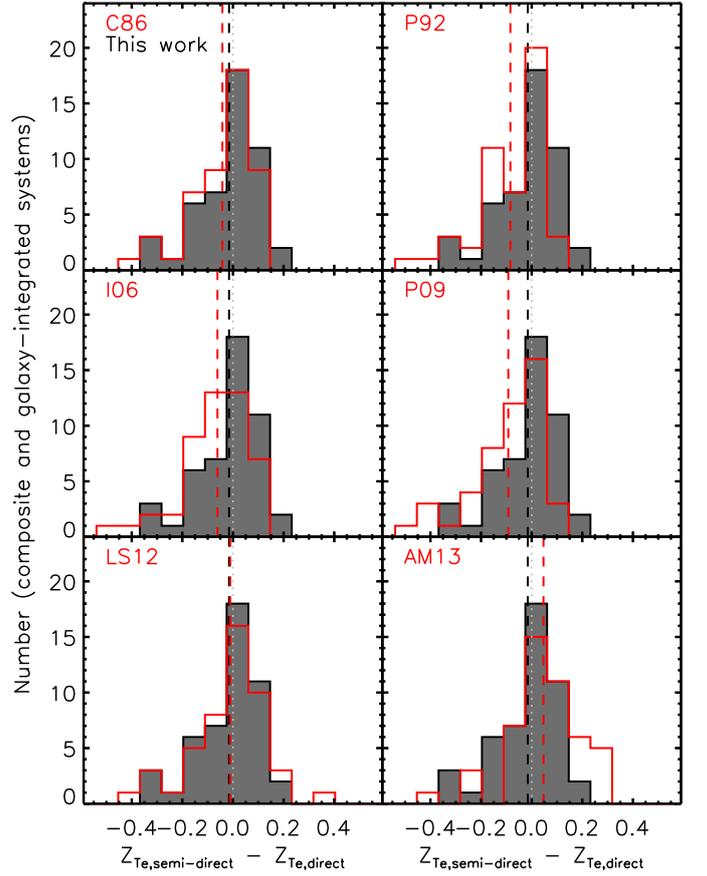}
 \caption{Same as Fig. \ref{fig:Zdiff_dists_highres}, but for composite and galaxy-integrated systems, as defined in \S \ref{sec:Composite HII regions}.}
 \label{fig:Zdiff_dists_lowintres}
\end{figure}

\subsection{Correcting the semi-direct \texorpdfstring{\ZTe{}}{ZTe} deficit at low \texorpdfstring{\Opp{}/\Op{}}{O++/O+}}
\label{sec:correcting_ZTe}
Given the findings above, we here derive an empirical correction to the semi-direct \ZTe{} deficit using the easily-observable ratio of oxygen nebular lines, \OIII{}$\lambda\lambda4959,5007$/\OII{}$\lambda\lambda3726,3729$ [see panel (b) of Fig. \ref{fig:TOII-Zdiff}]. A linear fit to this relation yields a correction factor, $f\sub{cor}$, given by,
\begin{align} \label{eqn:correction_factor}
f\sub{cor} = \bigg{\{} & \begin{array}{ll}
													0.77\,(x - 0.20) & \textnormal{for } x \leq 0.26\\
													0.0 & \textnormal{for } x > 0.26
									 \end{array} \;\;\;\;\tn{\footnotesize{(Least squares)}}\;\;,\nonumber\\
\nonumber\\
f\sub{cor} = \bigg{\{} & \begin{array}{ll}
													0.71\,(x - 0.29) & \textnormal{for } x \leq 0.29\\
													0.0 & \textnormal{for } x > 0.29
									 \end{array} \;\;\;\;\tn{\footnotesize{(Bayesian)}}\;\;,
\end{align}
where $x=\tn{log(\OIII{}}\lambda\lambda4959,5007/\tn{\OII{}}\lambda\lambda3726,3729)$. The corrected semi-direct oxygen abundance is therefore given by,
\begin{equation}\label{eqn:corrected_ZTe}
Z\sub{Te,cor} = \tn{\ZTe{}} - f\sub{cor}\;\;.
\end{equation}

\begin{figure*}
 \includegraphics[angle=0,width=\linewidth]{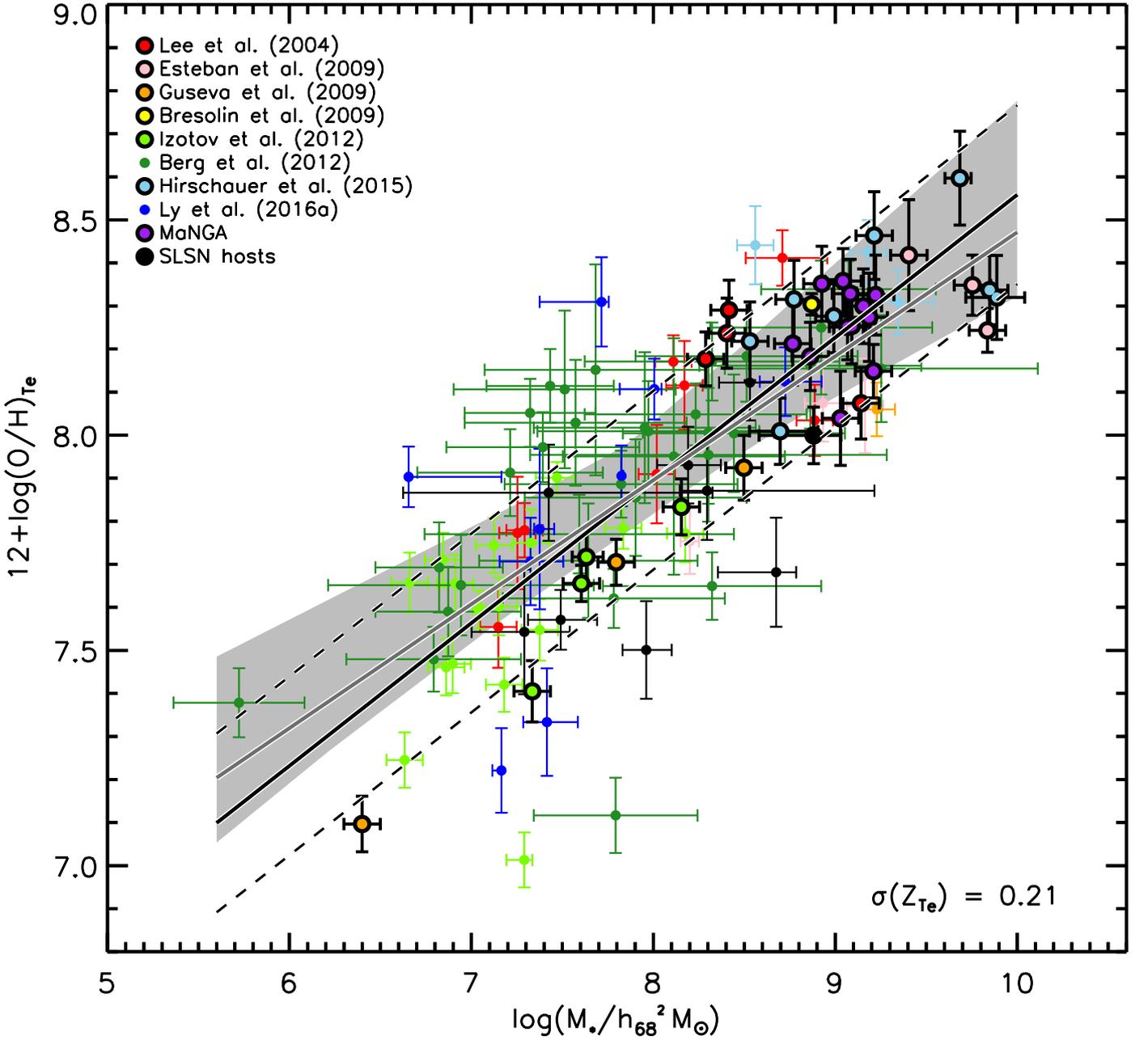}
  \caption{Mass-metallicity relation for the \NumLowZGals{} galaxies in our dataset. Galaxies are coloured according to the sample from which their fluxes were taken. Points with thick black rings have direct \ZTe{} estimates [\ie{}\TOIII{} and \TOII{} independently calculated], while the remaining points have semi-direct \ZTe{} estimates, \ie{}using our new semi-direct method, corrected for the observed \Opp{}/\Op{} bias using our least-squares-derived $f\sub{cor}$ (see \S \ref{sec:correcting_ZTe}). The least-squares and Bayesian linear fits to all the points (Eq. \ref{eqn:lowz_MZR_fit}) are shown by the solid black line and solid dark grey line, respectively, with the 1$\sigma$ dispersions given by dashed lines.}
\label{fig:lowz_MZR}
\end{figure*}

Fig. \ref{fig:Zcomp_new_relation_corrected} shows the comparison between direct $Z\sub{Te}$ and the corrected semi-direct $Z\sub{Te}$ using our new \TOIII{} -- \TOII{} relation. There is now a much more consistent correspondence between these two methods across a large range of oxygen abundances. However, despite this clear improvement, we do caution that empirical corrections of this nature do not directly relate the physics linking the two variables in question to the discrepancy observed and are, therefore, liable to provide spurious corrections when applied to inappropriate systems.

We also compare the accuracy of our new semi-direct method with those from the literature. We do this by splitting our sample into individual \HII{} region spectra and composite/integrated-galaxy spectra (as defined in \S \ref{sec:Composite HII regions}), and plotting the semi-direct \ZTe{} deficit distributions from each semi-direct method separately for these two sub-samples in Figs. \ref{fig:Zdiff_dists_highres} and \ref{fig:Zdiff_dists_lowintres}. These figures clearly show that those literature relations calibrated to single \HII{} regions (\eg{}\citealt{Izotov+06} or \citealt{Pilyugin+09}) perform relatively well for our individual \HII{} region spectra, but poorly for composite and global spectra. Conversely, those relations calibrated to composite \HII{} regions (\eg{}\citealt{Lopez-Sanchez+12} or \citealt{Andrews&Martini13}) perform relatively well for our composite or integrated-galaxy spectra, but poorly for the highly-resolved systems. Our new relation, on the other hand, calibrated to a varied dataset, works equally well for all systems of any physical size, with a mean semi-direct \ZTe{} deficit of almost zero in both cases.

When using our dataset as a whole, the standard deviation around the peak of the distribution for our new semi-direct method is 0.08 dex, and there are no systems with \ZTe{} discrepancies greater than $-0.35$ dex, regardless of their ionisation state.

%----------------------------------------------------------------
\section{The local MZR} \label{sec:Local MZR}
In Fig. \ref{fig:lowz_MZR}, we plot the $M_{*}$-\ZTe{} relation (MZR) for the \NumLowZGals{} galaxies that make up our dataset. These galaxies contain the \NumDirectHIISystems{} individual and composite \HII{} regions discussed in the preceding sections, as well as additional systems for which semi-direct \ZTe{} estimates can be obtained. For galaxies with electron temperature measurements for more than one H\textsc{ii} region, the H$\alpha$-flux-weighted mean \ZTe{} is used, except for our MaNGA galaxies for which the \ZTe{} obtained from the global spectra are used. Points with thick black outlines in Fig. \ref{fig:lowz_MZR} represent galaxies with direct \ZTe{} estimates, while the other points represent galaxies with our semi-direct \ZTe{} estimates. 

Our new MZR represents an improvement on previous \Te{}-based relations in the literature predominantly because (a) our dataset is not biased to strongly star-forming systems, and (b) we have corrected for the semi-direct \ZTe{} deficit in low-\Opp{}/\Op{} systems discussed in \S \ref{sec:Accuracy of TT relations} and \S \ref{sec:correcting_ZTe}.

We find the following linear fit to our MZR, using our least-squares and Bayesian analyses:
\begin{align}\label{eqn:lowz_MZR_fit}
\tn{\ZTe{}} & = 0.332\error{0.021}\;\tn{log}(M_{*}) + 5.242\error{0.171} & \tn{\footnotesize{(Least squares)}}\;,\nonumber\\
\nonumber\\
\tn{\ZTe{}} & = 0.293\error{0.023}\;\tn{log}(M_{*}) + 5.575\error{0.192} & \tn{\footnotesize{(Bayesian)}}\;,
\end{align}
in the range $5.67 < \tn{log}(M_{*}) < 9.87$, where $M_{*}$ is in units of $h^{2}_{68}\Msun$. The 1$\sigma$ dispersion in \ZTe{} is 0.21 dex for the least-squares fit from residuals. This spread is larger than the 1$\sigma$ dispersion of $\sim 0.10$ dex obtained from most strong-line-based MZRs (\eg \citealt{Tremonti+04,Yates+12}), suggesting that the local MZR has a wider scatter than usually assumed. The scatter above our relation is partly due to our low-\OIII{}/\OII{} correction to semi-direct \ZTe{} estimates (see \S \ref{sec:correcting_ZTe}). For example, the \citet{Berg+12} system CGCG035-007A with log$(M_{*}/h_{68}^{2}\Msun) = 7.51$ and \ZTe{}$ = 8.11$ has a low log(\OIII{}/\OII{}) ratio of -0.14, which has caused a \ZTe{} correction upwards by $\sim{}0.31$ dex. However, the other systems close to CGCG035-007A on the MZR actually have relatively high log(\OIII{}/\OII{}) ratio, so their high \ZTe{} estimates are not due to any correction. We find the scatter below the MZR is driven by unusually-high electron temperatures. For example, the low-metallicity outlier, UGC5340A from the \citet{Berg+12} sample (which is believed to be undergoing a merger), has measured \TOIII{} $= 18830$ K and \TOII{} $= 20359$ K, leading to \ZTe{} $= 7.12$.

\begin{figure}
 \includegraphics[angle=0,width=\linewidth]{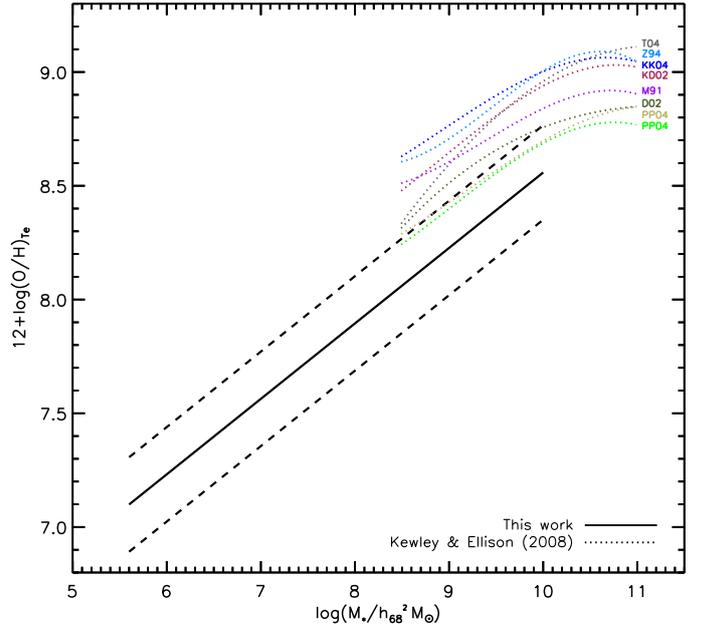}
 \caption{Comparison between our new \Te{}-based MZR (solid black line, Eq. \ref{eqn:lowz_MZR_fit}) and various strong-line-based MZRs presented by \citeauthor{Kewley&Ellison08} (2008, dotted lines). As for many other \ZTe{} studies, we find a lower normalisation for our MZR compared to the strong-line-based relations, although the slope is quite similar among most cases below $\tn{log(}M_{*}/\Msun) \sim{}10.0$.}
\label{fig:MZR_strongLines_comp}
\end{figure}

\begin{figure}
 \includegraphics[angle=0,width=\linewidth]{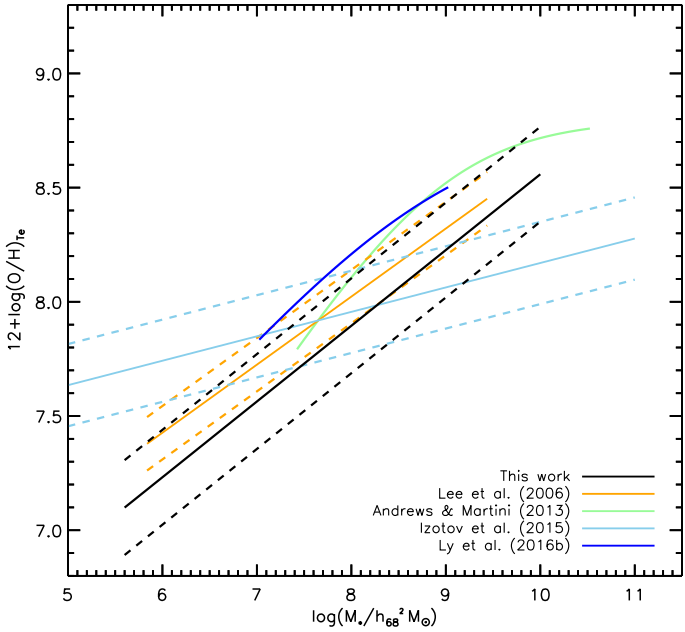}
 \caption{Comparison between our new \Te{}-based MZR \Te{}-based MZR (solid black line, Eq. \ref{eqn:lowz_MZR_fit}) and other recent \Te{}-based MZRs from the literature. Fits to the relations from \citeauthor{Lee+06} (2006, orange), \citeauthor{Andrews&Martini13} (2013, green), \citeauthor{Izotov+15} (2015, light blue), and \citeauthor{Ly+16b} (2016b, dark blue) are shown. Stellar masses have been corrected for differences in the assumed dimensionless Hubble parameter where possible.}
\label{fig:MZR_Tes_comp}
\end{figure}

\subsection{Comparison with other MZRs}\label{sec:Comparison with other methods}
\subsubsection{Strong-line MZRs}\label{sec:Strong-line MZRs}
In Fig. \ref{fig:MZR_strongLines_comp}, our MZR is shown alongside fits to the MZRs from the various strong-line metallicity diagnostics presented by \citet{Kewley&Ellison08}. Our new \Te{}-based MZR has a lower normalisation at fixed mass than MZRs based on strong lines, by $0.15-0.55$ dex at $\tn{log}(M_{*}/\Msun) \sim 9.0$, although the slopes are found to be similar.

A lower normalisation for \Te{}-based MZRs has been seen many times before (\eg{} \citealt{Stasinska05,Lee+06,Andrews&Martini13,Ly+14,Izotov+15}). This is unlikely to be caused by a preferential selection of low-metallicity systems at fixed mass in our case because our dataset contains a significant contribution of typically star-forming systems (see \S \ref{sec:Sample selection effects}). We therefore concur with most previous studies that the oxygen abundance in the ISM of low-redshift galaxies is lower than indicated by traditional strong-line diagnostics applied to large star-forming galaxy samples such as the SDSS.

Our findings are qualitatively consistent with the fundamental metallicity relation (FMR, \citealt{Mannucci+10}), although we are unable to draw any firm conclusions on the anti-correlation between \ZTe{} and SFR at low-mass, due to the relatively low number of systems at fixed mass and metallicity. We do, however, find a clear increase in star-formation rate (SFR) with both \ZTe{} and $M_{*}$ for our dataset. 

\subsubsection{\texorpdfstring{\Te{}}{Te}-based MZRs}\label{sec:Te-based MZRs}
Fig. \ref{fig:MZR_Tes_comp} compares our new MZR (black solid line) with those from other recent studies of \Te{}-based metallicities using different samples.

Our MZR appears to be in best agreement with that of \citet{Lee+06}, who utilised \textit{Spitzer} IR photometry to determine stellar masses and a variety of different semi-direct \Te{}-based metallicities from the literature for their smaller sample of 27 dwarf irregular galaxies. We determine a larger scatter around the MZR for our larger sample of \NumLowZGals{} galaxies, and a lower normalisation.

The slope of the MZR derived by \citet{Izotov+15} (light blue) for 3607 star-forming galaxies from the SDSS-DR7 is significantly shallower than the other \Te{}-based MZRs considered here. This is likely due to the different sample selection criteria applied. \citet{Izotov+15} selected compact ($R_{50}<6$ arcsec) galaxies with high nebular excitation as given by log(\OIII{}$\lambda5007/\tn{H}\beta) \gtrsim 0.5$. They also required a detection of the auroral \OIII{}$\lambda4363$ line from the relatively shallow SDSS spectroscopy in order to estimate \TOIII{}. These criteria all limit the number of higher-metallicity galaxies in their sample at higher mass, effectively selecting systems with similar physical conditions to higher-redshift galaxies. This is likely causing the lower typical metallicities seen in the \citet{Izotov+15} sample above $\tn{log}(M_{*}/\Msun) \sim 8.0$, and therefore the shallower slope compared to other works.

The MZR from \citet{Andrews&Martini13} has a normalisation $\sim{}0.3$ dex higher than ours at $\tn{log}(M_{*}/\Msun) \sim{} 9.0$. Their study utilised \Te{} measurements from stacked SDSS-DR7 spectra, containing around 2,000 galaxies in each bin at $\tn{log}(M_{*}/\Msun) \sim{} 9.0$.

Although Fig. \ref{fig:Zdiff_dists_highres} shows that the \TOIII{} -- \TOII{} relation fit to the \citet{Andrews&Martini13} data returns semi-direct \ZTe{} estimates around 0.2 dex higher than our relation, we note that this is the case when applied to spectra of individual \HII{} regions. For the more global galaxy spectra that make up the bulk of our MZR, we actually find a relatively close correspondence between the \ZTe{} estimated via the AM13 relation and our own (see also Fig. \ref{fig:Zdiff_dists_lowintres}).

The range of SFRs found in each sample at $\tn{log}(M_{*}/\Msun) \sim{} 9.0$ are also similar, with  $-1.0 < \tn{log(SFR}/\Msun\,\tn{yr}^{-1}) < 0.0$ in both cases. It is therefore unlikely that differences in the \TOIII{} -- \TOII{} relations or sample selection biases are responsible for the MZR normalisation difference seen here.

Rather, the predominant cause is likely differences in the estimated direct \OpH{} obtained from each analysis; \citet{Andrews&Martini13} find a value of $12 + \tn{log(\OpH{})} \sim{}8.5$ for their $\tn{log}(M_{*}/\Msun) \sim{} 9.0$ stack (their fig. 5), compared to values between 7.2 and 8.4 for our direct \ZTe{} systems of a similar mass. \OppH{} estimates at fixed mass, on the other hand, are more similar between the two studies.

The higher \OpH{} estimates they obtain could be partly due to the composite spectra issue discussed by \citet{Pilyugin+10} (see \S \ref{sec:Composite HII regions}). However this is unlikely to explain the entire difference in \ZTe{} seen, as stacked spectra should not be more prone to such issues than global spectra of individual galaxies \citep{Pilyugin+10, Andrews&Martini13, Curti+17}. Limitations with the assumption that galaxies of the same stellar mass should have similar metallicities (see \citealt{Curti+17}), as well as SDSS fibre aperture effects (see \citealt{Ellison+08}), could also play a role.

Finally, the results presented by \citet{Ly+16b} on the low-redshift MZR from the Metal Abundances Across Cosmic Time (MACT) survey also suggests higher metallicities than most other \Te{}-based MZRs considered here. Although \OppH{} abundances were determined in a very similar way to our methodology, the lack of direct \OpH{} measurements and reliance on the \TOIII{} -- \TOII{} relation calibrated to the \citet{Andrews&Martini13} stacked data is likely causing the higher normalisation, as discussed above.

\begin{figure}
 \includegraphics[angle=0,width=\linewidth]{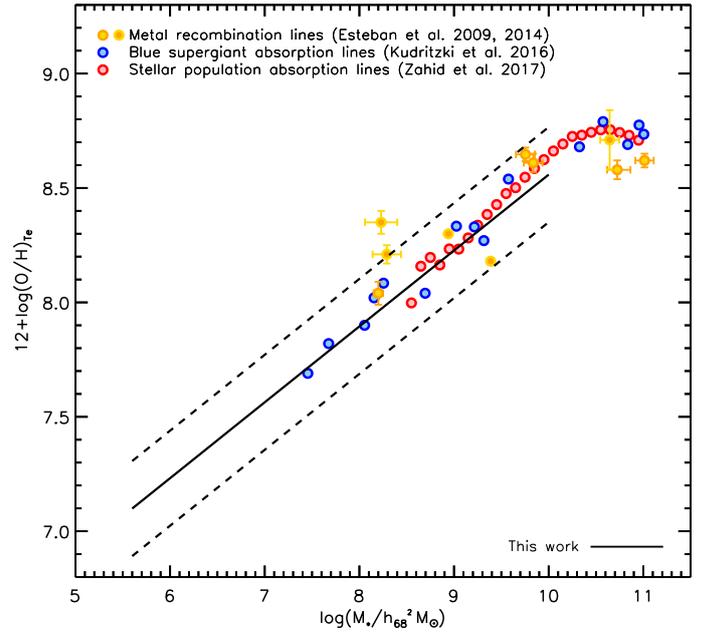}
 \caption{Fit to our \Te{}-based MZR (solid black line, Eq. \ref{eqn:lowz_MZR_fit}), plotted alongside the MZR from oxygen recombination lines (yellow/orange points, \citealt{Esteban+09,Esteban+14}), the MZR from blue supergiant star photospheres (blue points, \citealt{Kudritzki+16}), and the MZR from the integrated stellar populations of SDSS-DR7 star-forming galaxies (red points, \citealt{Zahid+17}).}
\label{fig:MZR_otherDirect_comp}
\end{figure}

\subsubsection{Alternative direct methods}\label{sec:Other direct methods}
In Fig. \ref{fig:MZR_otherDirect_comp}, the fit to our low-redshift MZR given by Eq. \ref{eqn:lowz_MZR_fit} is shown (black line), compared to MZRs formed using the two methods for obtaining gas-phase oxygen abundances that are generally considered to be most accurate. Large yellow points denote 10 nearby ($z<0.023$) galaxies for which O/H has been calculated in the brightest \HII{} regions from faint metal recombination lines by \citet{Esteban+09} and \citet{Esteban+14}. Large blue points denote 15 galaxies for which O/H has been calculated from absorption lines in the photospheres of blue supergiant stars by \citeauthor{Kudritzki+16} (2016, and references therein), taking the metallicity at two disc scale lengths from the measured blue-supergiant abundance gradient.

Stellar masses for the \citet{Esteban+09,Esteban+14} samples are taken from the literature (\citealt{Sick+14,Lelli+14,vanDokkum+14,Lopez-Sanchez+10,Skibba+11,Ostlin+03,Woo+08}; and the NSA catalogue). Stellar masses for the \citet{Kudritzki+16} sample are taken from various other works, as listed in table 4 of \citet{Kudritzki+12}. Where possible, we have ensured that these are corrected for any differences in the assumed IMF or cosmology.

A key finding of this study is that there is remarkably good agreement between the MZR formed from our new \Te{} analysis and those formed from metal recombination lines and blue supergiant absorption lines. There has been long-standing evidence that \Te{} methods typically under-predict the \OppH{} abundance by $0.26-0.43$ dex compared to RL methods for the same individual \HII{} regions (\citealt{Esteban+09,Esteban+14}). However, it appears from Fig. \ref{fig:MZR_otherDirect_comp} that, in an overall sense, our \Te{}-based MZR is consistent with that formed using RLs and stellar absorption-line spectra.

Additionally, \citet{Zahid+17} have also produced an MZR based on stacked absorption-line spectra from the SDSS-DR7, using sequential single-burst (SSB) SPS models. Their luminosity-weighted MZR is also plotted in Fig. \ref{fig:MZR_otherDirect_comp} as red points. There is also very good agreement between the \citet{Zahid+17} MZR and those of \citet{Kudritzki+16}, \citet{Esteban+09,Esteban+14}, and this work. This convergence of various different direct metallicity methods on a consistent MZR is a promising sign than the true metal content of nearby galaxies is being correctly probed by our new analysis.

\section{Summary \& Conclusions}\label{sec:Conclusions}
Electron temperatures (\Te{}) and gas-phase oxygen abundances (\ZTe{}) have been obtained for \NumLowZHIISystems{} emission-line systems in the local Universe ($z < \MaxRedshift$). This dataset comprises a mix of individual, composite, and whole-galaxy spectra, belonging to both starbursting galaxies and galaxies on the star-forming main sequence. The \NumDirectHIISystems{} systems with direct measurements of both \TOIII{} and \TOII{} are utilised to calibrate a new \TOIII{} -- \TOII{} relation, which can be used to estimate \ZTe{} using the \OIII{}$\lambda4363$ auroral line. The resulting mass -- metallicity relation (MZR) for \NumLowZGals{} low-redshift, star-forming galaxies with $5.5 \lesssim \tn{log}(M_{*}/\Msun) \lesssim 10.0$ is then compared to previous works. Our key findings and conclusions are as follows:

\begin{itemize}
\item Due to its hyperbolic functional form, our new metallicity-dependent \TOIII{} -- \TOII{} relation (Eq. \ref{eqn:ZTe}) allows for a wider range of \TOII{}/\TOIII{} ratios than previous relations. The semi axis of the hyperbola is tightly constrained by \ZTe{} (see \S \ref{sec:New TT relation}). Both \TOII{} and \ZTe{} can therefore be obtained iteratively for any system with a robust \OIII{}$\lambda{}4363$ auroral line detection.

\item We find that all the literature \TOIII{} -- \TOII{} relations considered here, as well as our own, underestimate \ZTe{} for systems with log(\Opp{}/\Op{}) $\lesssim 0.0$ (see Fig. \ref{fig:TOII-Zdiff}a). After investigating the possible causes for this semi-direct \ZTe{} deficit, we determine that it is most likely due to the physical dominance of \Op{} ions over \Opp{} in the \HII{} regions of these systems, making \ZTe{} estimates difficult to obtain from measurements of \TOIII{} alone (see \S \ref{sec:Accuracy of TT relations}).

\item We therefore provide an empirically-calibrated correction to our semi-direct \ZTe{} estimates for low-\Opp{}/\Op{} systems, based on the easily-observable nebular oxygen line ratio \OIII{}/\OII{}. Our new method can then return accurate \ZTe{} estimates for systems of either high or low ionisation, regardless of their spatial resolution. Overall, our semi-direct \ZTe{} estimates are within a standard deviation of 0.08 dex from the directly-measured values, which is comparable to or better than any of the literature \TOIII{} -- \TOII{} relations considered here (see Figs. \ref{fig:Zdiff_dists_highres} and \ref{fig:Zdiff_dists_lowintres}).

\item The low-redshift MZR formed using our new \ZTe{} estimates has a similar slope to most strong-line based MZRs but a lower normalisation, as found by most previous studies (see \S \ref{sec:Strong-line MZRs}). The scatter of $\sigma{}(\tn{\ZTe}) \sim{}0.21$ we find is larger than is typically found when using strong-line diagnostics.

\item When comparing to other \Te{}-based MZRs, we deduce that any differences are mainly due to sample selection biases or differences in the direct determination of \OpH{}, rather than the particular \TOIII{} -- \TOII{} relation used when obtaining semi-direct \ZTe{} estimates. The inclusion of many `star-forming main sequence' galaxies in our dataset makes our MZR more representative of the typically-star-forming population at low redshift.

\item Encouragingly, our new \Te{}-based MZR is in very good agreement with the MZRs obtained via direct metallicity measurements from metal recombination lines or blue supergiant absorption lines (see \S \ref{sec:Other direct methods}). This is a strong indication that our study is accurately probing the true range of metallicities present in the star-forming galaxy population at low redshift.
\end{itemize}

In a follow-up paper, we will compare our new \Te{}-based MZR with MZRs derived from alternative direct methods at higher redshifts, in order to ascertain the true evolution of the gas-phase metallicity in galaxies over cosmic time.

\begin{acknowledgements}
       \label{sec:Acknowledgements}
The authors would like to thank the anonymous referee for very helpful comments and suggestions, as well as Danielle Berg, Fabio Bresolin, I-Ting Ho, Rolf-Peter Kudritzki, Guinevere Kauffmann, Thomas Kr\"{u}hler, Brent Miszalski, Gwen Rudie, Alice Shapley, Martin Yates, and Jabran Zahid for valuable discussions during the undertaking of this work. We would also like to thank Ricardo Amor\'{i}n, Fabio Bresolin, Alec Hirschauer, Janice Lee, Matt Nicholl, and John Salzer for providing additional data and guidance, and Christophe Morisset for help with running the \texttt{pyneb} package. This research was partly supported by the Munich Institute for Astro- and Particle Physics (MIAPP) of the DFG cluster of excellence \textit{``Origin and Structure of the Universe''}. The authors would also like to acknowledge the \textit{TOPCAT} interactive graphical viewer and editor \citep{Taylor05} which was used for quick analysis and visualisation of our tabulated data. R.M.Y., T.-W.C., and P.W. acknowledge the support through the Sofia Kovalevskaja Award to P. Schady from the Alexander von Humboldt Foundation of Germany.
\end{acknowledgements}

%-------------------------------------------------------------------

%\bibliography{robyates.bib}
\bibliography{robyates}
\bibliographystyle{aa}

%%%%%%%%%%%%%%%%%%%%%%%%%%%%%%%%%%
%APPENDICES:
%%%%%%%%%%%%%%%%%%%%%%%%%%%%%%%%%%
\appendix

\section{An alternative nitrogen-based calibration}\label{sec:Appendix A}

In addition to the calibration of our new \TOIII{} -- \TOII{} relation presented in \S \ref{sec:New TT relation}, we also present here a complimentary calibration to the \TOIII{} -- \TNII{} relation, using a subset of 53 systems (hereafter, the `nitrogen sub-sample') for which measurements of both the \OIII{}$\lambda4363$ and \NII{}$\lambda5755$ auroral lines are available.

\NII{}$\lambda5755$ is a particularly weak auroral line, with a mean S/N of only 8.2 for our nitrogen sub-sample, compared to S/N(\OIII{}$\lambda4363)=17.4$ and S/N(\OII{}$\lambda\lambda7320,7330)=27.3$ for the same systems. Nonetheless, due to the similar ionisation energies required for \OII{} and \NII{}, both are expected to trace the same ionisation zone within an \HII{} region, and return similar electron temperatures. Therefore, \TNII{} can be a useful alternative to \TOII{} when estimating \Op{}, in cases where \OII{}$\lambda\lambda7320,7330$ is not available or believed to be contaminated (see \S \ref{sec:OII line contamination}).

\begin{figure}
\centering
 \includegraphics[angle=0,width=0.9\linewidth]{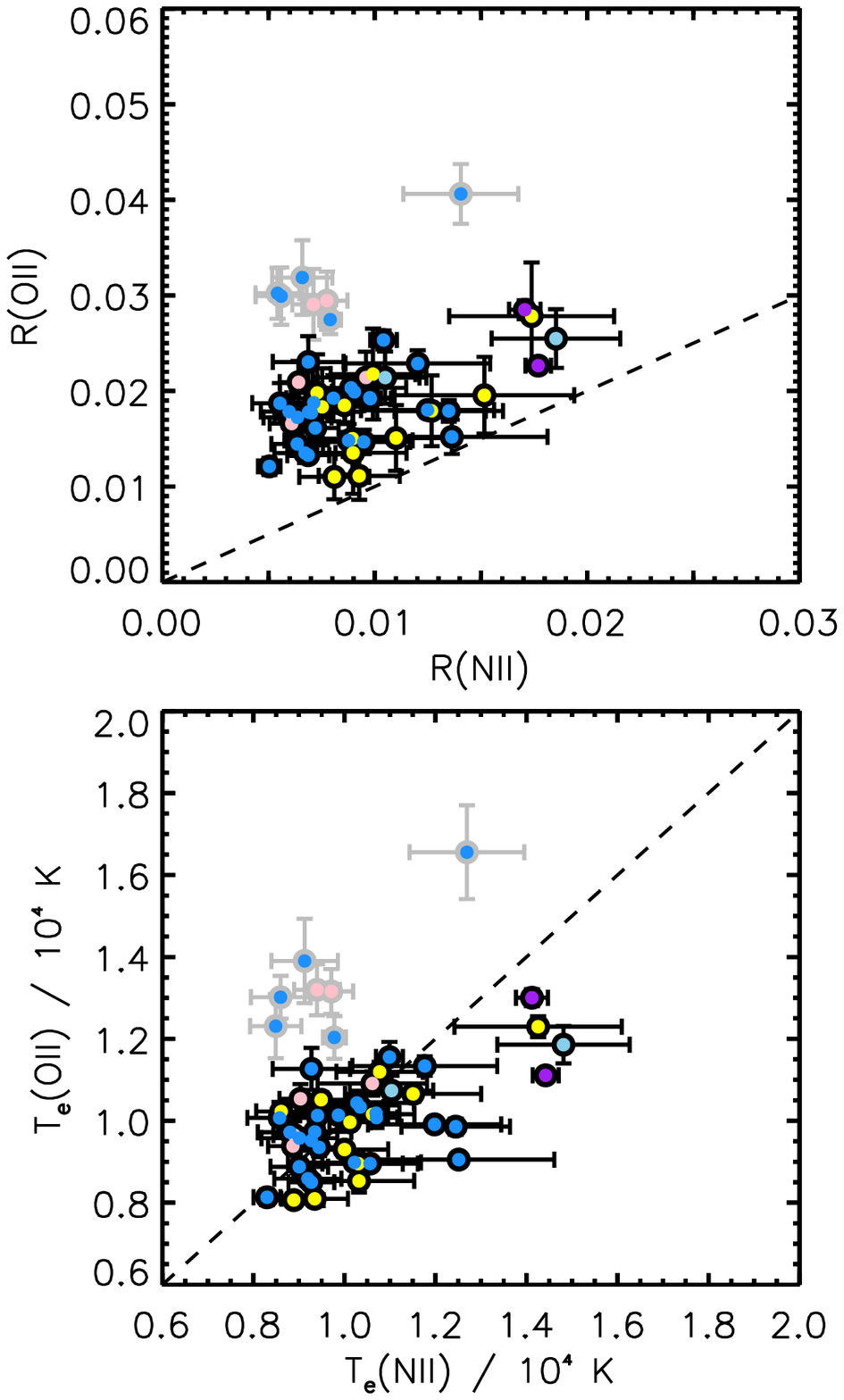}
 \caption{\textit{Top panel:} Relation between $R(\tn{N\textsc{ii}}) = [\tn{N\textsc{ii}}]\lambda5755 / (1.33\cdot{}\tn{N\textsc{ii}}]\lambda6584)$ and $R(\tn{O\textsc{ii}}) = [\tn{O\textsc{ii}}]\lambda\lambda7320,7330 / [\tn{O\textsc{ii}}]\lambda\lambda3726,3729$, for our nitrogen sub-sample of 53 systems. \textit{Bottom panel:} Relation between \TNII{} and \TOII{} for the same nitrogen sub-sample. These two temperatures are expected to be roughly the same for a given \HII{} region, as indicated by the line of equality (dashed line). Points with grey outlines (in both panels) are outliers with $T(\tn{\textsc{Oii}}) > T(\tn{\textsc{Nii}}) + 2000$ K, or equivalently, $R(\tn{\textsc{Oii}}) > R(\tn{\textsc{Nii}}) + 0.0175$ K.}
 \label{fig:ROII-RNII_and_TOII-TNII}
\end{figure}

Fig. \ref{fig:ROII-RNII_and_TOII-TNII} shows the \RNII{} - \ROII{} relation and \TNII{} - \TOII{} relation for our nitrogen sub-sample. With the exception of a few systems with unexpectedly high \ROII{} and \TOII{} (outlined in grey), we find a good linear relationship between these line ratios and temperatures, with a median difference between \TOII{} and \TNII{} of only 165 K. In particular, the good one-to-one correspondence between \TOII{} and \TNII{} for the CHAOS sample (blue points) is a clear improvement on that reported by the original works \citep{Berg+15,Croxall+16}. This is because we estimate higher \TNII{} at fixed \TOII{} than those studies, bringing these two temperatures more in line with each other.

\begin{figure}
\centering
\includegraphics[width=1.0\linewidth]{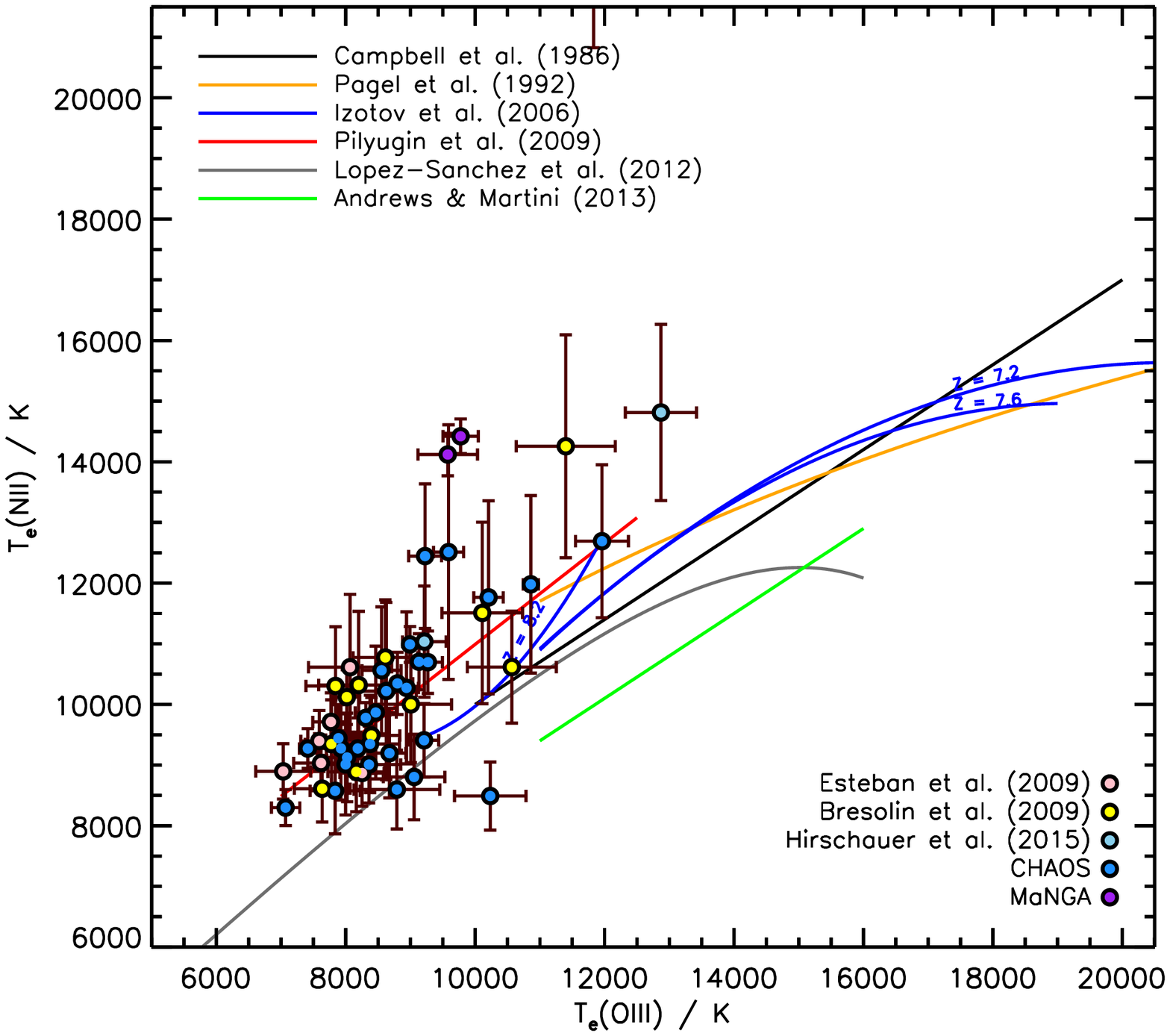} \\
\includegraphics[width=1.0\linewidth]{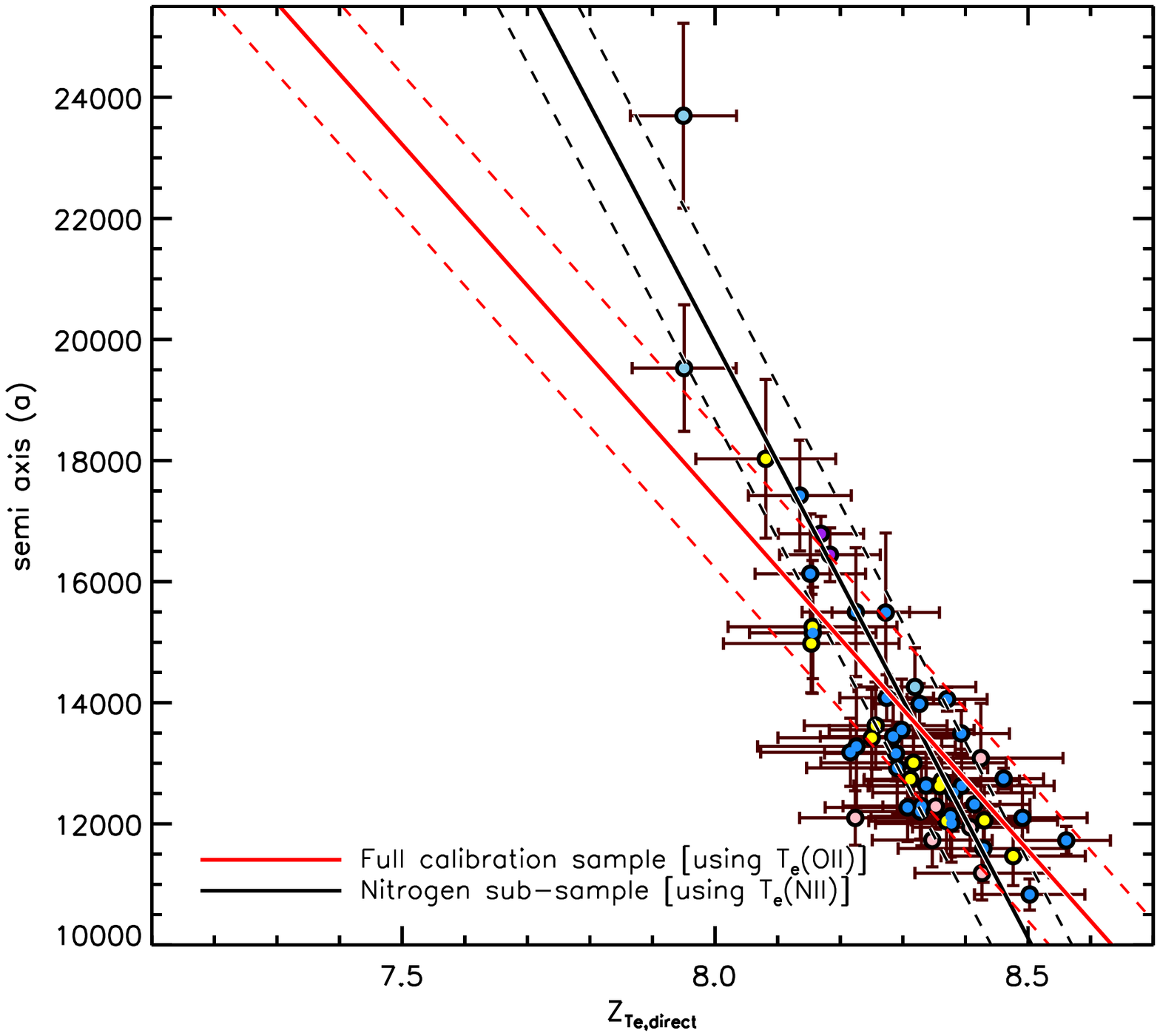} \\
\includegraphics[width=1.0\linewidth]{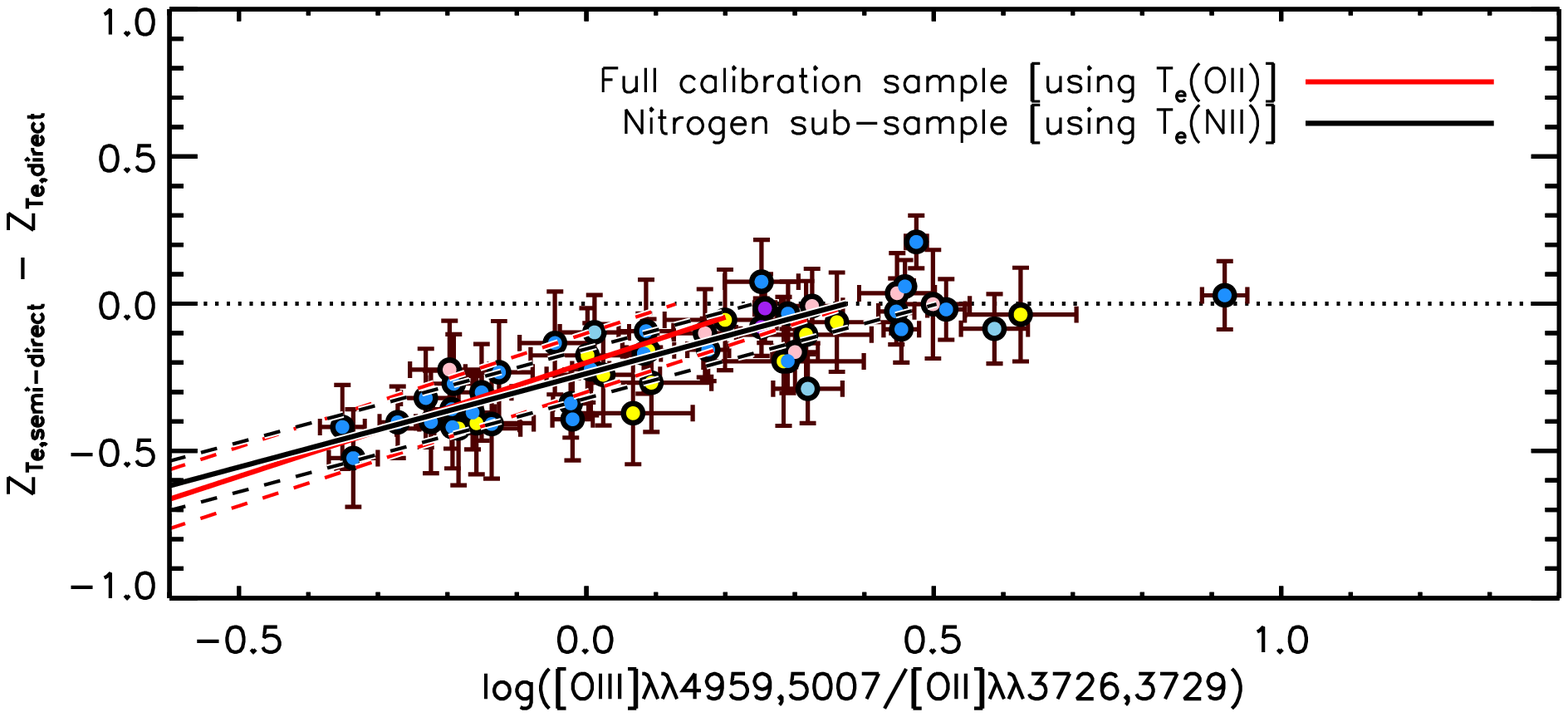}
 \caption{\textit{Top panel:} \TOIII{} -- \TNII{} relation for our nitrogen sub-sample of 53 systems (points). Fits to the \TOIII{} -- \TOII{} relation from the literature are also shown. \textit{Centre panel:} The relation between direct \ZTe{} and the hyperbolic semi-axis, $a$. Red lines indicate the least-squares fit to our full calibration sample, using \TOII{} (see Eq. \ref{eqn:semiaxis}). Black lines indicate the least-squares fit to our nitrogen sub-sample when using \TNII{}. \textit{Bottom panel:} Relation between the semi-direct \ZTe{} deficit and \OIII{}/\OII{}. Again, red lines denote our least-squares \OII{}-based fit (see Fig. \ref{fig:TOII-Zdiff}), black lines denote our least-squares alternative \NII{}-based fit.}
 \label{fig:Nitrogen_plots_1}
\end{figure}

The \TOIII{} -- \TNII{} relation is shown in the top panel of Fig. \ref{fig:Nitrogen_plots_1}. There are fewer systems with large differences between these temperatures than seen for the \TOIII{} -- \TOII{} relation of our full calibration sample (see Fig. \ref{fig:TOIII-TOII_old_relations}). The lack of systems in the upper part of this plane (\ie{}in region A of the schematic shown in Fig. \ref{fig:TOIII-TOII_schematic}) is mainly due to measured \TNII{} being lower than \TOII{} for those systems (outlined in grey in Fig. \ref{fig:ROII-RNII_and_TOII-TNII}). The lack of systems in the lower part of this plane (\ie{}region B in Fig. \ref{fig:TOIII-TOII_schematic}) is chiefly a selection bias: systems with higher metallicity have particularly weak \NII{}$\lambda5755$, and are therefore typically removed from samples selected on this line.

The centre panel of Fig. \ref{fig:Nitrogen_plots_1} shows the \ZTe{} -- $a$ relation. The red lines represent the least-squares fit for our full calibration sample, as shown in Fig. \ref{fig:ZTe-rT}. We find there is very little difference in the fit obtained when removing those systems that have $T(\tn{\textsc{Oii}}) > T(\tn{\textsc{Nii}}) + 2000$ K (shown with grey outlines in Fig. \ref{fig:ROII-RNII_and_TOII-TNII}). This further demonstrates that our new calibration performs well for systems on and off the expected relations for individual \HII{} regions.

The black lines in the centre panel of Fig. \ref{fig:Nitrogen_plots_1} represent the least-squares fit to our nitrogen sub-sample when using \TNII{} to obtain \OpH{}. This relation is steeper than that using \TOII{} to obtain \OpH{}. The steepness is driven by the small number of systems with $a \gtrsim 17000$, which have higher measured \TNII{} than \TOII{}. It is important to note that all these systems have $\tn{log(\Opp{}/\Op{})} > 0.5$, so are relatively insensitive to how \Op{} is estimated anyway. The majority of the \TNII{} sub-sample is still well fit by our \OII{}-based calibration. Also, the semi-direct \ZTe{} deficit correction we derive is very similar for both the full calibration sample and nitrogen sub-sample (bottom panel). Nonetheless, we still provide the revised least-squares fits to our hyperbolic semi axis, $a$, and correction factor, $f\sub{cor}$, when using \TNII{}:
\begin{align}\label{eqn:NII_calib}
a\sub{N\textsc{ii}} = & -19649.06\error{2504.29}\;\tn{\ZTe{}} + 177133.93\error{20827.90}\;\;,
\end{align}

\begin{align} \label{eqn:NII_correction_factor}
f\sub{cor,N\textsc{ii}} = \bigg{\{} & \begin{array}{ll}
													0.64\,(x - 0.24) & \textnormal{for } x \leq 0.37\\
													0.0 & \textnormal{for } x > 0.37
									 \end{array} \;\;.
\end{align}

As with our \OII{}-based calibration, our \NII{}-based calibration returns relatively accurate \ZTe{} estimates for the full range of systems available (as discussed in \ref{sec:correcting_ZTe}).

%%%%%%%%%%%%%%%%%%%%
\section{MaNGA \texorpdfstring{EW(H$\alpha$)}{EW(Ha)} maps and tables} \label{sec:Appendix B}

\begin{figure*}
\centering
\begin{tabular}{@{}c@{} @{}c@{} @{}c@{}}%{ccc}
 \includegraphics[angle=0,width=0.33\linewidth]{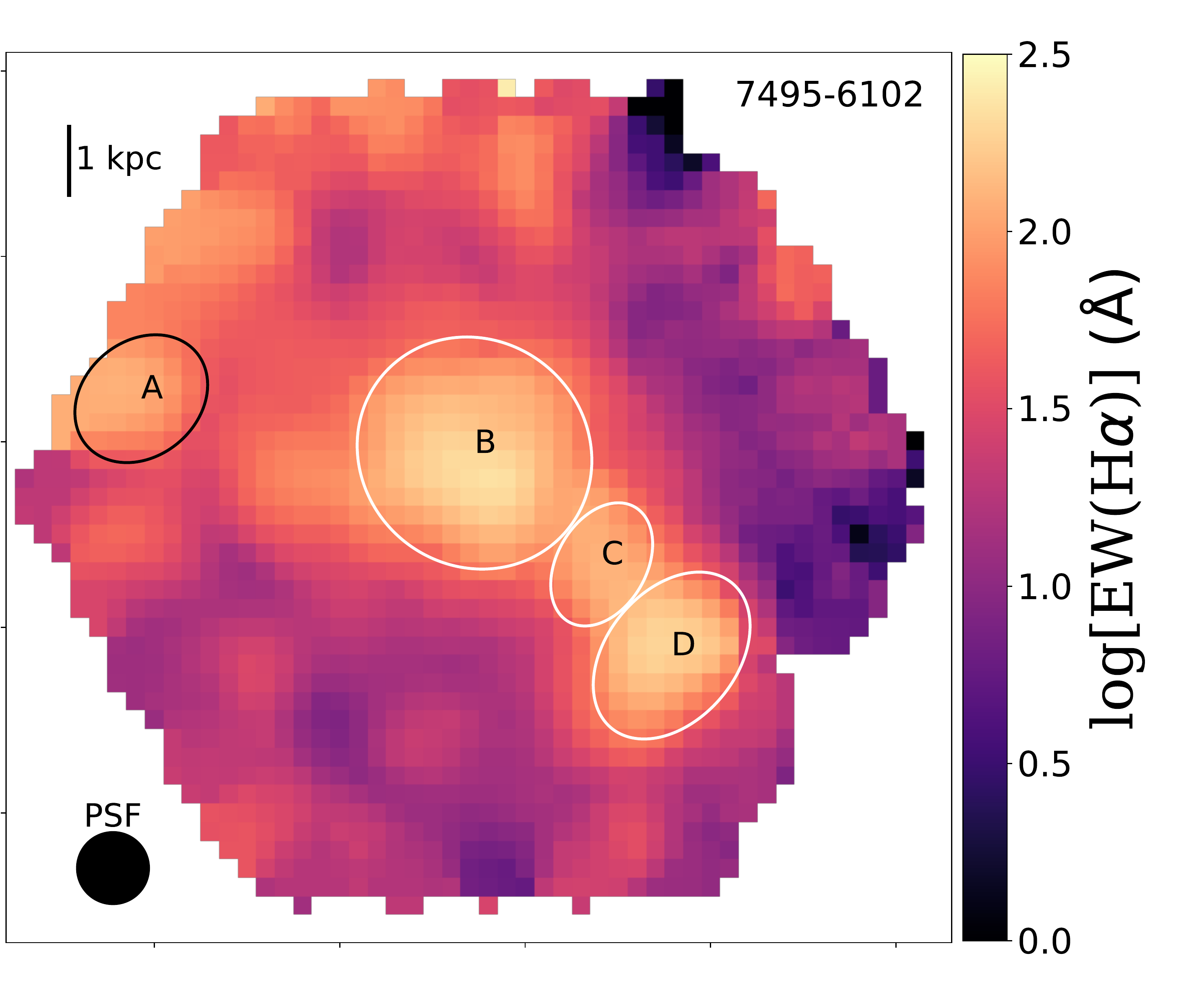} &
 \includegraphics[angle=0,width=0.33\linewidth]{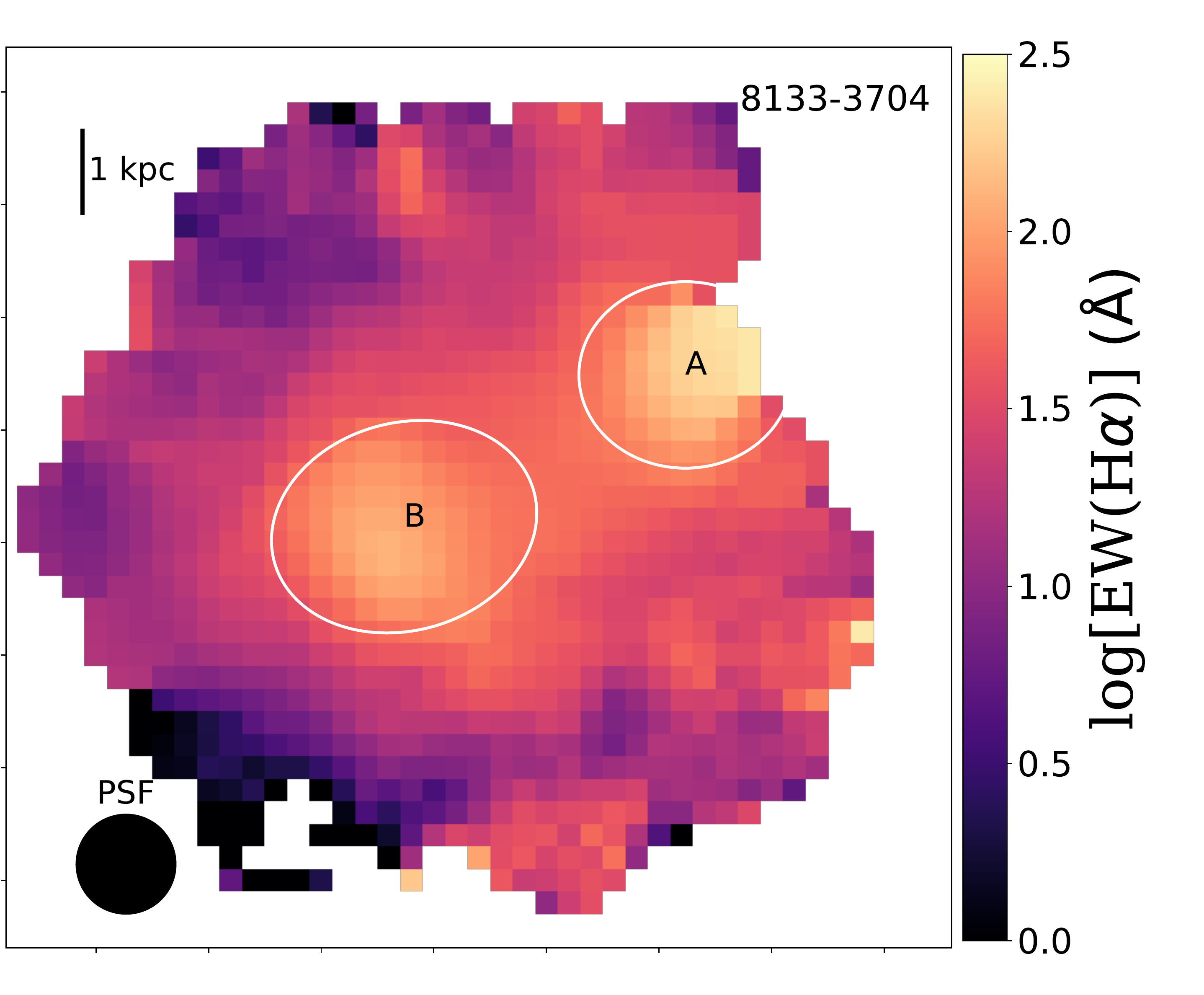} &
  \includegraphics[angle=0,width=0.33\linewidth]{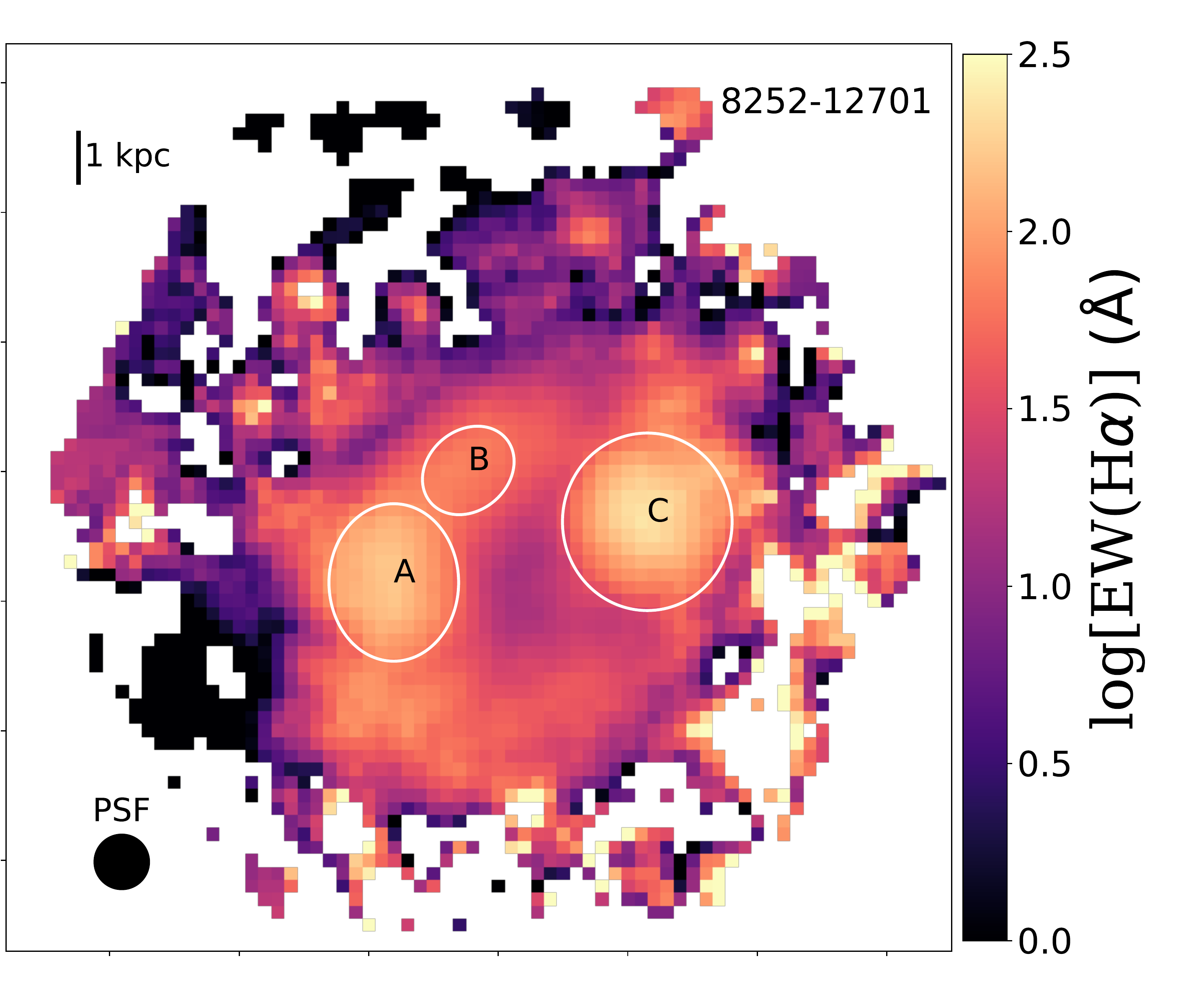} \\
 \includegraphics[angle=0,width=0.33\linewidth]{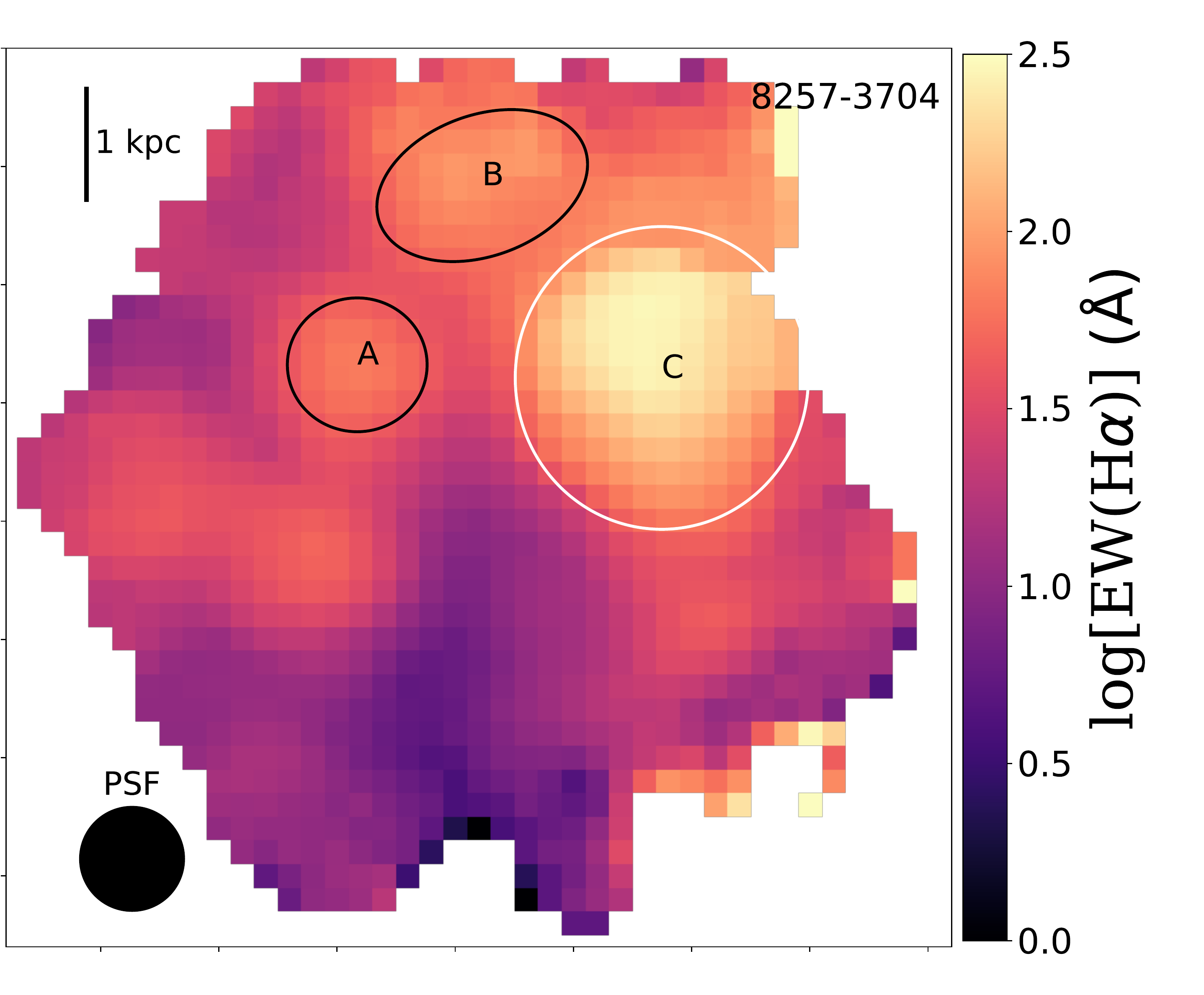} &
  \includegraphics[angle=0,width=0.33\linewidth]{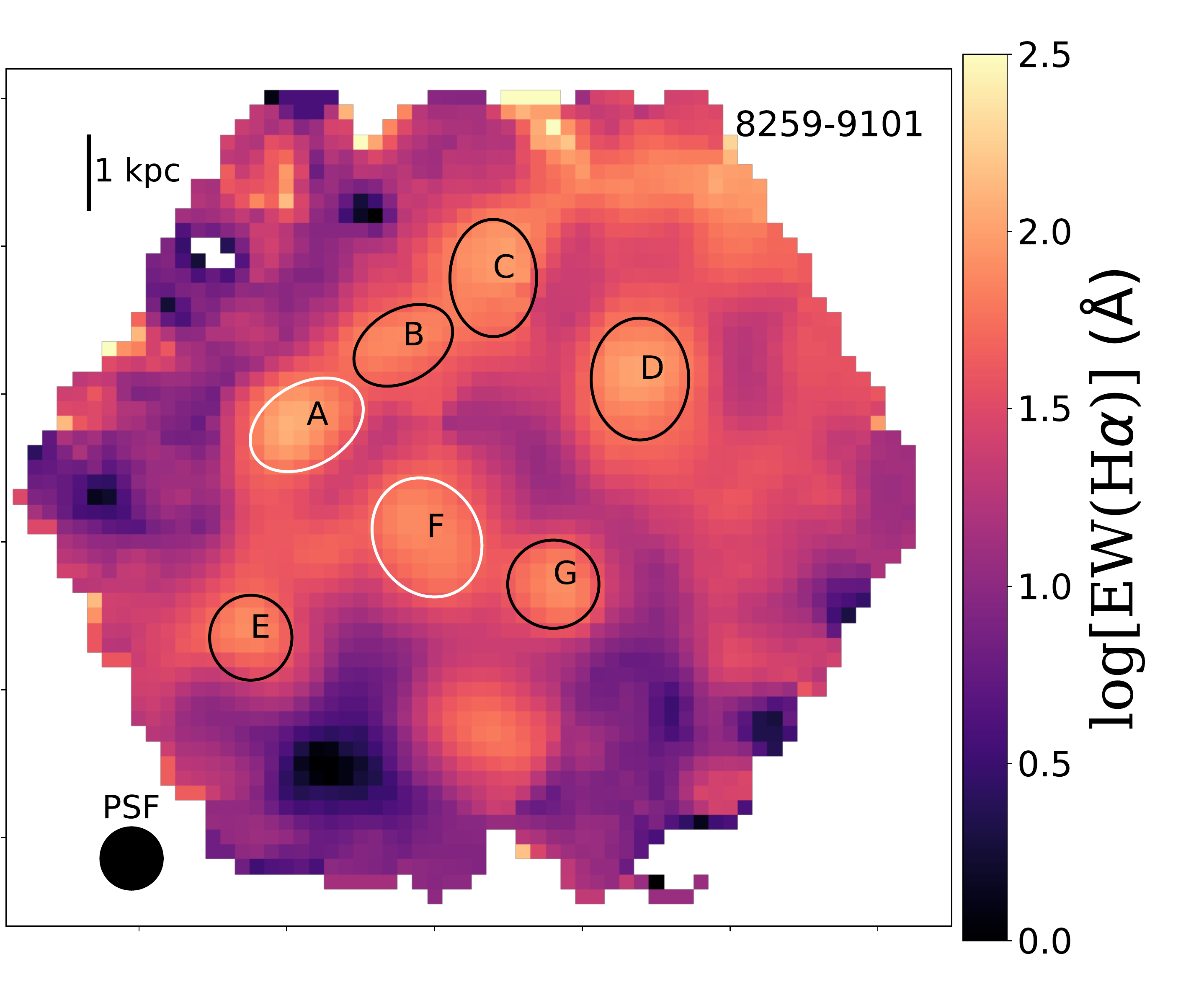} &
 \includegraphics[angle=0,width=0.33\linewidth]{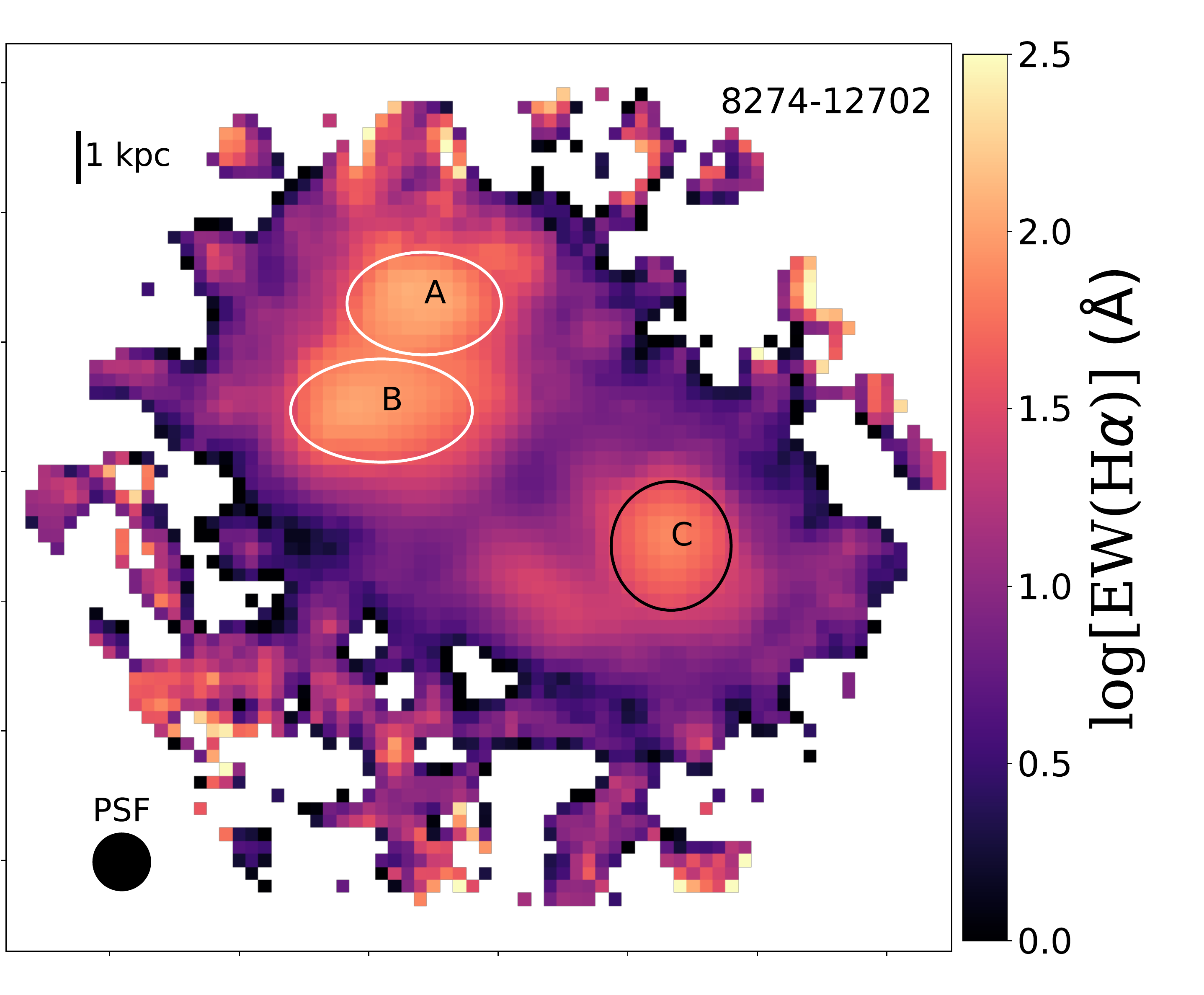} \\
 \includegraphics[angle=0,width=0.33\linewidth]{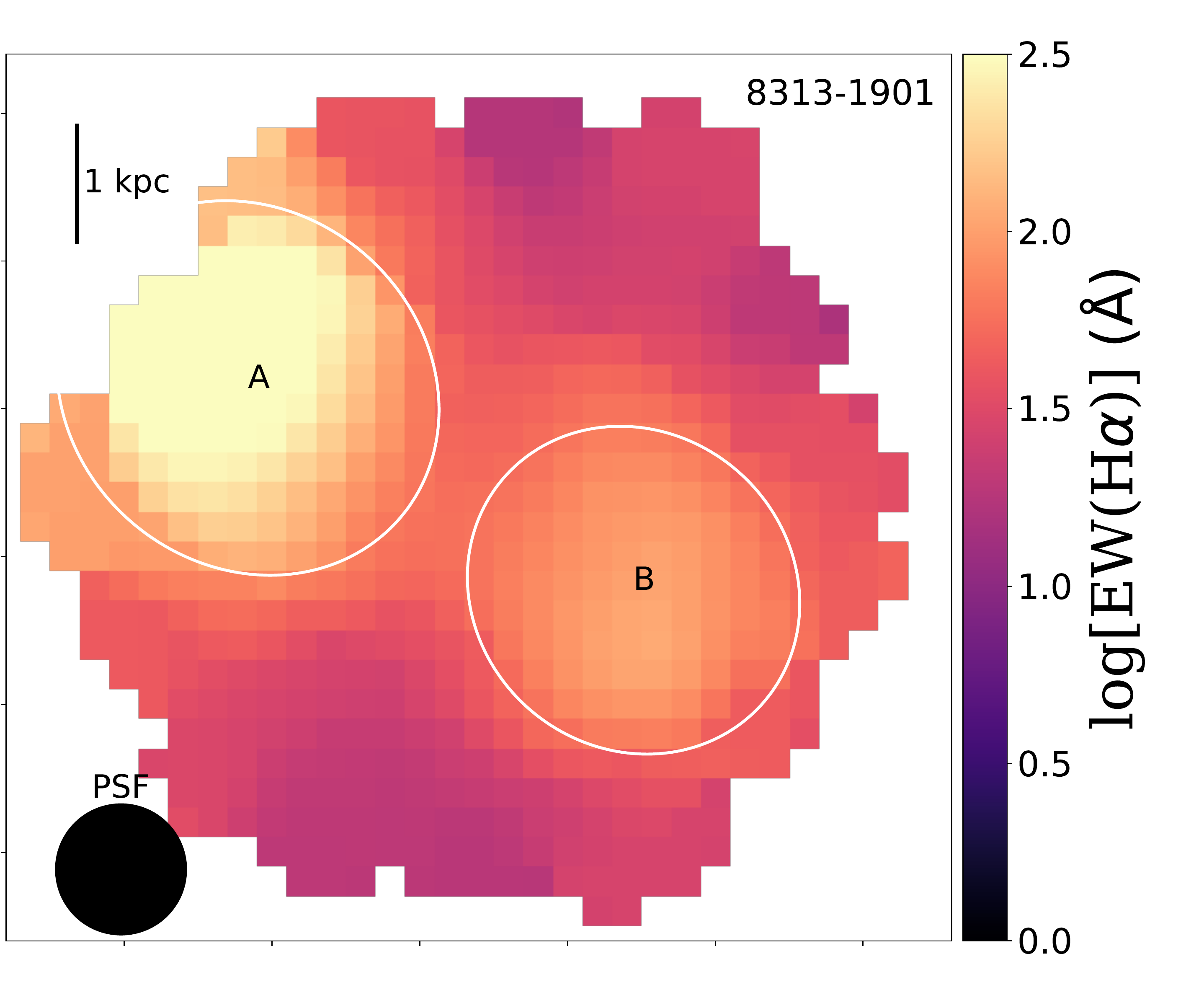} &
 \includegraphics[angle=0,width=0.33\linewidth]{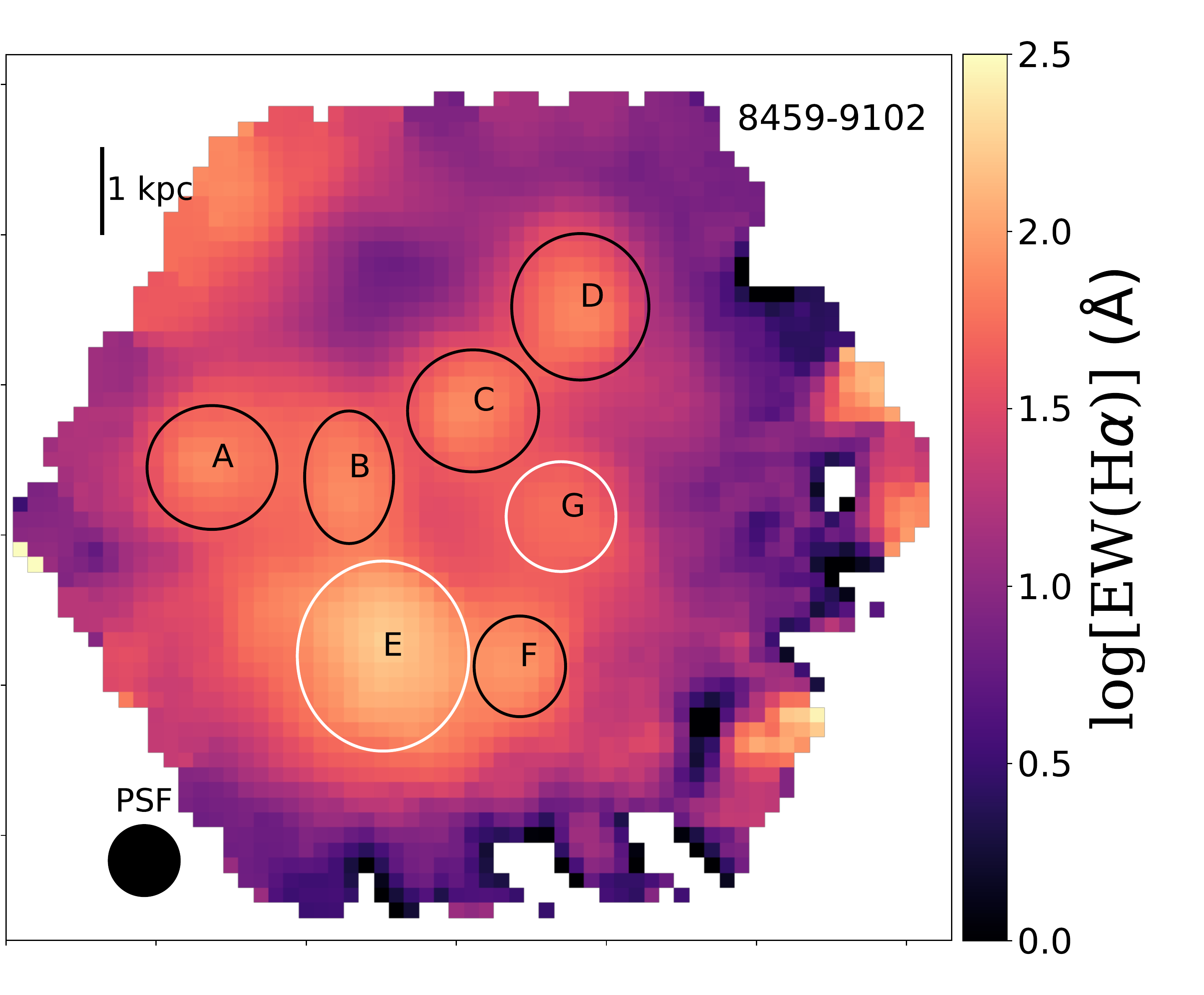} &
 \includegraphics[angle=0,width=0.33\linewidth]{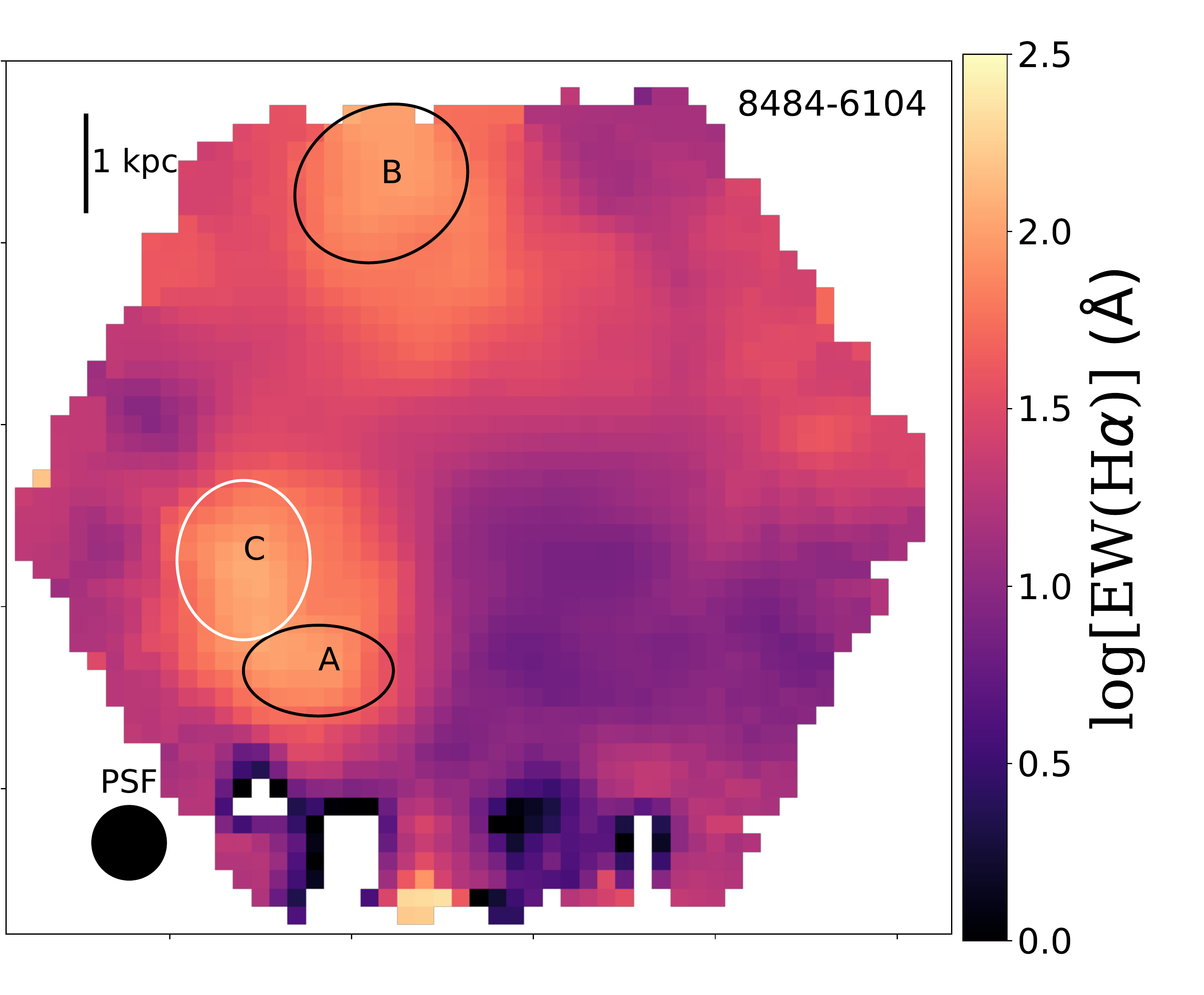} \\
 \includegraphics[angle=0,width=0.33\linewidth]{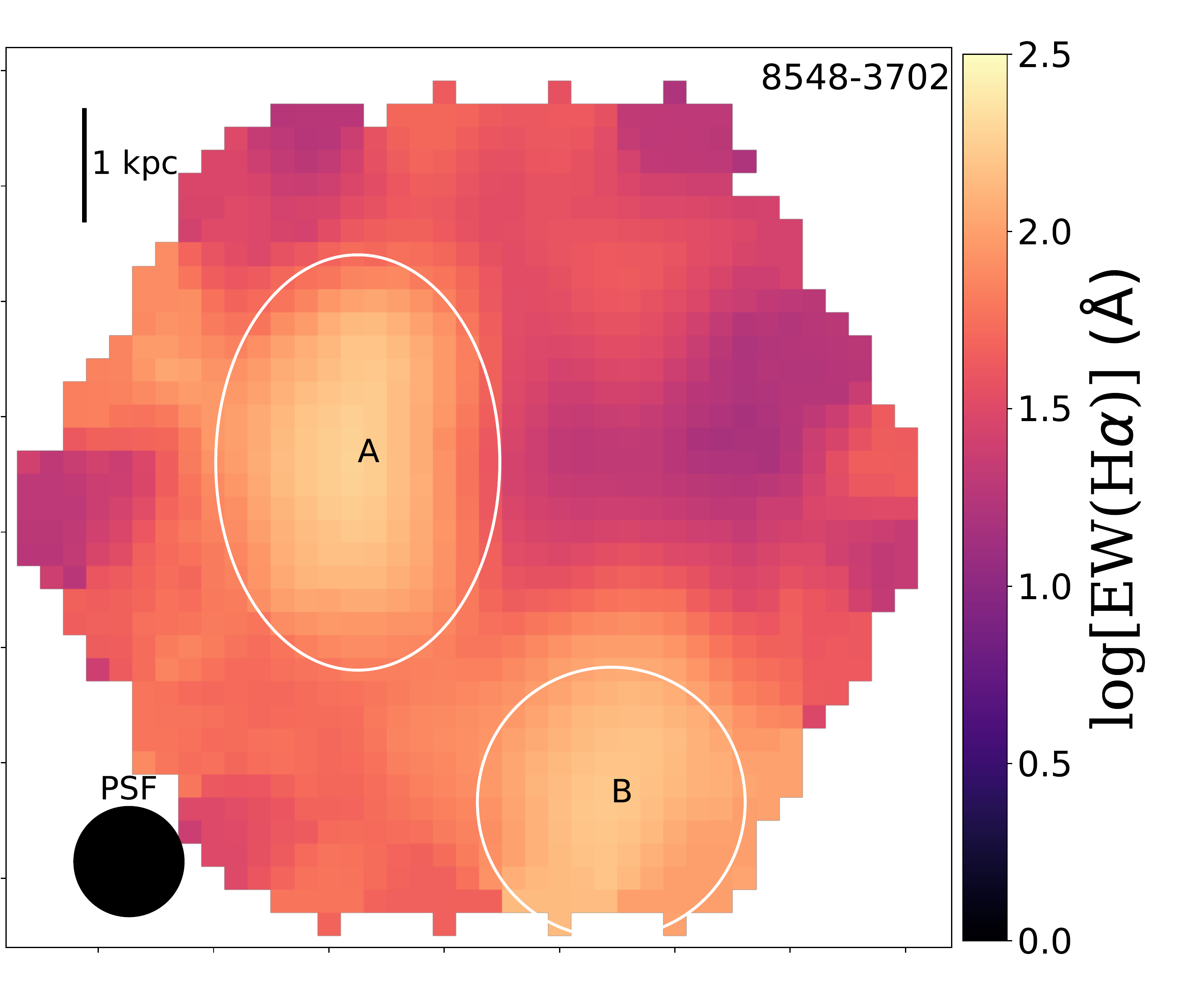} &
 \includegraphics[angle=0,width=0.33\linewidth]{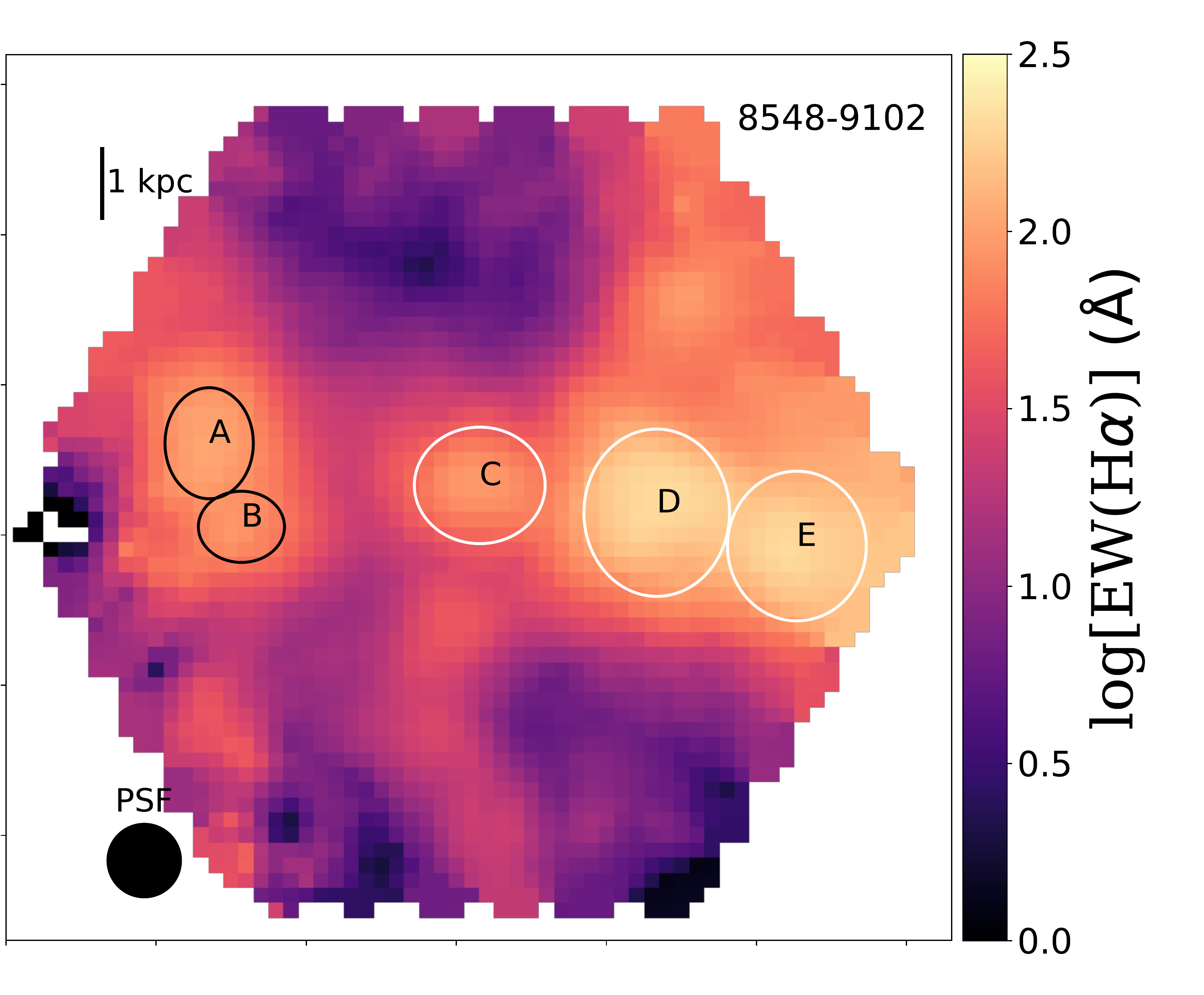} &
 \includegraphics[angle=0,width=0.33\linewidth]{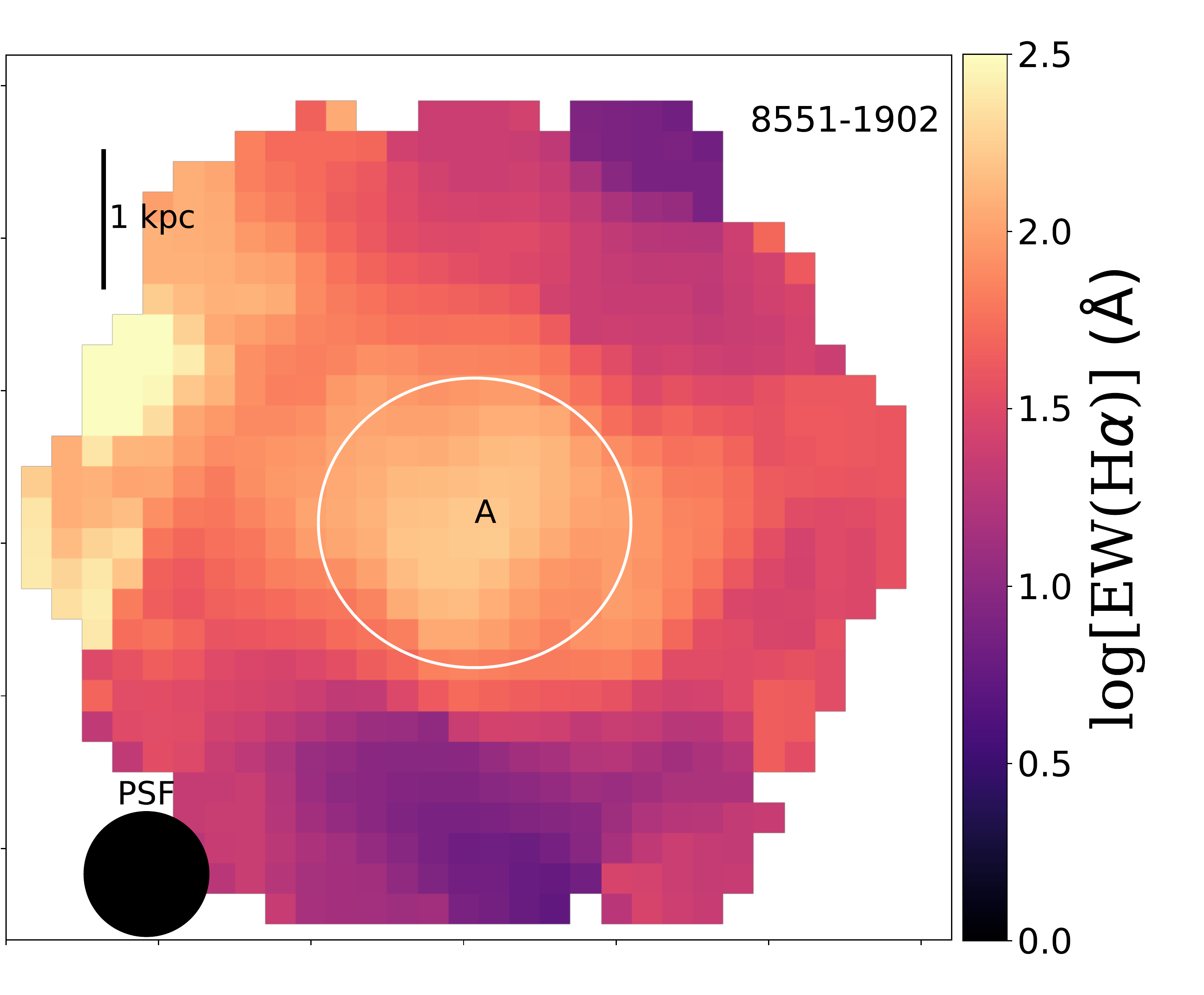} \\
\end{tabular}
 \caption{EW(H$\alpha$) maps for the \NumMaNGAs{} galaxies from our MaNGA sample containing \HII{} blobs with auroral line detections (see \S \ref{sec:MaNGA sample}). \HII{} blobs are shown and numbered in each panel. White ellipses signify \HII{} blobs with S/N\OIII{}$\lambda{}4363 \geq 3.0$, which are used for our \Te{} analysis in \S \ref{sec:Te_and_ZTe}.}
 \label{fig:HII_blob_maps}
\end{figure*}

In Fig. \ref{fig:HII_blob_maps}, we present the H$\alpha$ EW maps for our sample of MaNGA galaxies (see \S \ref{sec:MaNGA sample}). White ellipses signify \HII{} blobs with S/N\OIII{}$\lambda{}4363 \geq 3.0$. All spaxels within an ellipse with EW(H$\alpha$) $>50$\AA{} and S/N(H$\alpha$) $>50$ are used to determine emission line fluxes. Table \ref{tab:MaNGA_HIIblob_data} provides the emission line intensities measured for each of the MaNGA \HII{} blobs in our sample, along with their derived electron temperatures and oxygen abundances. Table \ref{tab:MaNGA_global_data} provides similar data from our global MaNGA galaxy spectra, and includes information on their position, stellar mass, extinction, and electron density.

% %------------------
\setlength{\rotFPtop}{0pt plus 1fil} %To keep a table on the page when using \sidewaystable
\setlength{\rotFPbot}{0pt plus 1fil} %To keep a table on the page when using \sidewaystable
\renewcommand{\arraystretch}{1.2} %To adapt row height

%------------------
%HII BLOB DATA:
\onecolumn
\begin{longtable}{p{1.1in}cccccc}
\caption{Fluxes in units of $10^{-15}$ erg s$^{-1}$ cm$^{2}$, corrected for Galactic foreground extinction, for the 24 \HII{} blobs with S/N([\textsc{Oiii}]$\lambda$4363) $\geq3.0$ from our MaNGA galaxies. Electron densities, temperatures, and oxygen abundances are also provided.}
\endfirsthead
\caption{continued.}
\endhead
\hline \hline
System \dotfill & 7495-6102B  & 7495-6102C  & 7495-6102D  & 8133-3704A  & 8133-3704B
 & 8252-12701A \\
\hline
$[$O\textsc{ii}]$\lambda$3726 \dotfill & 14.77$\pm$0.24 & 1.68$\pm$0.05 &
1.42$\pm$0.03 & 2.50$\pm$0.04 & 3.13$\pm$0.03 & 1.02$\pm$0.01 \\
$[$O\textsc{ii}]$\lambda$3729 \dotfill & 21.81$\pm$0.24 & 2.43$\pm$0.05 &
2.09$\pm$0.03 & 3.73$\pm$0.04 & 5.22$\pm$0.03 & 1.60$\pm$0.01 \\
$[$S\textsc{ii}]$\lambda$4073 \dotfill & 0.09$\pm$0.03 & 0.01$\pm$0.01 & -- &
0.05$\pm$0.04 & 0.02$\pm$0.03 & -- \\
$[$O\textsc{iii}]$\lambda$4363 \dotfill & 0.34$\pm$0.03 & 0.03$\pm$0.01 &
0.03$\pm$0.01 & 0.14$\pm$0.04 & 0.09$\pm$0.01 & 0.05$\pm$0.01 \\
H$\beta\ \lambda$4861 \dotfill & 14.79$\pm$0.05 & 1.38$\pm$0.01 &
1.26$\pm$0.02 & 3.08$\pm$0.04 & 2.87$\pm$0.02 & 0.96$\pm$0.01 \\
$[$O\textsc{iii}]$\lambda$4959 \dotfill & 16.99$\pm$0.05 & 1.28$\pm$0.01 &
1.37$\pm$0.01 & 4.90$\pm$0.05 & 2.77$\pm$0.01 & 1.03$\pm$0.01 \\
$[$O\textsc{iii}]$\lambda$5007 \dotfill & 48.39$\pm$0.22 & 3.71$\pm$0.02 &
4.08$\pm$0.02 & 14.09$\pm$0.04 & 8.16$\pm$0.04 & 3.01$\pm$0.02 \\
$[$N\textsc{ii}]$\lambda$5755 \dotfill & 0.070$\pm$0.002 & -- & -- & -- & -- &
-- \\
$[$S\textsc{iii}]$\lambda$6312 \dotfill & 0.22$\pm$0.01 & 0.026$\pm$0.001 &
0.03$\pm$0.01 & 0.06$\pm$0.02 & 0.043$\pm$0.005 & 0.038$\pm$0.004 \\
H$\alpha\ \lambda$6563 \dotfill & 42.37$\pm$0.16 & 4.10$\pm$0.01 &
3.99$\pm$0.01 & 11.80$\pm$0.04 & 9.22$\pm$0.03 & 2.86$\pm$0.01 \\
$[$N\textsc{ii}]$\lambda$6584 \dotfill & 2.992$\pm$0.001 & 0.287$\pm$0.001 &
0.207$\pm$0.001 & 0.449$\pm$0.001 & 0.653$\pm$0.001 & 0.138$\pm$0.001 \\
$[$S\textsc{ii}]$\lambda$6716 \dotfill & 5.46$\pm$0.01 & 0.63$\pm$0.01 &
0.48$\pm$0.01 & 1.13$\pm$0.02 & 1.33$\pm$0.02 & 0.32$\pm$0.01 \\
$[$S\textsc{ii}]$\lambda$6731 \dotfill & 4.01$\pm$0.01 & 0.47$\pm$0.01 &
0.35$\pm$0.01 & 0.70$\pm$0.02 & 1.00$\pm$0.02 & 0.26$\pm$0.01 \\
$[$O\textsc{ii}]$\lambda$7320 \dotfill & 0.43$\pm$0.01 & 0.06$\pm$0.01 &
0.052$\pm$0.002 & 0.14$\pm$0.01 & 0.11$\pm$0.01 & 0.020$\pm$0.004 \\
$[$O\textsc{ii}]$\lambda$7330 \dotfill & 0.40$\pm$0.01 & 0.05$\pm$0.01 &
0.033$\pm$0.001 & 0.15$\pm$0.01 & 0.15$\pm$0.01 & 0.022$\pm$0.004 \\
$[$S\textsc{iii}]$\lambda$9069 \dotfill & 2.79$\pm$0.03 & 0.37$\pm$0.01 &
0.190$\pm$0.004 & 0.36$\pm$0.03 & 0.34$\pm$0.01 & 0.19$\pm$0.01 \\
\hline
$N\sub{e}\,/\,\tn{cm}^{-3}$ \dotfill & 100$\pm$13 & 110$\pm$25 & 95$\pm$38 &
31$\pm$32 & 115$\pm$40 & 181$\pm$57 \\
\TOII{}$\,/\,$K \dotfill & 11110$\pm$82 & 12085$\pm$204 & 10890$\pm$226 &
15636$\pm$431 & 12647$\pm$276 & 8795$\pm$193 \\
\TNII{}$\,/\,$K \dotfill & 14421$\pm$286 & -- & -- & -- & -- & -- \\
\TOIII{}$\,/\,$K \dotfill & 9775$\pm$273 & 10267$\pm$1130 & 10022$\pm$1087 &
11227$\pm$1248 & 11528$\pm$517 & 13486$\pm$1186 \\
(\OpH{}) $\times 10^{4}$ \dotfill & 0.68$\pm$0.16 & 0.63$\pm$0.15 &
0.83$\pm$0.21 & 0.24$\pm$0.06 & 0.51$\pm$0.12 & 1.28$\pm$0.32 \\
(\OppH{}) $\times 10^{4}$ \dotfill & 1.13$\pm$0.22 & 0.78$\pm$0.28 &
1.01$\pm$0.37 & 0.98$\pm$0.35 & 0.56$\pm$0.12 & 0.39$\pm$0.10 \\
\ZTe{} \dotfill & 8.26$\pm$0.06 & 8.15$\pm$0.10 & 8.26$\pm$0.10 &
8.09$\pm$0.13 & 8.03$\pm$0.07 & 8.22$\pm$0.09 \\
 \\
\hline \hline
%-----------------------------
System \dotfill & 8252-12701B  & 8252-12701C  & 8257-3704C  & 8259-9101A  & 8259-9101F
 & 8274-12702A \\
\hline
$[$O\textsc{ii}]$\lambda$3726 \dotfill & 0.39$\pm$0.01 & 0.52$\pm$0.02 &
8.36$\pm$0.13 & 0.176$\pm$0.001 & 2.35$\pm$0.01 & 0.56$\pm$0.01 \\
$[$O\textsc{ii}]$\lambda$3729 \dotfill & 0.65$\pm$0.01 & 0.81$\pm$0.02 &
12.85$\pm$0.13 & 0.287$\pm$0.001 & 3.48$\pm$0.01 & 0.86$\pm$0.01 \\
$[$S\textsc{ii}]$\lambda$4073 \dotfill & 0.01$\pm$0.01 & -- & 0.12$\pm$0.03 &
0.002$\pm$0.003 & 0.05$\pm$0.02 & -- \\
$[$O\textsc{iii}]$\lambda$4363 \dotfill & 0.013$\pm$0.004 & 0.021$\pm$0.005 &
0.21$\pm$0.02 & 0.016$\pm$0.003 & 0.05$\pm$0.01 & 0.020$\pm$0.003 \\
H$\beta\ \lambda$4861 \dotfill & 0.397$\pm$0.005 & 0.56$\pm$0.01 &
8.56$\pm$0.03 & 0.247$\pm$0.005 & 2.06$\pm$0.01 & 0.46$\pm$0.01 \\
$[$O\textsc{iii}]$\lambda$4959 \dotfill & 0.42$\pm$0.01 & 0.674$\pm$0.003 &
9.99$\pm$0.04 & 0.361$\pm$0.003 & 2.00$\pm$0.01 & 0.35$\pm$0.01 \\
$[$O\textsc{iii}]$\lambda$5007 \dotfill & 1.22$\pm$0.01 & 1.96$\pm$0.01 &
28.33$\pm$0.10 & 1.101$\pm$0.003 & 5.95$\pm$0.02 & 1.062$\pm$0.005 \\
$[$N\textsc{ii}]$\lambda$5755 \dotfill & -- & -- & -- & -- & -- & -- \\
$[$S\textsc{iii}]$\lambda$6312 \dotfill & 0.007$\pm$0.005 & 0.005$\pm$0.005 &
0.15$\pm$0.01 & 0.003$\pm$0.004 & 0.03$\pm$0.01 & -- \\
H$\alpha\ \lambda$6563 \dotfill & 1.19$\pm$0.01 & 1.64$\pm$0.01 &
30.28$\pm$0.14 & 0.759$\pm$0.004 & 6.22$\pm$0.04 & 1.46$\pm$0.01 \\
$[$N\textsc{ii}]$\lambda$6584 \dotfill & 0.077$\pm$0.001 & 0.079$\pm$0.001 &
2.025$\pm$0.001 & 0.025$\pm$0.001 & 0.500$\pm$0.001 & 0.125$\pm$0.001 \\
$[$S\textsc{ii}]$\lambda$6716 \dotfill & 0.142$\pm$0.003 & 0.175$\pm$0.005 &
3.32$\pm$0.02 & 0.059$\pm$0.002 & 0.90$\pm$0.01 & 0.22$\pm$0.01 \\
$[$S\textsc{ii}]$\lambda$6731 \dotfill & 0.111$\pm$0.003 & 0.150$\pm$0.005 &
2.58$\pm$0.02 & 0.039$\pm$0.002 & 0.60$\pm$0.01 & 0.17$\pm$0.01 \\
$[$O\textsc{ii}]$\lambda$7320 \dotfill & 0.012$\pm$0.003 & 0.015$\pm$0.002 &
0.40$\pm$0.01 & 0.006$\pm$0.002 & 0.07$\pm$0.01 & 0.010$\pm$0.004 \\
$[$O\textsc{ii}]$\lambda$7330 \dotfill & 0.007$\pm$0.003 & 0.011$\pm$0.002 &
0.41$\pm$0.01 & 0.008$\pm$0.002 & 0.05$\pm$0.01 & 0.012$\pm$0.004 \\
$[$S\textsc{iii}]$\lambda$9069 \dotfill & 0.08$\pm$0.01 & 0.12$\pm$0.01 &
2.35$\pm$0.02 & 0.040$\pm$0.002 & 0.42$\pm$0.01 & 0.13$\pm$0.02 \\
\hline
$N\sub{e}\,/\,\tn{cm}^{-3}$ \dotfill & 145$\pm$41 & 243$\pm$71 & 140$\pm$73 &
50$\pm$29 & 53$\pm$17 & 136$\pm$75 \\
\TOII{}$\,/\,$K \dotfill & 9486$\pm$281 & 9563$\pm$256 & 13468$\pm$553 &
13168$\pm$463 & 10439$\pm$127 & 8548$\pm$343 \\
\TNII{}$\,/\,$K \dotfill & -- & -- & -- & -- & -- & -- \\
\TOIII{}$\,/\,$K \dotfill & 11305$\pm$1274 & 11307$\pm$994 & 10085$\pm$410 &
12839$\pm$1035 & 10435$\pm$706 & 14420$\pm$1094 \\
(\OpH{}) $\times 10^{4}$ \dotfill & 1.09$\pm$0.29 & 0.95$\pm$0.25 &
0.44$\pm$0.12 & 0.26$\pm$0.07 & 0.91$\pm$0.22 & 1.56$\pm$0.47 \\
(\OppH{}) $\times 10^{4}$ \dotfill & 0.65$\pm$0.22 & 0.74$\pm$0.21 &
1.01$\pm$0.23 & 0.64$\pm$0.16 & 0.78$\pm$0.20 & 0.24$\pm$0.06 \\
\ZTe{} \dotfill & 8.24$\pm$0.09 & 8.23$\pm$0.08 & 8.16$\pm$0.08 &
7.95$\pm$0.09 & 8.23$\pm$0.08 & 8.26$\pm$0.11 \\
\hline
\pagebreak
\hline \hline
System \dotfill & 8274-12702B  & 8313-1901A  & 8313-1901B  & 8459-9102E  & 8459-9102G
 & 8484-6104C \\
\hline
$[$O\textsc{ii}]$\lambda$3726 \dotfill & 1.81$\pm$0.05 & 17.13$\pm$0.34 &
17.04$\pm$0.19 & 6.69$\pm$0.11 & 0.54$\pm$0.01 & 0.81$\pm$0.03 \\
$[$O\textsc{ii}]$\lambda$3729 \dotfill & 2.71$\pm$0.05 & 25.40$\pm$0.34 &
24.38$\pm$0.19 & 10.05$\pm$0.11 & 0.82$\pm$0.01 & 1.24$\pm$0.03 \\
$[$S\textsc{ii}]$\lambda$4073 \dotfill & 0.000$\pm$0.020 & -- & -- &
0.07$\pm$0.03 & 0.006$\pm$0.005 & 0.02$\pm$0.02 \\
$[$O\textsc{iii}]$\lambda$4363 \dotfill & 0.030$\pm$0.003 & 0.64$\pm$0.04 &
0.18$\pm$0.03 & 0.10$\pm$0.02 & 0.009$\pm$0.003 & 0.046$\pm$0.004 \\
H$\beta\ \lambda$4861 \dotfill & 1.52$\pm$0.01 & 17.21$\pm$0.03 &
14.58$\pm$0.05 & 5.39$\pm$0.02 & 0.367$\pm$0.003 & 0.78$\pm$0.01 \\
$[$O\textsc{iii}]$\lambda$4959 \dotfill & 1.15$\pm$0.01 & 22.21$\pm$0.12 &
13.75$\pm$0.04 & 4.63$\pm$0.02 & 0.181$\pm$0.003 & 0.80$\pm$0.01 \\
$[$O\textsc{iii}]$\lambda$5007 \dotfill & 3.45$\pm$0.01 & 62.65$\pm$0.33 &
40.96$\pm$0.13 & 13.75$\pm$0.07 & 0.561$\pm$0.004 & 2.66$\pm$0.01 \\
$[$N\textsc{ii}]$\lambda$5755 \dotfill & -- & -- & -- & -- & -- & -- \\
$[$S\textsc{iii}]$\lambda$6312 \dotfill & 0.03$\pm$0.01 & 0.24$\pm$0.01 &
0.22$\pm$0.02 & 0.06$\pm$0.02 & 0.004$\pm$0.002 & 0.02$\pm$0.01 \\
H$\alpha\ \lambda$6563 \dotfill & 4.90$\pm$0.01 & 52.84$\pm$0.35 &
45.95$\pm$0.23 & 16.82$\pm$0.07 & 1.137$\pm$0.004 & 2.33$\pm$0.01 \\
$[$N\textsc{ii}]$\lambda$6584 \dotfill & 0.480$\pm$0.001 & 4.266$\pm$0.002 &
5.165$\pm$0.002 & 1.253$\pm$0.001 & 0.136$\pm$0.001 & 0.132$\pm$0.001 \\
$[$S\textsc{ii}]$\lambda$6716 \dotfill & 0.74$\pm$0.02 & 5.82$\pm$0.04 &
6.25$\pm$0.04 & 2.38$\pm$0.01 & 0.230$\pm$0.002 & 0.26$\pm$0.01 \\
$[$S\textsc{ii}]$\lambda$6731 \dotfill & 0.59$\pm$0.02 & 4.28$\pm$0.04 &
4.37$\pm$0.04 & 1.66$\pm$0.01 & 0.156$\pm$0.002 & 0.19$\pm$0.01 \\
$[$O\textsc{ii}]$\lambda$7320 \dotfill & 0.04$\pm$0.01 & 0.60$\pm$0.03 &
0.50$\pm$0.03 & 0.18$\pm$0.01 & 0.014$\pm$0.002 & 0.02$\pm$0.01 \\
$[$O\textsc{ii}]$\lambda$7330 \dotfill & 0.04$\pm$0.01 & 0.45$\pm$0.03 &
0.34$\pm$0.03 & 0.13$\pm$0.01 & 0.009$\pm$0.001 & 0.01$\pm$0.01 \\
$[$S\textsc{iii}]$\lambda$9069 \dotfill & 0.38$\pm$0.02 & 3.19$\pm$0.04 &
3.08$\pm$0.01 & 1.04$\pm$0.01 & 0.076$\pm$0.001 & 0.147$\pm$0.005 \\
\hline
$N\sub{e}\,/\,\tn{cm}^{-3}$ \dotfill & 163$\pm$63 & 100$\pm$24 & 73$\pm$23 &
72$\pm$21 & 59$\pm$20 & 97$\pm$45 \\
\TOII{}$\,/\,$K \dotfill & 8935$\pm$244 & 11290$\pm$132 & 10003$\pm$122 &
9595$\pm$102 & 9259$\pm$99 & 13005$\pm$546 \\
\TNII{}$\,/\,$K \dotfill & -- & -- & -- & -- & -- & -- \\
\TOIII{}$\,/\,$K \dotfill & 10613$\pm$402 & 11125$\pm$278 & 8599$\pm$405 &
9975$\pm$648 & 13420$\pm$1948 & 13925$\pm$584 \\
(\OpH{}) $\times 10^{4}$ \dotfill & 1.68$\pm$0.44 & 0.60$\pm$0.14 &
1.27$\pm$0.30 & 1.40$\pm$0.33 & 1.46$\pm$0.35 & 1.28$\pm$0.66 \\
(\OppH{}) $\times 10^{4}$ \dotfill & 0.58$\pm$0.12 & 0.81$\pm$0.15 &
1.53$\pm$0.37 & 0.81$\pm$0.21 & 0.19$\pm$0.07 & 0.38$\pm$0.08 \\
\ZTe{} \dotfill & 8.35$\pm$0.09 & 8.15$\pm$0.06 & 8.45$\pm$0.07 &
8.34$\pm$0.08 & 8.22$\pm$0.09 & 7.90$\pm$0.06 \\
 \\
\hline \hline
System \dotfill & 8548-3702A  & 8548-3702B  & 8548-9102C  & 8548-9102D  & 8548-9102E
 & 8551-1902A \\
\hline
$[$O\textsc{ii}]$\lambda$3726 \dotfill & 14.36$\pm$0.22 & 5.40$\pm$0.11 &
3.42$\pm$0.03 & 8.37$\pm$0.15 & 4.12$\pm$0.07 & 15.67$\pm$0.26 \\
$[$O\textsc{ii}]$\lambda$3729 \dotfill & 22.20$\pm$0.22 & 8.61$\pm$0.11 &
4.73$\pm$0.03 & 12.03$\pm$0.15 & 6.07$\pm$0.07 & 20.91$\pm$0.26 \\
$[$S\textsc{ii}]$\lambda$4073 \dotfill & 0.10$\pm$0.01 & 0.05$\pm$0.06 &
0.06$\pm$0.01 & 0.11$\pm$0.03 & 0.06$\pm$0.01 & 0.11$\pm$0.04 \\
$[$O\textsc{iii}]$\lambda$4363 \dotfill & 0.30$\pm$0.08 & 0.11$\pm$0.02 &
0.08$\pm$0.01 & 0.16$\pm$0.01 & 0.13$\pm$0.02 & 0.33$\pm$0.05 \\
H$\beta\ \lambda$4861 \dotfill & 12.71$\pm$0.04 & 4.50$\pm$0.02 &
2.84$\pm$0.01 & 6.77$\pm$0.04 & 3.54$\pm$0.01 & 16.14$\pm$0.07 \\
$[$O\textsc{iii}]$\lambda$4959 \dotfill & 11.68$\pm$0.03 & 4.02$\pm$0.01 &
3.22$\pm$0.01 & 7.08$\pm$0.01 & 4.06$\pm$0.02 & 17.06$\pm$0.05 \\
$[$O\textsc{iii}]$\lambda$5007 \dotfill & 34.66$\pm$0.08 & 11.84$\pm$0.04 &
9.63$\pm$0.05 & 21.10$\pm$0.09 & 12.08$\pm$0.04 & 52.14$\pm$0.24 \\
$[$N\textsc{ii}]$\lambda$5755 \dotfill & -- & -- & -- & -- & -- &
0.143$\pm$0.002 \\
$[$S\textsc{iii}]$\lambda$6312 \dotfill & 0.27$\pm$0.02 & 0.07$\pm$0.02 &
0.03$\pm$0.01 & 0.10$\pm$0.01 & 0.03$\pm$0.01 & 0.29$\pm$0.02 \\
H$\alpha\ \lambda$6563 \dotfill & 41.82$\pm$0.22 & 13.91$\pm$0.07 &
8.91$\pm$0.05 & 20.79$\pm$0.10 & 10.62$\pm$0.06 & 50.79$\pm$0.25 \\
$[$N\textsc{ii}]$\lambda$6584 \dotfill & 3.111$\pm$0.002 & 0.977$\pm$0.001 &
0.617$\pm$0.001 & 1.244$\pm$0.001 & 0.520$\pm$0.001 & 6.428$\pm$0.001 \\
$[$S\textsc{ii}]$\lambda$6716 \dotfill & 5.68$\pm$0.02 & 2.08$\pm$0.01 &
1.15$\pm$0.01 & 2.55$\pm$0.01 & 1.29$\pm$0.01 & 5.46$\pm$0.02 \\
$[$S\textsc{ii}]$\lambda$6731 \dotfill & 4.23$\pm$0.02 & 1.56$\pm$0.01 &
0.84$\pm$0.01 & 1.82$\pm$0.01 & 0.87$\pm$0.01 & 4.03$\pm$0.02 \\
$[$O\textsc{ii}]$\lambda$7320 \dotfill & 0.55$\pm$0.01 & 0.15$\pm$0.02 &
0.10$\pm$0.01 & 0.24$\pm$0.01 & 0.11$\pm$0.01 & 0.65$\pm$0.01 \\
$[$O\textsc{ii}]$\lambda$7330 \dotfill & 0.49$\pm$0.01 & 0.09$\pm$0.02 &
0.10$\pm$0.01 & 0.18$\pm$0.02 & 0.08$\pm$0.01 & 0.50$\pm$0.01 \\
$[$S\textsc{iii}]$\lambda$9069 \dotfill & 2.78$\pm$0.02 & 0.77$\pm$0.03 &
0.57$\pm$0.01 & 1.24$\pm$0.01 & 0.51$\pm$0.01 & 5.12$\pm$0.03 \\
\hline
$N\sub{e}\,/\,\tn{cm}^{-3}$ \dotfill & 108$\pm$40 & 113$\pm$27 & 96$\pm$28 &
83$\pm$21 & 57$\pm$16 & 103$\pm$29 \\
\TOII{}$\,/\,$K \dotfill & 11754$\pm$240 & 9161$\pm$113 & 11032$\pm$174 &
10176$\pm$112 & 9907$\pm$87 & 13004$\pm$208 \\
\TNII{}$\,/\,$K \dotfill & -- & -- & -- & -- & -- & 14120$\pm$350 \\
\TOIII{}$\,/\,$K \dotfill & 10603$\pm$983 & 10807$\pm$693 & 10428$\pm$456 &
10101$\pm$230 & 11373$\pm$647 & 9576$\pm$460 \\
(\OpH{}) $\times 10^{4}$ \dotfill & 0.67$\pm$0.16 & 1.54$\pm$0.37 &
0.79$\pm$0.19 & 1.09$\pm$0.26 & 1.02$\pm$0.24 & 0.43$\pm$0.10 \\
(\OppH{}) $\times 10^{4}$ \dotfill & 0.70$\pm$0.22 & 0.64$\pm$0.16 &
0.92$\pm$0.20 & 0.94$\pm$0.18 & 0.70$\pm$0.16 & 1.17$\pm$0.27 \\
\ZTe{} \dotfill & 8.14$\pm$0.09 & 8.34$\pm$0.08 & 8.23$\pm$0.07 &
8.31$\pm$0.07 & 8.24$\pm$0.07 & 8.20$\pm$0.08 \\
\\
\hline \hline
\label{tab:MaNGA_HIIblob_data}
\end{longtable}

%---------------------
% %GLOBAL GALAXY DATA:
\begin{sidewaystable*}
\caption{Fluxes in units of $10^{-15}$ erg s$^{-1}$ cm$^{2}$, corrected for Galactic foreground extinction, for our \NumMaNGAs{} MaNGA galaxies. These fluxes were taken from masks permitting only spaxels with EW(H$\alpha$) $> 30$\AA{}. Derived galaxy properties for these galaxies are also provided.}\label{tab:MaNGA_global_data}
\centering
\tabcolsep=8pt
\begin{tabular}{p{1.1in}cccccccc}
\hline \hline
System \dotfill & 7495-6102  & 8133-3704  & 8252-12701  & 8257-3704  & 8259-9101  & 8274-12702
 & 8313-1901  & 8459-9102 \\
\hline
$[$O\textsc{ii}]$\lambda$3726 \dotfill & 20.31$\pm$0.14 & 7.50$\pm$0.04 &
4.05$\pm$0.03 & 12.85$\pm$0.10 & 11.52$\pm$0.19 & 4.31$\pm$0.03 &
45.62$\pm$1.47 & 18.49$\pm$0.07 \\
$[$O\textsc{ii}]$\lambda$3729 \dotfill & 30.16$\pm$0.14 & 11.98$\pm$0.04 &
6.30$\pm$0.03 & 20.36$\pm$0.10 & 18.21$\pm$0.19 & 6.65$\pm$0.03 &
63.17$\pm$1.47 & 28.20$\pm$0.07 \\
$[$S\textsc{ii}]$\lambda$4073 \dotfill & 0.12$\pm$0.03 & 0.06$\pm$0.05 &
0.06$\pm$0.03 & 0.20$\pm$0.05 & 0.10$\pm$0.10 & 0.04$\pm$0.01 & 0.20$\pm$0.09 &
0.18$\pm$0.08 \\
$[$O\textsc{iii}]$\lambda$4363 \dotfill & 0.44$\pm$0.06 & 0.36$\pm$0.03 &
0.12$\pm$0.03 & 0.40$\pm$0.14 & 0.21$\pm$0.07 & 0.07$\pm$0.01 & 1.88$\pm$0.05 &
0.30$\pm$0.08 \\
H$\beta\ \lambda$4861 \dotfill & 18.53$\pm$0.12 & 7.28$\pm$0.04 &
3.60$\pm$0.03 & 11.83$\pm$0.11 & 9.17$\pm$0.13 & 3.50$\pm$0.01 &
46.50$\pm$0.08 & 13.32$\pm$0.06 \\
$[$O\textsc{iii}]$\lambda$4959 \dotfill & 20.65$\pm$0.09 & 9.09$\pm$0.05 &
3.72$\pm$0.04 & 12.77$\pm$0.06 & 7.84$\pm$0.04 & 2.68$\pm$0.01 &
64.53$\pm$0.23 & 10.13$\pm$0.03 \\
$[$O\textsc{iii}]$\lambda$5007 \dotfill & 59.59$\pm$0.20 & 26.75$\pm$0.10 &
10.87$\pm$0.04 & 37.76$\pm$0.10 & 23.79$\pm$0.04 & 8.05$\pm$0.02 &
178.21$\pm$0.78 & 30.51$\pm$0.12 \\
$[$N\textsc{ii}]$\lambda$5755 \dotfill & 0.11$\pm$0.07 & -- & -- & -- & -- &
-- & 0.45$\pm$0.04 & -- \\
$[$S\textsc{iii}]$\lambda$6312 \dotfill & 0.19$\pm$0.02 & 0.29$\pm$0.04 &
0.27$\pm$0.03 & 0.25$\pm$0.08 & 0.12$\pm$0.03 & 0.07$\pm$0.02 & 0.62$\pm$0.03 &
0.17$\pm$0.03 \\
H$\alpha\ \lambda$6563 \dotfill & 55.14$\pm$0.22 & 25.37$\pm$0.06 &
10.75$\pm$0.04 & 40.70$\pm$0.17 & 28.66$\pm$0.12 & 11.27$\pm$0.08 &
142.40$\pm$0.55 & 43.32$\pm$0.14 \\
$[$N\textsc{ii}]$\lambda$6584 \dotfill & 3.96$\pm$0.05 & -- & -- & -- & -- &
-- & 11.27$\pm$0.17 & -- \\
$[$S\textsc{ii}]$\lambda$6716 \dotfill & 7.69$\pm$0.04 & 2.95$\pm$0.07 &
1.51$\pm$0.02 & 5.11$\pm$0.03 & 4.32$\pm$0.05 & 1.84$\pm$0.06 & 14.05$\pm$0.07 &
7.41$\pm$0.05 \\
$[$S\textsc{ii}]$\lambda$6731 \dotfill & 5.50$\pm$0.04 & 2.04$\pm$0.07 &
1.13$\pm$0.02 & 3.88$\pm$0.03 & 2.84$\pm$0.05 & 1.37$\pm$0.06 & 10.42$\pm$0.07 &
5.15$\pm$0.05 \\
$[$O\textsc{ii}]$\lambda$7320 \dotfill & 0.55$\pm$0.01 & 0.29$\pm$0.04 &
0.10$\pm$0.03 & 0.57$\pm$0.05 & 0.33$\pm$0.05 & 0.11$\pm$0.05 & 1.54$\pm$0.04 &
0.46$\pm$0.03 \\
$[$O\textsc{ii}]$\lambda$7330 \dotfill & 0.29$\pm$0.01 & 0.13$\pm$0.04 &
0.06$\pm$0.03 & 0.63$\pm$0.05 & 0.27$\pm$0.05 & 0.11$\pm$0.05 & 1.12$\pm$0.04 &
0.36$\pm$0.03 \\
$[$S\textsc{iii}]$\lambda$9069 \dotfill & 4.49$\pm$0.03 & 0.89$\pm$0.05 &
0.72$\pm$0.05 & 3.04$\pm$0.17 & 1.43$\pm$0.06 & 0.98$\pm$0.05 & 9.18$\pm$0.03 &
2.83$\pm$0.12 \\
\hline
R.A. (J2000) / deg \dotfill & 204.5129 & 112.5148 & 144.2393 & 165.5536 &
178.3440 & 164.4108 & 240.2871 & 149.8889 \\
Dec. (J2000) / deg \dotfill & 26.3382 & 43.3792 & 48.2941 & 45.3039 & 44.9205 &
41.0549 & 41.8808 & 43.6605 \\
Redshift \dotfill & 0.0264 & 0.0267 & 0.0253 & 0.0205 & 0.0197 & 0.0258 &
0.0245 & 0.0174 \\
log$(M_{*}/\Msun)$ \dotfill & 9.07$\pm$0.10 & 8.86$\pm$0.10 & 9.22$\pm$0.10 &
9.03$\pm$0.10 & 9.09$\pm$0.10 & 9.19$\pm$0.10 & 9.21$\pm$0.10 & 8.93$\pm$0.10 \\
$E(B-V)\sub{gas}$ \dotfill & 0.027$\pm$0.007 & 0.126$\pm$0.021 &
0.022$\pm$0.010 & 0.152$\pm$0.009 & 0.060$\pm$0.015 & 0.092$\pm$0.008 &
0.049$\pm$0.006 & 0.102$\pm$0.006 \\
$N\sub{e}\,/\,\tn{cm}^{-3}$ \dotfill & 84$\pm$17 & 67$\pm$42 & 112$\pm$25 &
122$\pm$58 & 48$\pm$20 & 108$\pm$51 & 106$\pm$23 & 70$\pm$28 \\
\TOII{}$\,/\,$K \dotfill & 9260$\pm$74 & 9810$\pm$233 & 8875$\pm$302 &
13327$\pm$440 & 10155$\pm$135 & 9738$\pm$556 & 11213$\pm$124 & 9157$\pm$121 \\
\TNII{}$\,/\,$K \dotfill & 15964$\pm$6626 & -- & -- & -- & -- & -- &
21077$\pm$1656 & -- \\
\TOIII{}$\,/\,$K \dotfill & 9967$\pm$440 & 12535$\pm$573 & 11456$\pm$1061 &
11423$\pm$1501 & 10655$\pm$1227 & 10614$\pm$546 & 11230$\pm$156 &
11090$\pm$1073 \\
(\OpH{}) $\times 10^{4}$ \dotfill & 1.37$\pm$0.32 & 0.96$\pm$0.25 &
1.50$\pm$0.41 & 0.45$\pm$0.12 & 1.14$\pm$0.27 & 1.29$\pm$0.42 & 0.57$\pm$0.13 &
1.74$\pm$0.42 \\
(\OppH{}) $\times 10^{4}$ \dotfill & 1.03$\pm$0.22 & 0.56$\pm$0.12 &
0.61$\pm$0.18 & 0.65$\pm$0.25 & 0.65$\pm$0.24 & 0.59$\pm$0.13 & 0.84$\pm$0.15 &
0.51$\pm$0.16 \\
\ZTe{} \dotfill & 8.38$\pm$0.07 & 8.18$\pm$0.08 & 8.33$\pm$0.09 &
8.04$\pm$0.11 & 8.25$\pm$0.09 & 8.27$\pm$0.10 & 8.15$\pm$0.06 & 8.35$\pm$0.09 \\
\hline \hline
\end{tabular}
\end{sidewaystable*}

\begin{sidewaystable*}
\caption{continued.}\label{tab:MaNGA_global_data2}
\centering
\tabcolsep=8pt
\begin{tabular}{p{1.1in}cccc}
\hline \hline
System \dotfill & 8484-6104  & 8548-3702  & 8548-9102  & 8551-1902 \\
\hline
$[$O\textsc{ii}]$\lambda$3726 \dotfill & 8.07$\pm$0.08 & 20.67$\pm$0.17 &
21.01$\pm$0.08 & 17.52$\pm$0.21 \\
$[$O\textsc{ii}]$\lambda$3729 \dotfill & 12.51$\pm$0.08 & 32.27$\pm$0.17 &
29.82$\pm$0.08 & 23.70$\pm$0.21 \\
$[$S\textsc{ii}]$\lambda$4073 \dotfill & 0.15$\pm$0.07 & 0.16$\pm$0.05 &
0.24$\pm$0.08 & 0.13$\pm$0.03 \\
$[$O\textsc{iii}]$\lambda$4363 \dotfill & 0.15$\pm$0.05 & 0.40$\pm$0.06 &
0.48$\pm$0.06 & 0.38$\pm$0.10 \\
H$\beta\ \lambda$4861 \dotfill & 6.07$\pm$0.07 & 17.65$\pm$0.09 &
16.23$\pm$0.05 & 17.64$\pm$0.07 \\
$[$O\textsc{iii}]$\lambda$4959 \dotfill & 5.00$\pm$0.03 & 16.24$\pm$0.02 &
17.41$\pm$0.08 & 19.23$\pm$0.07 \\
$[$O\textsc{iii}]$\lambda$5007 \dotfill & 15.75$\pm$0.09 & 48.18$\pm$0.12 &
52.27$\pm$0.17 & 56.69$\pm$0.18 \\
$[$N\textsc{ii}]$\lambda$5755 \dotfill & -- & -- & -- & 0.14$\pm$0.03 \\
$[$S\textsc{iii}]$\lambda$6312 \dotfill & 0.08$\pm$0.02 & 0.48$\pm$0.13 &
0.22$\pm$0.06 & 0.33$\pm$0.04 \\
H$\alpha\ \lambda$6563 \dotfill & 18.96$\pm$0.06 & 58.40$\pm$0.22 &
51.12$\pm$0.14 & 54.70$\pm$0.17 \\
$[$N\textsc{ii}]$\lambda$6584 \dotfill & -- & -- & -- & 6.99$\pm$0.04 \\
$[$S\textsc{ii}]$\lambda$6716 \dotfill & 3.57$\pm$0.02 & 8.22$\pm$0.05 &
6.59$\pm$0.04 & 6.00$\pm$0.03 \\
$[$S\textsc{ii}]$\lambda$6731 \dotfill & 2.28$\pm$0.02 & 6.19$\pm$0.05 &
4.73$\pm$0.04 & 4.46$\pm$0.03 \\
$[$O\textsc{ii}]$\lambda$7320 \dotfill & 0.23$\pm$0.03 & 0.70$\pm$0.07 &
0.46$\pm$0.03 & 0.69$\pm$0.03 \\
$[$O\textsc{ii}]$\lambda$7330 \dotfill & 0.14$\pm$0.03 & 0.52$\pm$0.07 &
0.44$\pm$0.03 & 0.48$\pm$0.03 \\
$[$S\textsc{iii}]$\lambda$9069 \dotfill & 1.18$\pm$0.06 & 3.77$\pm$0.07 &
2.89$\pm$0.03 & 5.51$\pm$0.01 \\
\hline
R.A. (J2000) / deg \dotfill & 248.7371 & 243.3268 & 244.9146 & 234.5917 \\
Dec. (J2000) / deg \dotfill & 46.4470 & 48.3918 & 47.8732 & 45.8019 \\
Redshift \dotfill & 0.0184 & 0.0202 & 0.0212 & 0.0216 \\
log$(M_{*}/\Msun)$ \dotfill & 9.15$\pm$0.10 & 9.07$\pm$0.10 & 9.04$\pm$0.10 &
8.77$\pm$0.10 \\
$E(B-V)\sub{gas}$ \dotfill & 0.063$\pm$0.012 & 0.116$\pm$0.007 &
0.072$\pm$0.006 & 0.059$\pm$0.006 \\
$N\sub{e}\,/\,\tn{cm}^{-3}$ \dotfill & 39$\pm$17 & 116$\pm$45 & 86$\pm$26 &
107$\pm$26 \\
\TOII{}$\,/\,$K \dotfill & 9606$\pm$107 & 10298$\pm$205 & 9268$\pm$112 &
12220$\pm$167 \\
\TNII{}$\,/\,$K \dotfill & -- & -- & -- & 13332$\pm$1504 \\
\TOIII{}$\,/\,$K \dotfill & 10975$\pm$1301 & 10459$\pm$564 & 10797$\pm$489 &
9711$\pm$802 \\
(\OpH{}) $\times 10^{4}$ \dotfill & 1.40$\pm$0.34 & 1.05$\pm$0.26 &
1.50$\pm$0.36 & 0.51$\pm$0.12 \\
(\OppH{}) $\times 10^{4}$ \dotfill & 0.59$\pm$0.21 & 0.73$\pm$0.17 &
0.78$\pm$0.17 & 1.12$\pm$0.34 \\
\ZTe{} \dotfill & 8.30$\pm$0.09 & 8.25$\pm$0.08 & 8.36$\pm$0.08 &
8.21$\pm$0.10 \\
\hline \hline
\end{tabular}
\end{sidewaystable*}

\twocolumn[]

%%%%%%%%%%%%%%%%%%%
\section{Literature emission-line samples} \label{sec:Appendix C}
Here, we provide tables listing the electron temperatures and oxygen abundances measured for direct systems (Table \ref{tab:LitSamp_TeandAbund_data_direct}) and semi-direct systems (Table \ref{tab:LitSamp_TeandAbund_data_semidirect}) from the literature samples utilised in this work.

\begin{figure}
 \includegraphics[angle=0,width=1.0\linewidth]{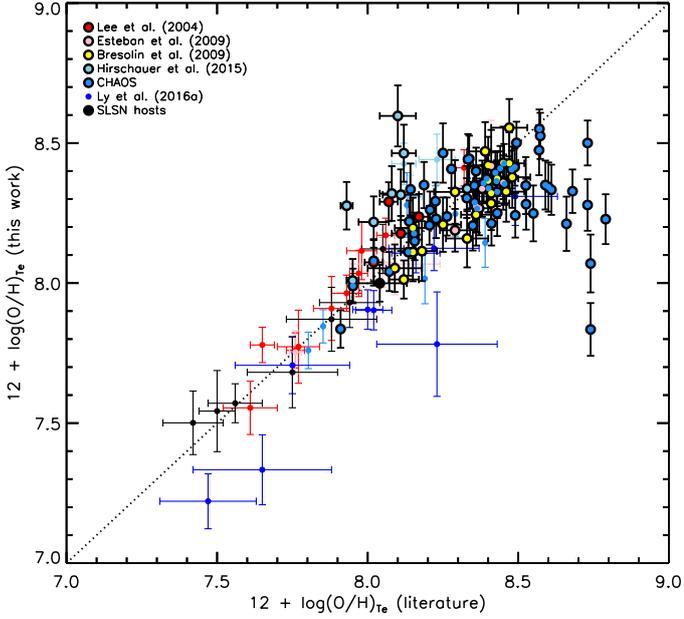}
 \caption{A comparison of the \Te{}-based oxygen abundances obtained from our study and those of previous studies for the same systems. Points with thick black rings represent systems for which we made direct \ZTe{} estimates, while the remaining points have semi-direct \ZTe{} estimates.}
 \label{fig:ZTe_comp_lit}
\end{figure}

We also systematically compare the \ZTe{} estimates we obtain following the procedures described in \S \ref{sec:Te_and_ZTe} and \ref{sec:New TT relation} with those provided by the original papers. This comparison is illustrated in Fig. \ref{fig:ZTe_comp_lit}. We find a median difference in \ZTe{} of only 0.026 dex on a system-to-system basis.

The values of \ZTe{} we obtain for some of the \citet{Hirschauer+15} sample when using the \OII{} fluxes provided are significantly higher than those obtained when assuming only \OIII{} lines and any \TOIII{} -- \TOII{} relation, with a mean increase of $+0.27$ dex. This is due to the apparently dominant contribution of \OpH{} to the total oxygen abundance in these systems (see \S \ref{sec:Accuracy of TT relations}), which we consider to be robust given the relatively high SNR of the \OII{} lines, at least compared to that of \OIII{}$\lambda$4363.

The few metal-rich \HII{} regions from the CHAOS sample with discrepancies greater than $\sim{}0.2$ dex are exclusively from NGC628 \citep{Berg+15}. These discrepancies could be due to the use of \TNII{} to estimate \OpH{} in the original work, which they find is systematically lower than their measured \TOII{}, with little correlation between the two temperatures. This issue, and a similar issue between \TOIII{} and \TSIII{} also reported by \citet{Berg+15}, is not seen in the later CHAOS data for NGC5457 \citep{Croxall+16}, whose \ZTe{} estimates agree more closely with ours.

%--------------------------
\setlength{\rotFPtop}{0pt plus 1fil} %To keep a table on the page when using \sidewaystable
\setlength{\rotFPbot}{0pt plus 1fil} %To keep a table on the page when using \sidewaystable
\renewcommand{\arraystretch}{1.2} %To adapt row height

\onecolumn
\begin{longtable}{ccccccccc}
\caption{Electron densities, temperatures, and oxygen abundances for direct systems from the literature samples.}
\endfirsthead
\caption{continued.}
\endhead
 \hline \hline
System  & Sample  & $N\sub{e}$  & \TOII{}  & \TNII{}  & \TOIII{}  & \OpH{}
 & \OppH{}  & $Z\sub{Te,direct}$ \\
   &    & cm$^{-3}$  & K  & K  & K  & $\times 10^{4}$  & $\times 10^{4}$  &   \\
\hline
KISSR116 & L04  & 69$\error{38}$  & 8781$\error{157}$  & --
 & 11520$\error{199}$  & 1.08$\error{0.27}$  & 0.87$\error{0.16}$
 & 8.29$\error{0.07}$  \\
KISSR286 & L04  & 87$\error{61}$  & 10729$\error{367}$  & --
 & 10460$\error{450}$  & 0.57$\error{0.17}$  & 1.15$\error{0.27}$
 & 8.24$\error{0.08}$  \\
KISSB171 & L04  & 69$\error{23}$  & 10710$\error{135}$  & --
 & 11353$\error{184}$  & 0.50$\error{0.12}$  & 1.00$\error{0.18}$
 & 8.18$\error{0.06}$  \\
KISSB175 & L04  & 157$\error{100}$  & 11299$\error{537}$  & --
 & 12822$\error{647}$  & 0.36$\error{0.12}$  & 0.82$\error{0.19}$
 & 8.07$\error{0.08}$  \\
NGC595 & E09  & 80$\error{70}$  & 9651$\error{335}$  & 8896$\error{457}$
 & 7034$\error{425}$  & 1.11$\error{0.33}$  & 1.18$\error{0.39}$
 & 8.36$\error{0.10}$  \\
NGC604 & E09  & 86$\error{70}$  & 13198$\error{623}$  & 9397$\error{503}$
 & 7590$\error{284}$  & 0.22$\error{0.07}$  & 1.95$\error{0.50}$
 & 8.34$\error{0.10}$  \\
NGC1741-C & E09  & 71$\error{48}$  & 10912$\error{296}$  & 10613$\error{1202}$
 & 8066$\error{642}$  & 0.45$\error{0.12}$  & 2.17$\error{0.77}$
 & 8.42$\error{0.13}$  \\
VS24 & E09  & 131$\error{77}$  & 10537$\error{360}$  & 9034$\error{462}$
 & 7618$\error{422}$  & 0.53$\error{0.15}$  & 1.32$\error{0.38}$
 & 8.27$\error{0.10}$  \\
VS38 & E09  & 95$\error{39}$  & 9383$\error{174}$  & 8869$\error{547}$
 & 8251$\error{341}$  & 0.56$\error{0.14}$  & 0.98$\error{0.23}$
 & 8.19$\error{0.08}$  \\
VS44 & E09  & 126$\error{75}$  & 13158$\error{547}$  & 9712$\error{479}$
 & 7771$\error{280}$  & 0.26$\error{0.07}$  & 1.61$\error{0.39}$
 & 8.27$\error{0.09}$  \\
J0014-0044-1 & G09  & 87$\error{39}$  & 15157$\error{459}$  & --
 & 12454$\error{366}$  & 0.09$\error{0.02}$  & 0.98$\error{0.19}$
 & 8.03$\error{0.08}$  \\
J0202-0047 & G09  & 109$\error{61}$  & 12679$\error{423}$  & --
 & 13686$\error{608}$  & 0.24$\error{0.07}$  & 0.60$\error{0.13}$
 & 7.92$\error{0.08}$  \\
J2104-0035-1 & G09  & 112$\error{59}$  & 14046$\error{515}$  & --
 & 19349$\error{1279}$  & 0.02$\error{0.01}$  & 0.15$\error{0.03}$
 & 7.24$\error{0.09}$  \\
J2302+0049-1 & G09  & 143$\error{61}$  & 11739$\error{352}$  & --
 & 16338$\error{694}$  & 0.08$\error{0.02}$  & 0.44$\error{0.09}$
 & 7.72$\error{0.08}$  \\
J2324-0006 & G09  & 112$\error{51}$  & 13418$\error{397}$  & --
 & 13519$\error{488}$  & 0.18$\error{0.05}$  & 0.68$\error{0.14}$
 & 7.93$\error{0.07}$  \\
NGC300-1 & B09  & 253 \scriptsize{\textit{(lit)}}  & 12680$\error{69}$  & --
 & 11078$\error{369}$  & 0.35$\error{0.09}$  & 0.68$\error{0.14}$
 & 8.01$\error{0.07}$  \\
NGC300-2 & B09  & 182 \scriptsize{\textit{(lit)}}  & 12298$\error{256}$
 & 14256$\error{1838}$  & 11400$\error{766}$  & 0.28$\error{0.09}$
 & 1.01$\error{0.30}$  & 8.11$\error{0.10}$  \\
NGC300-3 & B09  & 20 \scriptsize{\textit{(lit)}}  & 11259$\error{56}$  & --
 & 10947$\error{682}$  & 0.86$\error{0.21}$  & 0.27$\error{0.07}$
 & 8.05$\error{0.08}$  \\
NGC300-4 & B09  & 25 \scriptsize{\textit{(lit)}}  & 11330$\error{305}$  & --
 & 10387$\error{788}$  & 0.64$\error{0.23}$  & 0.66$\error{0.23}$
 & 8.11$\error{0.11}$  \\
NGC300-5 & B09  & 20 \scriptsize{\textit{(lit)}}  & 9792$\error{162}$  & --
 & 8773$\error{554}$  & 1.26$\error{0.40}$  & 1.11$\error{0.35}$
 & 8.37$\error{0.10}$  \\
NGC300-6 & B09  & 20 \scriptsize{\textit{(lit)}}  & 9351$\error{176}$  & --
 & 8708$\error{548}$  & 1.39$\error{0.47}$  & 1.24$\error{0.40}$
 & 8.42$\error{0.10}$  \\
NGC300-7 & B09  & 247 \scriptsize{\textit{(lit)}}  & 11429$\error{222}$  & --
 & 11111$\error{729}$  & 0.31$\error{0.10}$  & 2.00$\error{0.59}$
 & 8.37$\error{0.11}$  \\
NGC300-8 & B09  & 20 \scriptsize{\textit{(lit)}}  & 9298$\error{182}$
 & 10001$\error{960}$  & 9006$\error{629}$  & 1.49$\error{0.51}$
 & 0.63$\error{0.21}$  & 8.33$\error{0.11}$  \\
NGC300-9 & B09  & 115 \scriptsize{\textit{(lit)}}  & 10301$\error{214}$  & --
 & 8186$\error{462}$  & 0.63$\error{0.21}$  & 1.49$\error{0.48}$
 & 8.33$\error{0.11}$  \\
NGC300-10 & B09  & 91 \scriptsize{\textit{(lit)}}  & 8532$\error{282}$
 & 10320$\error{1214}$  & 8199$\error{617}$  & 1.23$\error{0.53}$
 & 1.46$\error{0.60}$  & 8.43$\error{0.13}$  \\
NGC300-11 & B09  & 20 \scriptsize{\textit{(lit)}}  & 11847$\error{251}$  & --
 & 9082$\error{651}$  & 0.58$\error{0.19}$  & 0.86$\error{0.29}$
 & 8.16$\error{0.10}$  \\
NGC300-14 & B09  & 57 \scriptsize{\textit{(lit)}}  & 10513$\error{192}$
 & 9488$\error{625}$  & 8397$\error{438}$  & 0.87$\error{0.28}$
 & 1.06$\error{0.32}$  & 8.28$\error{0.10}$  \\
NGC300-17 & B09  & 20 \scriptsize{\textit{(lit)}}  & 8096$\error{132}$
 & 9346$\error{731}$  & 7778$\error{406}$  & 2.07$\error{0.70}$
 & 1.52$\error{0.48}$  & 8.55$\error{0.10}$  \\
NGC300-19 & B09  & 20 \scriptsize{\textit{(lit)}}  & 8065$\error{123}$
 & 8888$\error{655}$  & 8163$\error{439}$  & 1.87$\error{0.62}$
 & 1.08$\error{0.34}$  & 8.47$\error{0.10}$  \\
NGC300-20 & B09  & 91 \scriptsize{\textit{(lit)}}  & 8954$\error{195}$
 & 10304$\error{977}$  & 7842$\error{456}$  & 0.96$\error{0.34}$
 & 1.72$\error{0.59}$  & 8.43$\error{0.11}$  \\
NGC300-23 & B09  & 30 \scriptsize{\textit{(lit)}}  & 10227$\error{209}$
 & 8610$\error{548}$  & 7636$\error{430}$  & 0.70$\error{0.24}$
 & 1.69$\error{0.55}$  & 8.38$\error{0.11}$  \\
NGC300-24 & B09  & 88 \scriptsize{\textit{(lit)}}  & 9966$\error{37}$
 & 10121$\error{451}$  & 8016$\error{200}$  & 0.86$\error{0.21}$
 & 1.24$\error{0.25}$  & 8.32$\error{0.07}$  \\
NGC300-26 & B09  & 43 \scriptsize{\textit{(lit)}}  & 11195$\error{248}$
 & 10774$\error{951}$  & 8612$\error{499}$  & 0.43$\error{0.14}$
 & 1.33$\error{0.42}$  & 8.24$\error{0.11}$  \\
NGC300-27 & B09  & 20 \scriptsize{\textit{(lit)}}  & 10159$\error{184}$
 & 10615$\error{923}$  & 10567$\error{687}$  & 1.12$\error{0.37}$
 & 0.45$\error{0.13}$  & 8.20$\error{0.11}$  \\
NGC300-28 & B09  & 20 \scriptsize{\textit{(lit)}}  & 10656$\error{198}$
 & 11508$\error{1496}$  & 10110$\error{623}$  & 0.90$\error{0.29}$
 & 0.71$\error{0.21}$  & 8.21$\error{0.10}$  \\
J1044+0353 & I12  & 141$\error{44}$  & 15759$\error{462}$  & --
 & 19036$\error{377}$  & 0.02$\error{0.00}$  & 0.23$\error{0.04}$
 & 7.40$\error{0.07}$  \\
J1132+5722No.1 & I12  & 166$\error{46}$  & 10887$\error{232}$  & --
 & 15048$\error{580}$  & 0.21$\error{0.05}$  & 0.20$\error{0.04}$
 & 7.61$\error{0.07}$  \\
J1132+5722No.2 & I12  & 65$\error{31}$  & 16223$\error{470}$  & --
 & 14102$\error{589}$  & 0.11$\error{0.03}$  & 0.37$\error{0.07}$
 & 7.69$\error{0.07}$  \\
J1244+3212No.2 & I12  & 125$\error{42}$  & 9384$\error{161}$  & --
 & 13475$\error{391}$  & 0.61$\error{0.15}$  & 0.56$\error{0.11}$
 & 8.07$\error{0.07}$  \\
J1244+3212No.3 & I12  & 73$\error{18}$  & 12899$\error{152}$  & --
 & 13493$\error{185}$  & 0.26$\error{0.06}$  & 0.40$\error{0.07}$
 & 7.82$\error{0.06}$  \\
J1248+4823 & I12  & 101$\error{36}$  & 10681$\error{182}$  & --
 & 15365$\error{343}$  & 0.17$\error{0.04}$  & 0.51$\error{0.09}$
 & 7.83$\error{0.07}$  \\
J1331+4151 & I12  & 94$\error{27}$  & 11759$\error{192}$  & --
 & 16292$\error{348}$  & 0.10$\error{0.02}$  & 0.43$\error{0.08}$
 & 7.72$\error{0.07}$  \\
KISSR242 & H15  & 108$\error{67}$  & 9215$\error{250}$  & 23730$\error{2904}$
 & 11831$\error{476}$  & 1.14$\error{0.33}$  & 0.75$\error{0.17}$
 & 8.28$\error{0.08}$  \\
KISSR451 & H15  & 98$\error{66}$  & 8606$\error{250}$  & --
 & 10803$\error{501}$  & 2.21$\error{0.66}$  & 0.70$\error{0.17}$
 & 8.46$\error{0.10}$  \\
KISSR475 & H15  & 116$\error{71}$  & 7734$\error{182}$  & --
 & 10385$\error{541}$  & 3.37$\error{0.98}$  & 0.58$\error{0.14}$
 & 8.60$\error{0.11}$  \\
KISSR590 & H15  & 83$\error{55}$  & 9986$\error{285}$  & --
 & 11452$\error{626}$  & 1.12$\error{0.32}$  & 0.54$\error{0.13}$
 & 8.22$\error{0.09}$  \\
KISSR653 & H15  & 115$\error{73}$  & 9075$\error{258}$  & --
 & 9862$\error{606}$  & 1.50$\error{0.44}$  & 0.59$\error{0.16}$
 & 8.32$\error{0.10}$  \\
KISSR1379 & H15  & 91$\error{61}$  & 9347$\error{272}$  & --
 & 10096$\error{505}$  & 1.34$\error{0.39}$  & 0.73$\error{0.18}$
 & 8.32$\error{0.09}$  \\
\hline
\pagebreak
\hline \hline
System  & Sample  & $N\sub{e}$  & \TOII{}  & \TNII{}  & \TOIII{}  & \OpH{}
 & \OppH{}  & $Z\sub{Te,direct}$ \\
   &    & cm$^{-3}$  & K  & K  & K  & $\times 10^{4}$  & $\times 10^{4}$  &   \\
\hline
KISSR1734 & H15  & 136$\error{76}$  & 11858$\error{457}$  & 14813$\error{1453}$
 & 12871$\error{550}$  & 0.31$\error{0.09}$  & 0.71$\error{0.16}$
 & 8.01$\error{0.08}$  \\
KISSR2117 & H15  & 119$\error{72}$  & 10736$\error{361}$  & 11035$\error{917}$
 & 9215$\error{337}$  & 1.08$\error{0.32}$  & 1.09$\error{0.25}$
 & 8.34$\error{0.08}$  \\
NGC628+131.9+18.5 & CHAOS  & 302$\error{140}$  & 12311$\error{781}$
 & 8490$\error{562}$  & 10233$\error{552}$  & 0.40$\error{0.13}$
 & 0.28$\error{0.07}$  & 7.83$\error{0.09}$  \\
NGC628+151.0+22.3 & CHAOS  & 170$\error{75}$  & 13017$\error{522}$
 & 8597$\error{650}$  & 8789$\error{663}$  & 0.35$\error{0.10}$
 & 0.82$\error{0.26}$  & 8.07$\error{0.10}$  \\
NGC628+232.7+6.6 & CHAOS  & 87$\error{42}$  & 9868$\error{211}$  & --
 & 9829$\error{1072}$  & 0.91$\error{0.24}$  & 0.71$\error{0.27}$
 & 8.21$\error{0.10}$  \\
NGC628+237.6+3.0 & CHAOS  & 68$\error{21}$  & 8966$\error{96}$  & --
 & 9642$\error{1000}$  & 1.66$\error{0.41}$  & 0.49$\error{0.18}$
 & 8.33$\error{0.09}$  \\
NGC628+289.9-17.4 & CHAOS  & 85$\error{40}$  & 11822$\error{290}$  & --
 & 9832$\error{355}$  & 0.36$\error{0.09}$  & 1.39$\error{0.31}$
 & 8.24$\error{0.08}$  \\
NGC628+298.4+12.3 & CHAOS  & 104$\error{47}$  & 11668$\error{278}$  & --
 & 10161$\error{349}$  & 0.26$\error{0.07}$  & 1.52$\error{0.33}$
 & 8.25$\error{0.08}$  \\
NGC628+31.6-191.1 & CHAOS  & 92$\error{79}$  & 11268$\error{512}$
 & 9274$\error{851}$  & 7926$\error{492}$  & 0.68$\error{0.21}$
 & 1.10$\error{0.35}$  & 8.25$\error{0.10}$  \\
NGC628-168.2+150.8 & CHAOS  & 78$\error{67}$  & 9320$\error{298}$  & --
 & 7968$\error{732}$  & 1.63$\error{0.49}$  & 1.28$\error{0.51}$
 & 8.46$\error{0.11}$  \\
NGC628-184.7+83.4 & CHAOS  & 99$\error{60}$  & 9720$\error{285}$
 & 8804$\error{705}$  & 9056$\error{476}$  & 1.32$\error{0.37}$
 & 0.58$\error{0.15}$  & 8.28$\error{0.09}$  \\
NGC628-200.6-4.2 & CHAOS  & 72$\error{40}$  & 8984$\error{170}$
 & 10220$\error{1462}$  & 8626$\error{464}$  & 1.52$\error{0.41}$
 & 0.72$\error{0.19}$  & 8.35$\error{0.09}$  \\
NGC628-206.5-25.7 & CHAOS  & 97$\error{49}$  & 9757$\error{238}$  & --
 & 7954$\error{282}$  & 0.89$\error{0.24}$  & 2.28$\error{0.53}$
 & 8.50$\error{0.08}$  \\
NGC628-42.8-158.2 & CHAOS  & 94$\error{24}$  & 9576$\error{114}$
 & 9011$\error{834}$  & 7998$\error{417}$  & 1.04$\error{0.25}$
 & 1.09$\error{0.28}$  & 8.33$\error{0.08}$  \\
NGC628-76.2-171.8 & CHAOS  & 78$\error{49}$  & 9583$\error{250}$  & --
 & 9228$\error{623}$  & 0.97$\error{0.27}$  & 0.72$\error{0.21}$
 & 8.23$\error{0.09}$  \\
NGC628-90.1+190.2 & CHAOS  & 106$\error{66}$  & 9730$\error{271}$
 & 9349$\error{803}$  & 8375$\error{455}$  & 1.32$\error{0.37}$
 & 0.88$\error{0.25}$  & 8.34$\error{0.09}$  \\
NGC5457+117.9-235.0 & CHAOS  & 90$\error{47}$  & 8772$\error{184}$  & --
 & 8503$\error{360}$  & 1.72$\error{0.45}$  & 0.79$\error{0.19}$
 & 8.40$\error{0.08}$  \\
NGC5457+164.6+9.9 & CHAOS  & 88$\error{55}$  & 8134$\error{184}$
 & 8299$\error{297}$  & 7070$\error{219}$  & 2.12$\error{0.57}$
 & 1.23$\error{0.28}$  & 8.53$\error{0.08}$  \\
NGC5457+17.3-235.4 & CHAOS  & 80$\error{75}$  & 10071$\error{391}$
 & 8575$\error{708}$  & 7836$\error{386}$  & 0.96$\error{0.29}$
 & 0.95$\error{0.27}$  & 8.28$\error{0.09}$  \\
NGC5457+189.2-136.3 & CHAOS  & 164$\error{67}$  & 9519$\error{248}$
 & 9273$\error{328}$  & 7414$\error{131}$  & 0.91$\error{0.24}$
 & 2.64$\error{0.52}$  & 8.55$\error{0.07}$  \\
NGC5457+225.6-124.1 & CHAOS  & 81$\error{32}$  & 8501$\error{117}$
 & 9274$\error{503}$  & 8185$\error{194}$  & 2.00$\error{0.50}$
 & 1.16$\error{0.24}$  & 8.50$\error{0.08}$  \\
NGC5457+249.3+201.9 & CHAOS  & 99$\error{41}$  & 10437$\error{222}$
 & 10273$\error{1254}$  & 8939$\error{185}$  & 0.89$\error{0.22}$
 & 1.05$\error{0.21}$  & 8.29$\error{0.07}$  \\
NGC5457+252.2-109.8 & CHAOS  & 75$\error{42}$  & 8588$\error{159}$
 & 9196$\error{738}$  & 8676$\error{252}$  & 1.70$\error{0.44}$
 & 0.86$\error{0.18}$  & 8.41$\error{0.08}$  \\
NGC5457+254.6-107.2 & CHAOS  & 209$\error{91}$  & 12035$\error{522}$
 & 9778$\error{252}$  & 8310$\error{66}$  & 0.39$\error{0.11}$
 & 2.16$\error{0.38}$  & 8.41$\error{0.07}$  \\
NGC5457+266.0+534.1 & CHAOS  & 71$\error{48}$  & 11134$\error{309}$  & --
 & 11455$\error{344}$  & 0.66$\error{0.17}$  & 0.95$\error{0.19}$
 & 8.21$\error{0.07}$  \\
NGC5457+281.4-71.8 & CHAOS  & 61$\error{53}$  & 9325$\error{241}$  & --
 & 8491$\error{320}$  & 1.42$\error{0.40}$  & 0.94$\error{0.23}$
 & 8.37$\error{0.08}$  \\
NGC5457+331.9+401.0 & CHAOS  & 72$\error{40}$  & 10347$\error{274}$  & --
 & 10282$\error{276}$  & 0.66$\error{0.17}$  & 1.03$\error{0.21}$
 & 8.23$\error{0.07}$  \\
NGC5457+354.1+71.2 & CHAOS  & 96$\error{63}$  & 11236$\error{403}$  & --
 & 9017$\error{250}$  & 0.65$\error{0.18}$  & 1.61$\error{0.35}$
 & 8.35$\error{0.07}$  \\
NGC5457+360.9+75.3 & CHAOS  & 84$\error{50}$  & 11338$\error{330}$  & --
 & 8732$\error{194}$  & 0.49$\error{0.13}$  & 1.83$\error{0.37}$
 & 8.36$\error{0.07}$  \\
NGC5457+509.5+264.1 & CHAOS  & 74$\error{36}$  & 11337$\error{236}$
 & 11765$\error{1592}$  & 10207$\error{227}$  & 0.50$\error{0.13}$
 & 1.23$\error{0.24}$  & 8.24$\error{0.07}$  \\
NGC5457+6.6+886.3 & CHAOS  & 65$\error{35}$  & 13896$\error{351}$  & --
 & 12559$\error{357}$  & 0.27$\error{0.07}$  & 0.41$\error{0.08}$
 & 7.84$\error{0.07}$  \\
NGC5457+667.9+174.1 & CHAOS  & 190$\error{105}$  & 16557$\error{1144}$
 & 12691$\error{1260}$  & 11959$\error{409}$  & 0.10$\error{0.03}$
 & 1.19$\error{0.25}$  & 8.11$\error{0.08}$  \\
NGC5457+67.5+277.0 & CHAOS  & 133$\error{126}$  & 13902$\error{1030}$
 & 9125$\error{730}$  & 8021$\error{320}$  & 0.35$\error{0.12}$
 & 1.88$\error{0.51}$  & 8.35$\error{0.10}$  \\
NGC5457+692.1+272.9 & CHAOS  & 76$\error{51}$  & 12906$\error{438}$  & --
 & 11441$\error{349}$  & 0.36$\error{0.10}$  & 0.74$\error{0.15}$
 & 8.04$\error{0.07}$  \\
NGC5457+96.7+266.9 & CHAOS  & 72$\error{50}$  & 8878$\error{204}$
 & 9008$\error{632}$  & 8359$\error{491}$  & 1.61$\error{0.43}$
 & 0.51$\error{0.14}$  & 8.33$\error{0.09}$  \\
NGC5457-164.9-333.9 & CHAOS  & 81$\error{47}$  & 10367$\error{257}$  & --
 & 7994$\error{163}$  & 0.87$\error{0.23}$  & 1.80$\error{0.36}$
 & 8.43$\error{0.07}$  \\
NGC5457-183.9-179.0 & CHAOS  & 70$\error{51}$  & 9033$\error{217}$  & --
 & 7237$\error{395}$  & 1.24$\error{0.33}$  & 1.75$\error{0.51}$
 & 8.47$\error{0.09}$  \\
NGC5457-200.3-193.6 & CHAOS  & 74$\error{31}$  & 9632$\error{154}$  & --
 & 7829$\error{189}$  & 1.01$\error{0.24}$  & 1.58$\error{0.31}$
 & 8.41$\error{0.07}$  \\
NGC5457-226.9-366.4 & CHAOS  & 76$\error{39}$  & 9857$\error{195}$
 & 12445$\error{1192}$  & 9226$\error{254}$  & 1.60$\error{0.41}$
 & 0.63$\error{0.13}$  & 8.35$\error{0.08}$  \\
NGC5457-243.0+159.6 & CHAOS  & 74$\error{48}$  & 9352$\error{216}$
 & 9442$\error{553}$  & 7887$\error{369}$  & 1.51$\error{0.40}$
 & 0.67$\error{0.17}$  & 8.34$\error{0.09}$  \\
NGC5457-297.7+87.1 & CHAOS  & 96$\error{69}$  & 10137$\error{362}$
 & 9867$\error{1094}$  & 8464$\error{318}$  & 1.14$\error{0.33}$
 & 0.70$\error{0.17}$  & 8.27$\error{0.09}$  \\
NGC5457-309.4+56.9 & CHAOS  & 74$\error{44}$  & 8224$\error{176}$  & --
 & 8384$\error{327}$  & 2.10$\error{0.55}$  & 0.66$\error{0.15}$
 & 8.44$\error{0.09}$  \\
NGC5457-345.5+273.8 & CHAOS  & 94$\error{81}$  & 16524$\error{1204}$  & --
 & 10315$\error{435}$  & 0.32$\error{0.10}$  & 1.28$\error{0.31}$
 & 8.20$\error{0.09}$  \\
NGC5457-368.3-285.6 & CHAOS  & 112$\error{70}$  & 10337$\error{319}$
 & 10346$\error{512}$  & 8798$\error{237}$  & 0.84$\error{0.23}$
 & 1.63$\error{0.35}$  & 8.39$\error{0.07}$  \\
NGC5457-371.1-280.0 & CHAOS  & 90$\error{51}$  & 10186$\error{267}$
 & 10696$\error{514}$  & 9270$\error{225}$  & 0.91$\error{0.24}$
 & 1.11$\error{0.23}$  & 8.30$\error{0.07}$  \\
NGC5457-377.9-64.9 & CHAOS  & 72$\error{43}$  & 10129$\error{229}$
 & 9409$\error{608}$  & 9210$\error{225}$  & 1.07$\error{0.28}$
 & 0.56$\error{0.11}$  & 8.21$\error{0.08}$  \\
NGC5457-392.0-270.1 & CHAOS  & 124$\error{60}$  & 10078$\error{256}$
 & 10704$\error{461}$  & 9132$\error{192}$  & 0.60$\error{0.16}$
 & 1.63$\error{0.32}$  & 8.35$\error{0.07}$  \\
NGC5457-405.5-157.7 & CHAOS  & 70$\error{37}$  & 8790$\error{157}$  & --
 & 10337$\error{362}$  & 1.79$\error{0.45}$  & 0.37$\error{0.08}$
 & 8.33$\error{0.09}$  \\
\hline
\pagebreak
\hline \hline
System  & Sample  & $N\sub{e}$  & \TOII{}  & \TNII{}  & \TOIII{}  & \OpH{}
 & \OppH{}  & $Z\sub{Te,direct}$ \\
   &    & cm$^{-3}$  & K  & K  & K  & $\times 10^{4}$  & $\times 10^{4}$  &   \\
\hline
NGC5457-410.3-206.3 & CHAOS  & 68$\error{39}$  & 8956$\error{162}$
 & 10563$\error{1047}$  & 8554$\error{302}$  & 1.96$\error{0.50}$
 & 0.82$\error{0.18}$  & 8.44$\error{0.08}$  \\
NGC5457-414.1-253.6 & CHAOS  & 81$\error{38}$  & 9053$\error{168}$
 & 12512$\error{2099}$  & 9589$\error{233}$  & 1.07$\error{0.27}$
 & 1.49$\error{0.30}$  & 8.41$\error{0.07}$  \\
NGC5457-453.8-191.8 & CHAOS  & 71$\error{30}$  & 9912$\error{150}$
 & 11980$\error{1463}$  & 10860$\error{119}$  & 0.93$\error{0.22}$
 & 0.90$\error{0.16}$  & 8.26$\error{0.07}$  \\
NGC5457-455.7-55.8 & CHAOS  & 54$\error{22}$  & 9647$\error{109}$  & --
 & 10585$\error{94}$  & 1.11$\error{0.26}$  & 0.85$\error{0.15}$
 & 8.29$\error{0.07}$  \\
NGC5457-464.7-131.0 & CHAOS  & 57$\error{30}$  & 11033$\error{196}$  & --
 & 11185$\error{292}$  & 0.68$\error{0.17}$  & 0.74$\error{0.14}$
 & 8.15$\error{0.07}$  \\
NGC5457-466.1-128.2 & CHAOS  & 68$\error{31}$  & 10476$\error{179}$  & --
 & 11009$\error{283}$  & 0.93$\error{0.23}$  & 0.57$\error{0.11}$
 & 8.18$\error{0.07}$  \\
NGC5457-479.7-3.9 & CHAOS  & 74$\error{40}$  & 9945$\error{209}$  & --
 & 11252$\error{168}$  & 1.08$\error{0.27}$  & 0.58$\error{0.10}$
 & 8.22$\error{0.07}$  \\
NGC5457-481.4-0.5 & CHAOS  & 88$\error{49}$  & 10812$\error{293}$  & --
 & 12373$\error{128}$  & 0.75$\error{0.19}$  & 0.46$\error{0.08}$
 & 8.08$\error{0.07}$  \\
NGC5457-540.5-149.9 & CHAOS  & 58$\error{20}$  & 11190$\error{131}$  & --
 & 13096$\error{144}$  & 0.41$\error{0.10}$  & 0.57$\error{0.10}$
 & 7.99$\error{0.06}$  \\
NGC5457-99.6-388.0 & CHAOS  & 183$\error{69}$  & 11555$\error{371}$
 & 10990$\error{295}$  & 8992$\error{70}$  & 0.48$\error{0.12}$
 & 1.80$\error{0.32}$  & 8.36$\error{0.06}$  \\
PTF12dam & SLSN  & 93$\error{30}$  & 13273$\error{270}$  & --
 & 13109$\error{233}$  & 0.22$\error{0.05}$  & 0.78$\error{0.14}$
 & 8.00$\error{0.07}$  \\
\hline \hline
\label{tab:LitSamp_TeandAbund_data_direct}
\end{longtable}

\newpage

\begin{longtable}{ccccccccc}
\caption{Electron densities, temperatures, and oxygen abundances for semi-direct systems from the literature samples.}
\endfirsthead
\caption{continued.}
\endhead
\hline \hline
System  & Sample  & $N\sub{e}$  & \TOII{}  & \TNII{}  & \TOIII{}  & \OpH{}
 & \OppH{}  & $Z\sub{Te,semi-direct}$ \\
   &    & cm$^{-3}$  & K  & K  & K  & $\times 10^{4}$  & $\times 10^{4}$  &   \\
\hline
KISSB23 & L04  & 76$\error{28}$  & 19389$\error{1386}$  & --
 & 14219$\error{1017}$  & 0.15$\error{0.02}$  & 0.15$\error{0.04}$
 & 7.78$\error{0.06}$  \\
KISSB61 & L04  & 53$\error{29}$  & 13502$\error{316}$  & --
 & 15108$\error{353}$  & 0.16$\error{0.01}$  & 0.42$\error{0.08}$
 & 7.76$\error{0.06}$  \\
KISSB86 & L04  & 61$\error{22}$  & 10163$\error{231}$  & --
 & 11900$\error{270}$  & 0.71$\error{0.07}$  & 0.81$\error{0.15}$
 & 8.17$\error{0.06}$  \\
KISSR73 & L04  & 85$\error{64}$  & 12255$\error{504}$  & --
 & 13068$\error{537}$  & 0.42$\error{0.07}$  & 0.50$\error{0.10}$
 & 7.96$\error{0.06}$  \\
KISSR85 & L04  & 2093$\error{1530}$  & 16692$\error{1658}$  & --
 & 15283$\error{1518}$  & 0.08$\error{0.02}$  & 0.27$\error{0.07}$
 & 7.55$\error{0.10}$  \\
KISSR87 & L04  & 105$\error{52}$  & 9525$\error{217}$  & --  & 9486$\error{217}$
 & 1.47$\error{0.18}$  & 1.69$\error{0.32}$  & 8.41$\error{0.06}$  \\
KISSR310 & L04  & 93$\error{98}$  & 11665$\error{1025}$  & --
 & 14591$\error{1282}$  & 0.16$\error{0.06}$  & 0.66$\error{0.20}$
 & 7.91$\error{0.11}$  \\
KISSR311 & L04  & 93$\error{99}$  & 10174$\error{825}$  & --
 & 13189$\error{1069}$  & 0.74$\error{0.29}$  & 0.66$\error{0.20}$
 & 8.12$\error{0.10}$  \\
KISSR666 & L04  & 31$\error{57}$  & 12701$\error{1244}$  & --
 & 15814$\error{1549}$  & 0.02$\error{0.01}$  & 0.58$\error{0.18}$
 & 7.77$\error{0.13}$  \\
KISSR814 & L04  & 113$\error{79}$  & 11108$\error{581}$  & --
 & 13060$\error{683}$  & 0.38$\error{0.09}$  & 0.70$\error{0.17}$
 & 8.03$\error{0.08}$  \\
NGC2363 & E09  & 154$\error{71}$  & 13339$\error{323}$  & 15440$\error{1626}$
 & 15400$\error{373}$  & 0.03$\error{0.00}$  & 0.54$\error{0.10}$
 & 7.75$\error{0.08}$  \\
NGC4395-70 & E09  & 88$\error{60}$  & 13780$\error{907}$  & --
 & 9998$\error{658}$  & 0.15$\error{0.03}$  & 1.02$\error{0.29}$
 & 8.07$\error{0.11}$  \\
NGC4861 & E09  & 116$\error{71}$  & 11106$\error{519}$  & 14082$\error{2501}$
 & 12302$\error{575}$  & 0.19$\error{0.04}$  & 0.99$\error{0.23}$
 & 8.07$\error{0.09}$  \\
NGC5461 & E09  & 224$\error{196}$  & 9022$\error{443}$  & 9928$\error{704}$
 & 7945$\error{390}$  & 0.60$\error{0.16}$  & 2.33$\error{0.69}$
 & 8.47$\error{0.11}$  \\
J0301-0059-1 & G09  & 72$\error{42}$  & 13044$\error{854}$  & --
 & 11972$\error{784}$  & 0.48$\error{0.12}$  & 0.48$\error{0.12}$
 & 8.14$\error{0.08}$  \\
J0301-0059-2 & G09  & 70$\error{42}$  & 14028$\error{1495}$  & --
 & 13386$\error{1427}$  & 0.34$\error{0.11}$  & 0.34$\error{0.11}$
 & 7.97$\error{0.10}$  \\
J0338+0013 & G09  & 127$\error{62}$  & 13871$\error{778}$  & --
 & 17071$\error{957}$  & 0.05$\error{0.01}$  & 0.37$\error{0.08}$
 & 7.63$\error{0.09}$  \\
G0405-3648-1 & G09  & 41$\error{23}$  & 19294$\error{1710}$  & --
 & 14440$\error{1280}$  & 0.08$\error{0.01}$  & 0.22$\error{0.05}$
 & 7.47$\error{0.09}$  \\
G0405-3648-2 & G09  & 124$\error{31}$  & 19859$\error{1026}$  & --
 & 17629$\error{911}$  & 0.08$\error{0.01}$  & 0.09$\error{0.02}$
 & 7.37$\error{0.05}$  \\
G0405-3648-3 & G09  & 53$\error{23}$  & 18797$\error{1425}$  & --
 & 20486$\error{1553}$  & 0.06$\error{0.01}$  & 0.06$\error{0.01}$
 & 7.24$\error{0.06}$  \\
J0519+0007 & G09  & 367$\error{161}$  & 15396$\error{1126}$
 & 23742$\error{3423}$  & 19272$\error{1409}$  & 0.02$\error{0.00}$
 & 0.24$\error{0.05}$  & 7.40$\error{0.09}$  \\
J2104-0035-3+4 & G09  & 22$\error{20}$  & 24512$\error{3360}$  & --
 & 19261$\error{2640}$  & 0.03$\error{0.01}$  & 0.04$\error{0.01}$
 & 7.02$\error{0.09}$  \\
J2302+0049-2 & G09  & 60$\error{32}$  & 15311$\error{861}$  & --
 & 14442$\error{812}$  & 0.16$\error{0.03}$  & 0.33$\error{0.07}$
 & 7.69$\error{0.07}$  \\
DDO68No.2 & I12  & 15$\error{18}$  & 23799$\error{1515}$  & --
 & 16560$\error{1054}$  & 0.02$\error{0.00}$  & 0.10$\error{0.02}$
 & 7.08$\error{0.08}$  \\
DDO68No.3a & I12  & 71$\error{54}$  & 26779$\error{3372}$  & --
 & 16448$\error{2071}$  & 0.02$\error{0.00}$  & 0.07$\error{0.02}$
 & 6.95$\error{0.10}$  \\
J0113+0052No.1 & I12  & --  & 18436$\error{2701}$  & --  & 22135$\error{3243}$
 & 0.04$\error{0.01}$  & 0.07$\error{0.02}$  & 7.04$\error{0.09}$  \\
J0113+0052No.3 & I12  & --  & 14006$\error{2190}$  & --
 & 25844$\error{4042}$  & 0.09$\error{0.04}$  & 0.06$\error{0.02}$
 & 7.19$\error{0.09}$  \\
J0851+8416 & I12  & 36$\error{27}$  & 16599$\error{1298}$  & --
 & 14430$\error{1129}$  & 0.09$\error{0.02}$  & 0.33$\error{0.08}$
 & 7.61$\error{0.09}$  \\
J0906+2528No.2 & I12  & 14$\error{22}$  & 18365$\error{2235}$  & --
 & 16485$\error{2007}$  & 0.07$\error{0.02}$  & 0.17$\error{0.05}$
 & 7.38$\error{0.10}$  \\
J0908+0517No.2 & I12  & 86$\error{29}$  & 15147$\error{507}$  & --
 & 15466$\error{517}$  & 0.07$\error{0.01}$  & 0.36$\error{0.07}$
 & 7.64$\error{0.07}$  \\
J1016+3754 & I12  & 120$\error{39}$  & 14599$\error{199}$  & --
 & 16630$\error{227}$  & 0.05$\error{0.00}$  & 0.35$\error{0.06}$
 & 7.60$\error{0.07}$  \\
J1016+5823No.1 & I12  & 57$\error{20}$  & 14662$\error{206}$  & --
 & 14596$\error{206}$  & 0.10$\error{0.00}$  & 0.42$\error{0.07}$
 & 7.72$\error{0.06}$  \\
J1016+5823No.2 & I12  & 46$\error{30}$  & 18221$\error{1490}$  & --
 & 14616$\error{1195}$  & 0.09$\error{0.02}$  & 0.24$\error{0.06}$
 & 7.51$\error{0.09}$  \\
J1016+5823No.3 & I12  & 25$\error{17}$  & 15055$\error{248}$  & --
 & 16179$\error{267}$  & 0.08$\error{0.00}$  & 0.32$\error{0.06}$
 & 7.60$\error{0.06}$  \\
J1056+3608No.1 & I12  & 675$\error{488}$  & 18824$\error{2334}$  & --
 & 21494$\error{2665}$  & 0.03$\error{0.01}$  & 0.09$\error{0.02}$
 & 7.05$\error{0.09}$  \\
J1056+3608No.2 & I12  & 15$\error{24}$  & 13358$\error{3762}$  & --
 & 25760$\error{7255}$  & 0.14$\error{0.11}$  & 0.04$\error{0.02}$
 & 7.48$\error{0.15}$  \\
J1056+3608No.3 & I12  & 25$\error{31}$  & 18067$\error{3111}$  & --
 & 19860$\error{3420}$  & 0.07$\error{0.03}$  & 0.08$\error{0.03}$
 & 7.27$\error{0.10}$  \\
J1119+0935No.2 & I12  & 74$\error{31}$  & 17828$\error{1253}$  & --
 & 13505$\error{949}$  & 0.11$\error{0.02}$  & 0.29$\error{0.07}$
 & 7.61$\error{0.08}$  \\
J1119+5130 & I12  & 166$\error{75}$  & 17683$\error{511}$  & --
 & 16492$\error{476}$  & 0.06$\error{0.00}$  & 0.20$\error{0.04}$
 & 7.42$\error{0.06}$  \\
J1132+5722No.3 & I12  & 30$\error{34}$  & 14000$\error{1007}$  & --
 & 16141$\error{1160}$  & 0.14$\error{0.03}$  & 0.33$\error{0.07}$
 & 7.67$\error{0.08}$  \\
J1154+4636 & I12  & 134$\error{40}$  & 14705$\error{380}$  & --
 & 14671$\error{379}$  & 0.13$\error{0.01}$  & 0.38$\error{0.07}$
 & 7.71$\error{0.06}$  \\
J1215+5223 & I12  & 73$\error{37}$  & 17772$\error{600}$  & --
 & 15775$\error{533}$  & 0.07$\error{0.01}$  & 0.22$\error{0.04}$
 & 7.46$\error{0.07}$  \\
J1224+3724 & I12  & 80$\error{27}$  & 12861$\error{288}$  & --
 & 15635$\error{350}$  & 0.11$\error{0.01}$  & 0.48$\error{0.09}$
 & 7.77$\error{0.07}$  \\
J1226-0115No.1 & I12  & 50$\error{16}$  & 12912$\error{239}$  & --
 & 15389$\error{284}$  & 0.11$\error{0.01}$  & 0.50$\error{0.09}$
 & 7.78$\error{0.07}$  \\
J1226-0115No.2 & I12  & 128$\error{37}$  & 13453$\error{372}$  & --
 & 14691$\error{406}$  & 0.10$\error{0.01}$  & 0.51$\error{0.10}$
 & 7.79$\error{0.07}$  \\
J1235+2755No.1 & I12  & 77$\error{39}$  & 13001$\error{1535}$  & --
 & 22649$\error{2675}$  & 0.17$\error{0.06}$  & 0.09$\error{0.02}$
 & 7.50$\error{0.07}$  \\
J1235+2755No.2 & I12  & 62$\error{18}$  & 14606$\error{389}$  & --
 & 13319$\error{355}$  & 0.19$\error{0.01}$  & 0.44$\error{0.08}$
 & 7.80$\error{0.06}$  \\
J1235+2755No.3 & I12  & 64$\error{24}$  & 13373$\error{564}$  & --
 & 14864$\error{627}$  & 0.27$\error{0.04}$  & 0.34$\error{0.07}$
 & 7.78$\error{0.06}$  \\
 \hline
\pagebreak
\hline \hline
System  & Sample  & $N\sub{e}$  & \TOII{}  & \TNII{}  & \TOIII{}  & \OpH{}
 & \OppH{}  & $Z\sub{Te,semi-direct}$ \\
   &    & cm$^{-3}$  & K  & K  & K  & $\times 10^{4}$  & $\times 10^{4}$  &   \\
\hline
J1241-0340 & I12  & 84$\error{28}$  & 12977$\error{373}$  & --
 & 15998$\error{459}$  & 0.15$\error{0.02}$  & 0.41$\error{0.08}$
 & 7.74$\error{0.06}$  \\
J1244+3212No.1 & I12  & 96$\error{21}$  & 13925$\error{199}$  & --
 & 14073$\error{201}$  & 0.21$\error{0.01}$  & 0.41$\error{0.07}$
 & 7.79$\error{0.06}$  \\
J1257+3341No.1 & I12  & 119$\error{71}$  & 20481$\error{2922}$  & --
 & 15367$\error{2193}$  & 0.08$\error{0.02}$  & 0.14$\error{0.05}$
 & 7.43$\error{0.10}$  \\
J1257+3341No.3 & I12  & 250$\error{133}$  & 19337$\error{2485}$  & --
 & 16135$\error{2073}$  & 0.10$\error{0.03}$  & 0.13$\error{0.04}$
 & 7.50$\error{0.09}$  \\
J1327+4022 & I12  & 95$\error{42}$  & 13976$\error{202}$  & --
 & 16356$\error{236}$  & 0.05$\error{0.00}$  & 0.41$\error{0.07}$
 & 7.66$\error{0.07}$  \\
J1355+4651 & I12  & --  & 13586$\error{218}$  & --
 & 18911$\error{304}$  & 0.02$\error{0.00}$  & 0.33$\error{0.06}$
 & 7.55$\error{0.07}$  \\
J1403+5804No.2 & I12  & 89$\error{58}$  & 16176$\error{1246}$  & --
 & 19336$\error{1489}$  & 0.07$\error{0.01}$  & 0.15$\error{0.03}$
 & 7.35$\error{0.07}$  \\
J1608+3528 & I12  & 773$\error{594}$  & 11876$\error{422}$  & --
 & 17369$\error{618}$  & 0.04$\error{0.01}$  & 0.52$\error{0.10}$
 & 7.75$\error{0.08}$  \\
PHL293B & I12  & 49$\error{21}$  & 13923$\error{400}$  & --
 & 16456$\error{473}$  & 0.10$\error{0.01}$  & 0.35$\error{0.06}$
 & 7.66$\error{0.07}$  \\
SBS1420+544 & I12  & 68$\error{27}$  & 12371$\error{207}$  & --
 & 16066$\error{269}$  & 0.08$\error{0.00}$  & 0.53$\error{0.09}$
 & 7.78$\error{0.07}$  \\
UGC521A & B12  & 38$\error{48}$  & 14333$\error{911}$  & --
 & 16611$\error{1056}$  & 0.15$\error{0.03}$  & 0.27$\error{0.06}$
 & 7.62$\error{0.07}$  \\
UGC695E & B12  & 36$\error{29}$  & 16264$\error{3649}$  & --
 & 15909$\error{3569}$  & 0.22$\error{0.12}$  & 0.13$\error{0.06}$
 & 7.85$\error{0.14}$  \\
UGC1056A & B12  & 139$\error{142}$  & 14878$\error{3407}$  & --
 & 13312$\error{3048}$  & 0.31$\error{0.20}$  & 0.31$\error{0.17}$
 & 7.98$\error{0.17}$  \\
UGC1056B & B12  & 46$\error{56}$  & 12719$\error{2906}$  & --
 & 11594$\error{2649}$  & 0.41$\error{0.30}$  & 0.64$\error{0.40}$
 & 8.03$\error{0.21}$  \\
UGC1176A & B12  & 78$\error{92}$  & 13152$\error{2154}$  & --
 & 12237$\error{2004}$  & 0.30$\error{0.15}$  & 0.60$\error{0.27}$
 & 7.95$\error{0.16}$  \\
NGC784A & B12  & 104$\error{126}$  & 11733$\error{1869}$  & --
 & 11435$\error{1821}$  & 0.37$\error{0.23}$  & 0.85$\error{0.43}$
 & 8.09$\error{0.18}$  \\
NGC784B & B12  & 85$\error{137}$  & 13632$\error{4175}$  & --
 & 12317$\error{3772}$  & 0.28$\error{0.27}$  & 0.55$\error{0.45}$
 & 7.92$\error{0.28}$  \\
UGC2716A & B12  & 73$\error{73}$  & 12012$\error{2157}$  & --
 & 12335$\error{2214}$  & 0.35$\error{0.22}$  & 0.69$\error{0.35}$
 & 8.02$\error{0.18}$  \\
NGC2537A & B12  & 114$\error{147}$  & 12291$\error{2321}$  & --
 & 8859$\error{1673}$  & 0.64$\error{0.42}$  & 1.06$\error{0.73}$
 & 8.42$\error{0.23}$  \\
NGC2537B & B12  & 79$\error{71}$  & 17340$\error{3522}$  & --
 & 10359$\error{2104}$  & 0.23$\error{0.12}$  & 0.50$\error{0.31}$
 & 8.07$\error{0.21}$  \\
UGC4278B & B12  & 90$\error{77}$  & 18670$\error{1872}$  & --
 & 13611$\error{1365}$  & 0.13$\error{0.03}$  & 0.23$\error{0.07}$
 & 7.67$\error{0.09}$  \\
UGC4278A & B12  & 101$\error{89}$  & 17432$\error{2122}$  & --
 & 13548$\error{1649}$  & 0.11$\error{0.04}$  & 0.32$\error{0.11}$
 & 7.63$\error{0.12}$  \\
NGC2552A & B12  & 101$\error{64}$  & 12281$\error{940}$  & --
 & 9807$\error{751}$  & 0.47$\error{0.13}$  & 0.98$\error{0.30}$
 & 8.18$\error{0.11}$  \\
UGC4393B & B12  & 60$\error{64}$  & 12443$\error{2439}$  & --
 & 10897$\error{2136}$  & 0.42$\error{0.27}$  & 0.78$\error{0.45}$
 & 8.08$\error{0.20}$  \\
UGC4393C & B12  & 87$\error{94}$  & 12208$\error{2067}$  & --
 & 12054$\error{2040}$  & 0.61$\error{0.36}$  & 0.45$\error{0.22}$
 & 8.23$\error{0.15}$  \\
CGCG035-007A & B12  & 57$\error{38}$  & 11938$\error{3789}$  & --
 & 17590$\error{5582}$  & 0.61$\error{0.69}$  & 0.13$\error{0.08}$
 & 8.11$\error{0.18}$  \\
UGC5139A & B12  & 59$\error{47}$  & 13631$\error{1227}$  & --
 & 12438$\error{1119}$  & 0.24$\error{0.07}$  & 0.58$\error{0.17}$
 & 7.91$\error{0.10}$  \\
IC559A & B12  & 48$\error{75}$  & 12998$\error{3296}$  & --
 & 10715$\error{2717}$  & 0.45$\error{0.35}$  & 0.71$\error{0.52}$
 & 8.15$\error{0.24}$  \\
UGC5272A & B12  & 71$\error{38}$  & 13067$\error{495}$  & --
 & 13411$\error{508}$  & 0.14$\error{0.02}$  & 0.63$\error{0.13}$
 & 7.89$\error{0.08}$  \\
UGC5340A & B12  & 522$\error{473}$  & 20359$\error{2088}$  & --
 & 18830$\error{1932}$  & 0.02$\error{0.01}$  & 0.11$\error{0.03}$
 & 7.12$\error{0.09}$  \\
UGC5423A & B12  & 80$\error{64}$  & 14273$\error{1239}$  & --
 & 14345$\error{1246}$  & 0.20$\error{0.06}$  & 0.37$\error{0.11}$
 & 7.76$\error{0.10}$  \\
UGC5423B & B12  & 88$\error{102}$  & 14339$\error{2467}$  & --
 & 13841$\error{2381}$  & 0.17$\error{0.09}$  & 0.44$\error{0.19}$
 & 7.78$\error{0.15}$  \\
UGC5797A & B12  & 62$\error{56}$  & 11104$\error{1727}$  & --
 & 13132$\error{2042}$  & 0.32$\error{0.18}$  & 0.75$\error{0.30}$
 & 8.03$\error{0.15}$  \\
UGC5923A & B12  & 81$\error{136}$  & 13516$\error{5481}$  & --
 & 14840$\error{6018}$  & 0.35$\error{0.43}$  & 0.25$\error{0.22}$
 & 7.95$\error{0.27}$  \\
NGC3741A & B12  & 89$\error{42}$  & 16527$\error{1931}$  & --
 & 14874$\error{1738}$  & 0.11$\error{0.03}$  & 0.28$\error{0.08}$
 & 7.59$\error{0.10}$  \\
NGC3738A & B12  & 87$\error{44}$  & 12707$\error{1053}$  & --
 & 11187$\error{927}$  & 0.48$\error{0.13}$  & 0.64$\error{0.18}$
 & 8.14$\error{0.09}$  \\
NGC3738B & B12  & 92$\error{71}$  & 11353$\error{2004}$  & --
 & 11854$\error{2092}$  & 0.72$\error{0.45}$  & 0.57$\error{0.28}$
 & 8.20$\error{0.16}$  \\
UGC6817A & B12  & 89$\error{46}$  & 17457$\error{772}$  & --
 & 15767$\error{698}$  & 0.06$\error{0.01}$  & 0.24$\error{0.05}$
 & 7.48$\error{0.08}$  \\
NGC4163A & B12  & 48$\error{48}$  & 18162$\error{3095}$  & --
 & 17072$\error{2909}$  & 0.19$\error{0.08}$  & 0.04$\error{0.01}$
 & 8.11$\error{0.09}$  \\
CGCG269-049C & B12  & --  & 18852$\error{4262}$  & --  & 18508$\error{4184}$
 & 0.08$\error{0.04}$  & 0.09$\error{0.04}$  & 7.35$\error{0.14}$  \\
CGCG269-049A & B12  & 87$\error{49}$  & 18265$\error{983}$  & --
 & 16115$\error{867}$  & 0.06$\error{0.01}$  & 0.20$\error{0.04}$
 & 7.41$\error{0.08}$  \\
UGC7577A & B12  & 114$\error{65}$  & 11028$\error{673}$  & --
 & 12829$\error{782}$  & 0.36$\error{0.08}$  & 0.77$\error{0.17}$
 & 8.05$\error{0.08}$  \\
NGC4449C & B12  & 84$\error{54}$  & 13373$\error{1525}$  & --
 & 9995$\error{1140}$  & 0.50$\error{0.18}$  & 0.74$\error{0.29}$
 & 8.29$\error{0.13}$  \\
NGC4449B & B12  & 84$\error{72}$  & 12439$\error{1312}$  & --
 & 9662$\error{1019}$  & 0.56$\error{0.20}$  & 0.91$\error{0.35}$
 & 8.29$\error{0.13}$  \\
NGC4449A & B12  & 90$\error{74}$  & 9398$\error{296}$  & --  & 9046$\error{285}$
 & 1.83$\error{0.33}$  & 1.56$\error{0.34}$  & 8.44$\error{0.07}$  \\
UGC7605A & B12  & 25$\error{32}$  & 17452$\error{2378}$  & --
 & 13604$\error{1853}$  & 0.14$\error{0.04}$  & 0.28$\error{0.10}$
 & 7.65$\error{0.12}$  \\
NGC4656A & B12  & 17$\error{70}$  & 10127$\error{1205}$  & --
 & 11885$\error{1414}$  & 0.22$\error{0.11}$  & 1.23$\error{0.46}$
 & 8.16$\error{0.15}$  \\
UGC8201A & B12  & 74$\error{117}$  & 16203$\error{2326}$  & --
 & 13343$\error{1915}$  & 0.13$\error{0.05}$  & 0.38$\error{0.14}$
 & 7.71$\error{0.13}$  \\
  \hline
\pagebreak
\hline \hline
System  & Sample  & $N\sub{e}$  & \TOII{}  & \TNII{}  & \TOIII{}  & \OpH{}
 & \OppH{}  & $Z\sub{Te,semi-direct}$ \\
   &    & cm$^{-3}$  & K  & K  & K  & $\times 10^{4}$  & $\times 10^{4}$  &   \\
\hline
UGC8508A & B12  & 53$\error{69}$  & 15753$\error{1668}$  & --
 & 13963$\error{1479}$  & 0.12$\error{0.03}$  & 0.37$\error{0.11}$
 & 7.69$\error{0.10}$  \\
UGC8638A & B12  & 84$\error{67}$  & 12586$\error{286}$  & --
 & 12276$\error{279}$  & 0.29$\error{0.02}$  & 0.68$\error{0.12}$
 & 7.98$\error{0.06}$  \\
UGC8638B & B12  & 189$\error{260}$  & 12970$\error{1915}$  & --
 & 12315$\error{1818}$  & 0.23$\error{0.11}$  & 0.68$\error{0.28}$
 & 7.96$\error{0.15}$  \\
UGC8837A & B12  & 84$\error{43}$  & 18654$\error{1582}$  & --
 & 12841$\error{1089}$  & 0.23$\error{0.05}$  & 0.18$\error{0.05}$
 & 8.05$\error{0.07}$  \\
NGC5477A & B12  & 111$\error{64}$  & 12431$\error{1199}$  & --
 & 12025$\error{1160}$  & 0.21$\error{0.07}$  & 0.81$\error{0.25}$
 & 8.01$\error{0.12}$  \\
KISSR148 & H15  & 70$\error{47}$  & 11158$\error{711}$  & --
 & 9586$\error{611}$  & 1.01$\error{0.27}$  & 0.94$\error{0.26}$
 & 8.44$\error{0.09}$  \\
KISSR258 & H15  & 168$\error{98}$  & 12073$\error{1375}$  & --
 & 22508$\error{2564}$  & 0.34$\error{0.14}$  & 0.05$\error{0.01}$
 & 7.92$\error{0.08}$  \\
KISSR1056 & H15  & 106$\error{67}$  & 11138$\error{488}$  & --
 & 10494$\error{460}$  & 0.89$\error{0.18}$  & 0.82$\error{0.19}$
 & 8.31$\error{0.08}$  \\
KISSR2132 & H15  & 90$\error{58}$  & 11316$\error{606}$  & --
 & 10993$\error{589}$  & 1.25$\error{0.30}$  & 0.64$\error{0.16}$
 & 8.43$\error{0.07}$  \\
NGC628+252.1-92.1 & CHAOS  & 88$\error{46}$  & 15262$\error{1176}$
 & 13829$\error{2034}$  & 11489$\error{885}$  & 0.29$\error{0.06}$
 & 0.48$\error{0.13}$  & 8.02$\error{0.09}$  \\
NGC628+254.3-42.8 & CHAOS  & 72$\error{38}$  & 16125$\error{1296}$  & --
 & 11291$\error{907}$  & 0.32$\error{0.08}$  & 0.40$\error{0.11}$
 & 8.14$\error{0.09}$  \\
NGC628+261.9-99.7 & CHAOS  & 56$\error{65}$  & 14681$\error{1569}$  & --
 & 10713$\error{1145}$  & 0.45$\error{0.15}$  & 0.49$\error{0.19}$
 & 8.28$\error{0.12}$  \\
NGC5457+1.0+885.8 & CHAOS  & 66$\error{37}$  & 18157$\error{519}$  & --
 & 12624$\error{361}$  & 0.17$\error{0.02}$  & 0.28$\error{0.05}$
 & 7.85$\error{0.06}$  \\
NGC5457+324.5+415.8 & CHAOS  & 74$\error{38}$  & 12078$\error{264}$  & --
 & 10563$\error{231}$  & 0.65$\error{0.05}$  & 0.70$\error{0.13}$
 & 8.29$\error{0.06}$  \\
NGC5457-397.4-71.7 & CHAOS  & 63$\error{26}$  & 15334$\error{712}$  & --
 & 9011$\error{418}$  & 0.31$\error{0.04}$  & 0.85$\error{0.20}$
 & 8.25$\error{0.08}$  \\
NGC5457-8.5+886.7 & CHAOS  & 40$\error{30}$  & 19612$\error{729}$  & --
 & 12505$\error{465}$  & 0.13$\error{0.01}$  & 0.26$\error{0.05}$
 & 7.76$\error{0.07}$  \\
MMT10 & L16  & 113$\error{77}$  & 12373$\error{534}$  & --  & 13873$\error{598}$
 & 0.21$\error{0.03}$  & 0.59$\error{0.12}$  & 7.90$\error{0.07}$  \\
MMT11 & L16  & 1147$\error{504}$  & 10133$\error{555}$  & --
 & 13209$\error{724}$  & 0.71$\error{0.16}$  & 0.63$\error{0.14}$
 & 8.11$\error{0.07}$  \\
MMT15 & L16  & 114$\error{83}$  & 11004$\error{1870}$  & --
 & 22255$\error{3782}$  & 0.37$\error{0.24}$  & 0.11$\error{0.03}$
 & 7.71$\error{0.10}$  \\
MMT16 & L16  & 279$\error{363}$  & 13513$\error{2909}$  & --
 & 23462$\error{5051}$  & 0.10$\error{0.06}$  & 0.12$\error{0.04}$
 & 7.33$\error{0.12}$  \\
MMT18 & L16  & 53$\error{38}$  & 15317$\error{2511}$  & --
 & 22886$\error{3752}$  & 0.07$\error{0.03}$  & 0.10$\error{0.03}$
 & 7.22$\error{0.10}$  \\
MMT19 & L16  & --  & 12045$\error{507}$  & --  & 14206$\error{598}$
 & 0.19$\error{0.03}$  & 0.61$\error{0.12}$  & 7.91$\error{0.07}$  \\
MMT26 & L16  & 53$\error{27}$  & 12121$\error{1259}$  & --
 & 14574$\error{1514}$  & 0.56$\error{0.20}$  & 0.25$\error{0.07}$
 & 8.12$\error{0.08}$  \\
MMT28 & L16  & --  & 11241$\error{2668}$  & --  & 17746$\error{4211}$
 & 0.23$\error{0.21}$  & 0.37$\error{0.19}$  & 7.78$\error{0.19}$  \\
MMT32 & L16  & 121$\error{73}$  & 9724$\error{798}$  & --  & 10349$\error{850}$
 & 1.05$\error{0.38}$  & 1.19$\error{0.36}$  & 8.31$\error{0.10}$  \\
SN2010kd & SLSN  & --  & 13352$\error{2792}$  & --  & 19336$\error{4043}$
 & 0.09$\error{0.06}$  & 0.26$\error{0.10}$  & 7.54$\error{0.14}$  \\
SN2011ke & SLSN  & --  & 14672$\error{647}$  & --  & 17091$\error{754}$
 & 0.08$\error{0.01}$  & 0.29$\error{0.06}$  & 7.57$\error{0.07}$  \\
LSQ14an & SLSN  & 102$\error{76}$  & 10477$\error{865}$  & --
 & 13310$\error{1099}$  & 0.88$\error{0.31}$  & 0.67$\error{0.18}$
 & 8.12$\error{0.09}$  \\
PTF10bfz & SLSN  & 22$\error{56}$  & 13730$\error{2157}$  & --
 & 16231$\error{2549}$  & 0.16$\error{0.08}$  & 0.32$\error{0.12}$
 & 7.68$\error{0.13}$  \\
SN2012il & SLSN  & 151$\error{125}$  & 12020$\error{940}$  & --
 & 13802$\error{1080}$  & 0.25$\error{0.07}$  & 0.60$\error{0.15}$
 & 7.93$\error{0.09}$  \\
PTF09as & SLSN  & 45$\error{62}$  & 11755$\error{1508}$  & --
 & 15227$\error{1953}$  & 0.25$\error{0.11}$  & 0.49$\error{0.16}$
 & 7.87$\error{0.11}$  \\
SN2010gx & SLSN  & 168$\error{236}$  & 13560$\error{1876}$  & --
 & 19868$\error{2749}$  & 0.09$\error{0.04}$  & 0.22$\error{0.07}$
 & 7.50$\error{0.11}$  \\
GAIA16apd & SLSN  & 36$\error{67}$  & 13235$\error{1174}$  & --
 & 13572$\error{1204}$  & 0.14$\error{0.04}$  & 0.60$\error{0.18}$
 & 7.87$\error{0.11}$  \\
\hline \hline
\label{tab:LitSamp_TeandAbund_data_semidirect}
\end{longtable}

\twocolumn[]

\section{Linear fitting methods} \label{sec:FittingMethods}
We adopt two distinct methods when making linear fits to the three key relations discussed in this work, namely, the \ZTe{} -- $a$ relation (Eqn. \ref{eqn:semiaxis}), our \OIII{}/\OII{}-based semi-direct \ZTe{} correction (Eqn. \ref{eqn:correction_factor}), and the galaxy MZR (Eqn. \ref{eqn:lowz_MZR_fit}). An overview of these two fitting methods is provided below.

\subsection{Least-squares fitting}\label{sec:A1}
The first method we adopt is the linear least-squares approximation provided by the IDL routine \texttt{MPFITEXY} \citep{Williams+10}, which depends on the \texttt{MPFIT} package \citep{Markwardt09}. This routine utilises the $\chi^{2}$ minimisation technique to obtain a straight-line best fit to the data, accounting for errors in both the ordinate and abscissa axes and a determination of the intrinsic scatter which is adjusted to ensure the reduced $\chi^{2} \sim{}1$ using the method outlined in \citet{Bedregal+06}.

\subsection{Nested Sampling fitting}\label{sec:A2}
In addition to the linear least-squares fitting method outlined above, we also applied a simple Bayesian fit using nested sampling.
For this we used the \texttt{dynesty} package \citep{Speagle19}. For deriving the posterior we used dynamic nested sampling \citep{Higson+17}.

\subsubsection{Linear models}\label{sec:A2.1}

For the comparison between the \ZTe{} -- $a$ relation and the \ZTe{} -- \TOIII{} and \ZTe{} -- \TOII{} relations (\S \ref{sec:New TT relation}), we use the same prior in all cases. We note that the derived evidence depends very strongly on the prior, and that improper priors can lead to inaccurate results.

In general, each relation combines one or more observable $X$ with the desired quantity $Z$, such that for each observable we have a dataset of $X_i$ and $Z_i$.

We assume that each measurement, $X_{i,\text{obs}}$, is drawn from a normal distribution centered at the value of the model function, $f_x$, for this measurement pair with an effective standard deviation value. This gives,
\begin{equation}
  Z_{i,\text{obs}} \sim \mathcal{N}(f_x(X_{i,\text{obs}}, \theta), \sigma_{Z,i,\text{eff}})
\end{equation}

The effective standard deviation $\sigma_{Z,i,\text{eff}}$ is defined as follows and includes the standard deviation of both observables as well as an additional scattering term $f$,

\begin{equation}
    \sigma_{Z,i,\text{eff}} = \sqrt{\sigma_{Z,i,\text{obs}}^2 + \left( \frac{\partial f_x(X_{i,\text{true}}, \theta)}{\partial X_{i,\text{true}}} \right)^2 \sigma_{X,i,\text{obs}}^2 + f_x(X_{i,\text{obs}},\theta)^2\, f^2 }
\end{equation}

Where $X_{i,\text{true}}$ is the true value of $X_i$, which $X_{i,\text{obs}}$ is sampled from (conceptually). Since sampling an $N+3$ dimensional cube is quite expensive, we approximated the partial derivative using the observed value. Since the range for values for $Z$ are typically quite small, other errors (systematic uncertainties, for instance) will dominate over the approximation error. In the case where $\partial f_x / \partial X_{i,\text{true}}$ does not depend on $X_{i,\text{true}}$, the solution is exact (as is the case for the simple linear relations).

For most cases in the paper we use the following model function:
\begin{equation}
    f_x (X, \theta\{\alpha,\beta\}) = \alpha X + \beta
\end{equation}

This is used for the various temperature -- metallicity fits, as well as the MZR fit. We use weak uninformative priors for the parameters:
\begin{align}
    \ln \alpha^{-1} &\sim \mathcal{N} (-8, 10) \\
    \beta &\sim \mathcal{N} (0, 1000) \\
    \ln f &\sim \mathcal{U} (-21, 1)
\end{align}

When using the information from the posterior, one has to keep in mind the whole covariance matrix,
\begin{equation}
  \mathbb{C_{\alpha,\beta}} = \left(\begin{array}{ll}
    \sigma_\alpha^2 & \sigma_\alpha \sigma_\beta \rho \\
    \sigma_\alpha \sigma_\beta \rho & \sigma_\beta^2
  \end{array} \right)
\end{equation}

While the off-axis elements between most parameters are mostly zero, the slope $\alpha$ and intercept $\beta$ are highly correlated.

\subsubsection{\ZTe{} correction factor}
To fit the semi-direct \ZTe{} correction factor discussed in \S \ref{sec:correcting_ZTe}, we used a fairly simple linear model that assumes all measurements below a certain critical \OIII{}/\OII{} ratio, $x_c \sim \mathcal{U} (-0.5, 1.0)\ [\text{dex}]$, under-estimate the metallicity by a \OIII{}/\OII{}-dependent factor. The location of the cutoff $x_c$ was left as a free parameter with a non-informative prior.

We assumed the following relation for this correction factor:
\begin{equation}
  f_\text{cor}(x, \theta\{\alpha, x_c\}) = \bigg{\{}  \begin{array}{ll}
    \alpha\,(x-x_c) & \text{for}\,x \leq x_c \\
    0 & \text{for}\,x > x_c\;\;\;,
    \end{array}
\end{equation}
which is equivalent to Eqn. \ref{eqn:correction_factor}.

\end{document}